\documentclass[fleqn,reqno]{amsart}

\usepackage{comment}
\usepackage{dirtytalk}
\usepackage{mathtools}
\usepackage{bbm}
\usepackage{dsfont}
\usepackage{relsize}
\usepackage{exscale}
\usepackage[version=4]{mhchem}

\usepackage{setspace}

\newcommand{\Tn}{T}
\newcommand{\Hbo}{H_{\rm bo}}
\newcommand{\oH}{\overline{H}}
\newcommand{\oK}{\overline{K}}
\newcommand{\oV}{\overline{V}}
\newcommand{\oU}{\overline{U}}

\newcommand{\oE}{\overline{E}}
\newcommand{\oT}{\overline{T}}
\newcommand{\heff}{h_{\rm eff}}

\newcommand{\cHel}{\mathcal{H}_{\rm el}}

\newcommand{\norm}[1]{\lVert{#1}\rVert}

\allowdisplaybreaks

\usepackage{tikz}

\newtheorem{innerassumption}{Assumption}
\newenvironment{assumption}[1]
  {\renewcommand\theinnerassumption{#1}\innerassumption}
  {\endinnerassumption}

\usepackage[hidelinks, colorlinks=true, linkcolor=blue, citecolor=red]{hyperref}
\usepackage{bm}
\usepackage{amsfonts,amssymb,amsthm,amsmath}
\usepackage{epsfig}
\usepackage{graphicx}
\usepackage{mathrsfs}
\usepackage{color} 
\usepackage[margin=1in]{geometry}
\usepackage{cases}

\theoremstyle{plain}
\newtheorem{theorem}{Theorem}
\newtheorem{proposition}[theorem]{Proposition}
\newtheorem{lemma}[theorem]{Lemma}

\newtheorem*{remark}{Remark}

\theoremstyle{definition}

\makeatletter
\newtheorem*{rep@theorem}{\rep@title}
\newcommand{\newreptheorem}[2]{%
\newenvironment{rep#1}[1]{%
 \def\rep@title{#2 \ref{##1}}%
 \begin{rep@theorem}}%
 {\end{rep@theorem}}}
\makeatother

\newreptheorem{theorem}{Theorem}
\newreptheorem{lemma}{Lemma}

\numberwithin{theorem}{section}
\numberwithin{equation}{section}

\newcommand{\R}{\mathbb{R}}
\newcommand{\C}{\mathbb{C}}

\newcommand{\N}{\mathbb{N}}

\newcommand{\cC}{{\mathcal{C}}}

\newcommand{\cD}{{\mathcal{D}}}

\newcommand{\cH}{{\mathcal{H}}}
\newcommand{\cK}{{\mathcal{K}}}
\newcommand{\cL}{{\mathcal{L}}}

\newcommand{\cR}{{\mathcal{R}}}

\newcommand{\nc}{\newcommand}
\nc{\e}{\epsilon}
\nc{\al}{\alpha}
\nc{\be}{\beta}
\nc{\del}{\delta}
\nc{\G}{\Gamma}
\nc{\g}{\gamma}
\nc{\ka}{\kappa}
\nc{\lam}{\lambda}
\nc{\Lam}{\Lambda}
\nc{\Om}{\Omega}
\nc{\om}{\omega} 
\nc{\ta}{\tau}
\nc{\w}{\omega}
\nc{\io}{\iota}
\nc{\h}{\theta}
\nc{\z}{\zeta}
\nc{\s}{\sigma}
\nc{\sig}{\sigma}
\nc{\Si}{\Sigma}
\nc\vphi{{\varphi}}

\nc\lb{\lambda}
\nc\eps{\epsilon}

\nc\cchi{{\check \chi}}

\newcommand{\oP}{{\overline{P}}}
\newcommand{\oR}{{\overline{R}}}

\nc{\1}{{\bf 1}}
\newcommand{\lan}{\langle}
\newcommand{\ran}{\rangle}

\newcommand{\one}{{\bf 1}}

\newcommand{\rRan}{{\rm{Ran\, }}}
\newcommand{\Ran}{{\rm{Ran\, }}}

\newcommand{\p}{\partial}

\newcommand{\n}{\nabla}

\newcommand{\rRe}{{\rm{Re\, }}}

\newcommand{\nrm}[1]{\left|#1\right|}

\renewcommand{\part}{{\rm part}}
\nc\matter{{\rm matter}}

\nc\hc{{\rm h.c.}}
\newcommand{\DETAILS}[1]{}

\usepackage[style=nature,sorting=nty]{biblatex}
\begin{document}

\title[Time-Dependent Born-Oppenheimer approximation]{On the Time-Dependent Born-Oppenheimer Approximation}
\author{S. Gherghe, I. Moyano and I.M. Sigal} 

\begin{abstract}
In this paper, we consider the time-dependent Born-Oppenheimer approximation (BOA) of a classical quantum molecule involving a possibly large number of nuclei and electrons, described by a Schr\"odinger equation. In the spirit of Born and Oppenheimer's original idea we study quantitatively the approximation of the molecular evolution. We obtain an iterable approximation of the molecular evolution to arbitrary order and we derive an effective equation for the reduced dynamics involving the nuclei equivalent to the original Schr{\"o}dinger equation and containing no electron variables. We estimate the coefficients of the new equation and find tractable approximations for the molecular dynamics going beyond the one corresponding to the original Born and Oppenheimer approximation.
\end{abstract}

\maketitle


\section{The Problem and Main Results} \label{Prelim_Problem_MainResult_Section}

The Born-Oppenheimer theory \cite{BoOp} is the bedrock of molecular physics and quantum chemistry. It is used, among other things, to compute the shapes of molecules and their spectra. Consider a molecule consisting of \(M\) nuclei with charges \(Z_1, \dots, Z_M\) and masses \(m_1, \dots, m_M\), and \(N\) electrons with charge \(-1\) and mass \(m\). The associated Schr{\"o}dinger operator (the quantum Hamiltonian) is given by
\begin{equation} \label{Hmol}
  H_{mol} = - \sum_{j=1}^N \frac{1}{2}\Delta_{x_j}
  - \sum_{j=1}^M \frac{1}{2m_j} \Delta_{y_j}
  + V_e(x) + V_{en}(x,y) + V_n(y),
\end{equation}
where \(x = (x_1,\ldots, x_N) \in \mathbb{R}^{3N}\) and \(y = (y_1, \ldots, y_M) \in \mathbb{R}^{3M}\) are, respectively, the coordinates of the electrons and nuclei. \(V_e(x)\) represents the potential of the electron-electron interaction, while \(V_{en}(x,y)\) and \(V_n(y)\) denote the electron-nuclei and nuclei-nuclei interaction potentials, respectively (see, for instance, \cite{SchiffQM, MessiahQM, alma991106901804106196, SO2012, alma991106292291806196}). We will explicitly write out \(V_e(x)\), \(V_{en}(x,y)\), and \(V_n(y)\) later on.

The operator $H_{mol}$ acts on the system's state space \(L^2_{p.sym.}(\mathbb{R}^{3(N+M)})\), the permutation symmetry subspace of \(L^2(\mathbb{R}^{3(N+M)})\).

According to the Born-Oppenheimer theory, to find the shapes and spectra of molecules, one proceeds as follows:
\begin{itemize}
    \item First, one finds the ground state (smallest) energy \(E(y)\) of the electrons in the external electrostatic potential of nuclei frozen at given positions \(y := (y_1, \dots, y_M) \in \R^{3M}\), i.e., the smallest eigenvalues of the operators 
    \begin{equation}
        H_{bo}(y) = - \sum_{j=1}^N \frac{1}{2}\Delta_{x_j} + V_e(\cdot) + V_{en}(\cdot,y) + V_n(y), \qquad \text{on }L^2_x = L^2_{sym}(\R^{3N}),
    \end{equation}
    depending on (frozen) nuclear positions \(y \in \R^{3M}\), for each \(y \in \R^{3M}\).
    \item Secondly, this energy \(E(y)\) is considered as the potential for the nuclear motion, leading to the second spectral problem associated with the operator
    \begin{equation}
        K = - \sum_{j=1}^M \frac{1}{2m_j} \Delta_{y_j} + E(y), \qquad \text{on }L^2_y = L^2(\R^{3M}).
    \end{equation} 
    \item The minima of \(E(y)\) give the possible equilibrium configurations of the molecule (see also \cite{LT}), and the nuclear oscillations around these equilibria give the low energy spectrum.
\end{itemize}

According to the Born-Oppenheimer approximation (BOA), the low energy spectrum of \(H_{mol}\) and therefore EM radiation frequencies should be close in some sense to the low energy spectrum of \(K\), which is a considerably simpler operator. (For example, for a molecule of ozone \ce{O3}, \(3(N+M) = 81\), while \(3M = 9\).)

The BO theory is based on the observation that, since the electrons are much lighter than the nuclei (the mass ratio \(\leq 1/1000\)), they move much faster and adjust to changing nuclear configurations \say{instantaneously}. It has been refined in numerous works, e.g., \cite{CDS1981, Ha-tindep-1, Ha-tindep-2, KMSW, GS}, where the authors obtain precise error estimates valid to higher-order approximations. This is a stationary, or time-independent, theory.

Proceeding to the time-dependent case, we recall that the evolution in time of the molecule is given by the Schrödinger equation,
\begin{equation} \label{SE}
	\left\{ \begin{array}{ll}
    i \partial_s \Psi = H_{mol} \Psi, & \\
	\Psi \vert_{s=0} = \Psi_0, &
	\end{array} \right.
\end{equation}
where \(\Psi_0 \in L^2_{sym}(\mathbb{R}^{3(N+M)})\) and \(H_{mol}\) is given by \eqref{Hmol}. The stationary states of this equation are called bound states. They correspond to various eigenfunctions of \(H_{mol}\) and are the primary objects of the stationary (time-independent) theory briefly mentioned above.

In the time-dependent case, the BOA relates—in some approximation—the evolution in time of the full molecular system (in \(x,y\) variables) with the decoupled evolution of the nuclear variables only. The solution to the full molecular dynamics (\ref{SE}), given by \(H_{mol}\), is reduced to the solutions to the dynamics associated with \(K\), i.e.,
\begin{equation} \label{SEreduced}
	\left\{ \begin{array}{ll}
    i \partial_s \psi = K \psi, & \text{in } L^2_{y}, \\
	\psi \vert_{s=0} = \psi_0 \in L^2_{y}. &
	\end{array} \right.
\end{equation}
The advantage of this approximation is obvious, as the dynamics described by (\ref{SEreduced}) deals with considerably fewer particles and, therefore, dimensions. Moreover, (\ref{SEreduced}) describes slow motion controlled by the small parameter \(\kappa^2\), which is the ratio of the electron and nuclear masses,
\begin{equation}
    \kappa^2 := \frac{m_e}{\min_j m_j} = \frac{1}{\min_j m_j},
\end{equation}
where \(m_e = 1\) in atomic units.

There is considerable literature on the time-dependent BO theory; see \cite{HagJoye_BO_overview, Teufel, PST, ST_2001adiabatic, ST2002, PST2003, Ha-tdep-1, Ha-tdep-2, Ha-tdep-3, HaJ-tdep-1, HaJ-tdep-2, GS, jecko:hal-00803621} for reviews.

In this paper, we pursue the original Born-Oppenheimer idea and extend it to molecular evolution. We derive an effective equation for the nuclear motion equivalent to the original Schrödinger equation, containing no electron variables. Then, we estimate the coefficients of the new equation and find tractable approximations for it.

In conclusion of this overview, we mention that the reduction described above is expected to break down at very large times. Thus, one has to choose a time scale in which to approximate the molecular dynamics. A standard time scale usually considered is \(O(1/\kappa)\). In this paper, we adopt this timescale. Hence, we write \(s = t/\kappa\), with \(t = O(1)\), and rescale (\ref{SE}) appropriately to obtain
\begin{equation} \label{rescaledSE}
\left\{ \begin{array}{ll}
    i \kappa \partial_t \Psi = H_{mol} \Psi, & \text{in } L^2_{x,y}, \\ 
    \Psi \vert_{t=0} = \Psi_0 \in L^2_{x,y}, &
    \end{array} \right.
\end{equation}
with \(\Psi = \Psi(x,y,t)\).

In what follows, we formulate our results first in an abstract setting, and then specify them to the concrete problem at hand.

\subsection{The abstract setting}



Let us denote the abstract \emph{nuclear} variables by $y= (y_1, \dots, y_m) \in \R^m$, for some $m \in \mathbb{N}$, fixed but arbitrary. Let us fix a Hilbert space $\cH_{el} $ for the electronic configuration. Consider the Hilbert space $\cH = L^2(\R^m, \cH_{el})$ for the whole (abstract) molecular dynamics. To separate the nuclear variable $y \in \R^m$ from the electronic variables in $\cH_{el}$, we use the formalism of direct integrals of Hilbert spaces (following, for instance, \cite[pp 15-17]{Nielson}, \cite[pp 280-287]{RSv4}, \cite[pp 48-68]{HoeverThesis}, and \cite[p. 15]{jecko:hal-00803621}). 

We consider the whole space $\cH = L^2(\R^m, \cH_{el})$ where the molecular dynamics take place as the fibered space of square integrable functions of $y \in \R^m$ with values in the Hilbert space $\cH_{el}$. A vector $\Psi \in \cH$ can be written as the vector direct integral
\begin{equation}
    \Psi = \int_{\R^m}^\oplus \Psi(y) dy, 
\end{equation}
so that the usual inner product in $\cH$ reads
\begin{equation}
    \langle \Psi, \Phi \rangle = \int_{\R^m} \langle \Psi(y), \Phi(y) \rangle_{\cH_{el}} dy, \qquad \qquad \Psi, \Phi \in \cH.
\end{equation}
With this notation, we can write $\cH$ as a constant fiber direct integral or a direct integral of Hilbert spaces,
\begin{equation}
    \cH = \int_{\R^m}^\oplus \cH_{el} dy.
\end{equation}
Given a family of operators $\left\{ A(y)\right\}_{y\in \R^m}$ on $\cH_{el}$, we can define an operator $A$ acting on $\cH$ by setting 
\begin{equation}
    (A\Psi)(y) = A(y)\Psi(y), \qquad y\in \R^m, \qquad \Psi \in \cH,
\end{equation}
whenever this definition makes sense. We call such operators \textit{fibered} (over $\R^m$) and write them as
\begin{equation}
    A = \int^{\oplus}_{\R^m} A(y) dy,
\end{equation}
with $A(y)$ called the \textit{fiber} operators.

In the spirit of the BOA described above, let $H_\ka$ be the operator family on the space $\cH$ given by
\begin{equation} \label{Hkappa_def}
    H_\ka = - \ka^2 \Delta_y + H_{bo}, \qquad \:\textrm{ with }\: H_{bo} = \int_{\R^{m}}^\oplus H(y) dy,
\end{equation}
where $H_{bo}$ is an abstract Born-Oppenheimer operator, meaning that we assume $H(y)$ is a smooth family of self-adjoint operators with isolated and simple lowest eigenvalues $E(y)$ and associated eigenvectors $\psi_\circ(y)$. For simplicity, we assume $H(y)$ is real. Let $P(y)$ denote the eigenprojection associated with $E(y)$, i.e., the orthogonal projection onto the ground state $\psi_\circ(y)$. \par

Now, consider the Schr{\"o}dinger equation associated with this abstract \emph{molecular} Hamiltonian in the \emph{molecular} space $\cH$, i.e.,
\begin{equation} \label{abstractSE}
\left\{ \begin{array}{ll}
    i \ka \partial_t \Psi = H_{\ka} \Psi, & \textrm{in } \cH,   \\
	\Psi \vert_{t=0} = \Psi_0 \in \cH. &
\end{array} \right.
\end{equation}
We seek the \textit{effective nuclear dynamics} in the nuclear variables $y\in \R^m$ only. To find an effective nuclear dynamics, we consider the reduction induced by the projections $\left\{P(y)\right\}_{y\in \R^m}$ on the solutions to (\ref{abstractSE}). This amounts to considering time-dependent paths in $L^2_y$ in the following way. Writing $ t \mapsto \Psi(t) = \int^{\oplus}_{\R^m} \Psi(y,t) dy$ in $\mathcal{H}$, we define a corresponding path $t \mapsto f(t)$ in $L^2_y$ as 
\begin{equation} \label{defoff}
    f(y, t) := \langle \psi_\circ(y), \Psi(y, t) \rangle_{\cH_{el}}.
\end{equation}
The function $f(y, t)$ can be interpreted as the state of the nuclei at time $t$, given that the electrons are in the ground state. We expect the function $f(y, t)$, associated with $\Psi(t)$, solution to the Schr{\"o}dinger equation (\ref{abstractSE}), to satisfy another Schr\"odinger equation, the reduced dynamics, that could provide insights into the BOA. More precisely, one expects $f$ to satisfy a Schr{\"o}dinger equation of the form 
\begin{equation}
\left\{ \begin{array}{ll}
    i \ka \partial_t f = \heff^\ka f, & \\ 
    f |_{t=0} = f_0\in L^2_y, & 
\end{array} \right.
\label{effdynamics_abstract}
\end{equation}
where the \textit{effective nuclear Hamiltonian} $\heff^\ka$ is to be determined in terms of the abstract molecular Hamiltonian (\ref{Hkappa_def}). It will turn out that $\heff^\ka$ is a non-local operator acting on functions $f \in C(\R, L^2(\R^m))$. Once the effective nuclear Hamiltonian is found, we explain how to reconstruct the original molecular wave function $\Psi(t)$ from the reduced effective nuclear wave function $f(t)$.


\subsection{Effective nuclear dynamics in the abstract setting}

Let $H_{\ka}$ be given by \eqref{Hkappa_def}, and let
\begin{equation}
    U_t = e^{- i H_\ka t / \ka},
\end{equation}
be the associated propagator, and 
\begin{equation}
    T = - \ka^2 \Delta_y, \qquad \textrm{in } \R^m,
\end{equation} 
where $\Delta_y$ is the usual (rescaled) Laplacian in nuclear variables.

Our first result is the time-dependent Born-Oppenheimer correction to first order. Let $(\psi_\circ f)(y) = \psi_\circ(y) f(y)$ and $\lan t \ran = (1 + \nrm{t}^2)^{1/2}$. For $s \in \N$ the spaces $H^s_{\ka,y}$ and $H^s_{\ka,y} \cH_{el}$ are defined in Section \ref{assumptions_section}. For $s,s' \in \N$, we shall say that the operator $F : H^s_{\ka,y} \cH_{el} \rightarrow H^{s'}_{\ka,y} \cH_{el}$ is $O_{\cL_{s,s'}}(a),$ for some $a \geq 0$ if there exists a constant $C>0$ such that 
\begin{equation}
    \Vert F \phi \Vert_{H^s_{k,y}\cH_{el}} \leq C \Vert \phi \Vert_{H^{s'}_{k,y}\cH_{el}}, \qquad \forall \phi \in H^s_{k,y}\cH_{el}.
\end{equation}

With the assumptions formulated in Section \ref{assumptions_section} and the previous notation, we have our first result.

\begin{theorem} \label{TDBO_FIRSTORDER_MAINTHEOREM}
(Time-dependent BOA to order $\ka$.) Let Assumptions \ref{assumption1}-\ref{assumption4} below hold. Then, for any $f \in H^{2}_{\ka,y}$, we have 
\begin{equation} \label{H=Hbo+T_propagator_duhamel}
     U_t \psi_\circ f = \psi_\circ \left( e^{-i \heff t /\ka}  f \right) + O_{\cL_{2,0}}(\ka \lan t \ran^3),
\end{equation}
where 
\begin{equation} \label{heff_def_0order}
    \heff = \Tn + E + \ka^2 v,
\end{equation}
with $E$ and $v$ being multiplication operators by the functions $E(y)$ and
\begin{equation} \label{v-expr1}
    v(y):= \frac{1}{2} \norm{\n_{y} \psi_\circ}^2_{\cH_{el}}.
\end{equation}
\end{theorem}

Theorem \ref{TDBO_FIRSTORDER_MAINTHEOREM} is proven in Section \ref{Section_proof of Thm-Propagators-Duhamel}. It states that the solution of the Schr{\"o}dinger IVP \eqref{abstractSE} with the initial condition $\Psi_0 = \psi_\circ f_0$ can be approximated, up to the order $O_{\cL_{2,0}}(\ka \lan t \ran^3)$, by $\Psi(t) = \psi_\circ f(t)$, where $f(t)$ is the solution of the effective equation
\begin{equation} \label{0ordereffectiveequation}
    i \ka \p_t f = \heff f, \qquad f |_{t=0} = f_0.
\end{equation}

Our next result gives the effective nuclear dynamics (with effective nuclear Hamiltonian $\heff^\ka$) to an arbitrary order beyond the leading order given in \eqref{heff_def_0order}. 

Let $\oR$ denote the reduced resolvent 
\begin{equation}
    \oR = \oP ((\Hbo - E)|_{\Ran \oP})^{-1} \oP, 
\end{equation}
where $\oP = 1 - P$, for the projections $P = \int^\oplus_{\R^m} P(y) dy$ defined by Assumptions \ref{assumption1} - \ref{assumption3}. The reduced resolvent $\oR$ is well-defined thanks to Assumptions \ref{assumption1} - \ref{assumption3} (see Lemma \ref{(Hbar - E)inverselemma}). Let us introduce the operator 
\begin{equation} \label{K_Rbar_Hbar_defs}
\oH = \oP H_{\ka} \oP,
\end{equation} 
which, by self-adjointness (see Appendix \ref{SelfAdjointness_Section}), generates the propagator
\begin{equation} \label{unitaries_Ut_def}
    \oU_t = e^{- i \oH t / \ka}, \qquad t \in \R.
\end{equation} 
Given an interval $I \subseteq \mathbb{R}$, let us define the map $Q_P : C(I,L^2_y) \to C(I, \cH)$ as
\begin{align}
    &(Q_P f)(t) := \psi_\circ f(t) - \frac{i}{ \kappa } \int_0^t \oU_{t-s} \overline{P} H_{\ka} P\psi_\circ f(s) ds, \qquad t \in I, \label{Qpexpr}
\end{align}
where we denote $f(t)(y) = f(y,t)$ and $\Psi(t)(y) = \Psi(y,t) \in \cH_{el}$. The main result of this paper is the following theorem.

\begin{theorem} \label{MAINTHEOREM_ABSTRACT-PREVIOUS}
(Effective nuclear dynamics.) Let Assumptions \ref{assumption1}-\ref{assumption4} below hold. Let $H_{\ka}$ be given by \eqref{Hkappa_def}. Let $\Psi_0 = \psi_\circ f_0$ for some $f_0 \in L^2_y$. We have the following:
\par 
\vspace{0.5em}   
(a) If $\Psi = \Psi(t)$ satisfies the Schr{\"o}dinger equation (\ref{abstractSE}) on $\cH = L^2(\R^m, \cH_{el})$ with $\Psi(0) = \Psi_0$, then $f(t) := \langle \psi_\circ, \Psi(t) \rangle_{\mathcal{H}_{el}}$ (see \eqref{defoff}) obeys the Schr{\"o}dinger equation (\ref{effdynamics_abstract}),
\begin{equation} \label{effectiveeqforf}
    i \ka \p f = \heff^\ka f, \qquad f(0) = f_0,
\end{equation}
with the effective (non-local) nuclear Hamiltonian
\begin{equation} \label{heff_maintheorem}
    \heff^\ka = T + E + \ka^2 v + w^\ka,
\end{equation} 
where $E$ and $v$ are multiplication operators by the functions $E(y)$ and \eqref{v-expr1},
and $w^{\ka}$ is the non-local operator acting on functions $f \in C(\R, L^2(\R^m))$ given by
\begin{equation}
w^\ka[f](t) = - \frac{i}{ \kappa} \int_0^t \langle \psi_\circ, P \Tn \overline{P} \:\overline{U}_{t-s} \overline{P} \Tn P \psi_\circ f(s) \rangle_{\cH_{el}} ds.  \label{rewritew}
\end{equation} 
Conversely, if $f(t)$ obeys (\ref{effectiveeqforf}) with the effective nuclear Hamiltonian $\heff^\ka$, then $Q_Pf(t)$ obeys (\ref{rescaledSE}) with initial conditions $Q_Pf(0) = \psi_\circ f_0$.

\par 
\vspace{0.5em}   
(b) Let $n \geqslant 2$ be an integer. Assume that $f_0 \in H^{s+2n}_{\ka ,y}$, for some integer $0 \leqslant s \leqslant k_A - 2n - 2$, and let $f$ be the corresponding solution of \eqref{effectiveeqforf} in $B^{s+2n+2}_\tau := L^\infty(0,\tau; H^{s+2n+2}_{\ka,y})$ (see \eqref{BsT_norm_def}) for some $\tau>0$. Then, for every $0 \leq t \leq \tau$, the electronic feedback operator $w^\ka$, defined in \eqref{rewritew}, acting on $f$ admits the following expansion:
\begin{equation} \label{wf(t)_expansion_Oka3}
     w^\ka[f](t) = \sum_{j=1}^{n-1} (- i \ka)^{j+1} \left(w_jf(t) - \tilde w_j(t) f_0 \right) + (-i\ka)^{n+1} w^\ka_n[f](t),
\end{equation} 
where, for $j \geqslant 1, w_j$ is an operator satisfying the estimate
\begin{equation} \label{wjf(t)_estimate}
    \norm{w_j u}_{H^s_{\ka ,y}} \lesssim \norm{u}_{H^{s+j+1}_{\ka,y}}, \qquad \forall u \in H^{s+j+1}_{\ka,y},
\end{equation} 
$\tilde w_j(t)$ is a local operator family on $L^2(\R^m)$ satisfying the estimates
\begin{equation} \label{tilde wjf(t)_estimate}
    \norm{\tilde w_j(t) u}_{H^s_{\ka ,y}} \lesssim e^{C \tau} \norm{u}_{H^{s+2j}_{\ka,y}}, \qquad \forall u \in H^{s+2j}_{\ka,y}, \qquad 0 \leq t \leq \tau,
\end{equation} 
and $w^{\ka}_n[f]$ is a non-local in $t$ operator satisfying the estimate
\begin{equation} \label{w3f(t)_estimate}
    \norm{w^{\ka}_n [f](t)}_{H^s_{\ka,y}} \lesssim e^{C \tau} \left( \norm{f}_{B^{s+2n+2}_\tau} + \norm{f_0}_{H^{s+2n}_{\ka ,y}} \right), \qquad 0 \leq t \leq \tau.
\end{equation} 
The operators $w_1$, $\tilde w_1(t)$, and $w_2^\ka$ are given explicitly in \eqref{w1_def} - \eqref{wkappa2_def}, respectively. The higher order operators can be computed explicitly using the expansion scheme described in Section \ref{sec:Expansion Atf}.
\end{theorem}

In part (b) of Theorem \ref{MAINTHEOREM_ABSTRACT-PREVIOUS} above, we have assumed that if $f_0 \in H^{s+2n}_{\ka ,y}$, for some $s>0$ and $n \geqslant 1$, the corresponding solution $f$ of \eqref{effectiveeqforf} exists and satisfies $ f \in {B^{s+2n+2}_\tau}$ for some $\tau > 0$. While the existence of such $f$ is clear from the existence of the full dynamics $\Psi$, the existence of the effective dynamics stemming from $f_0$ alone and satisfying $ f \in {B^{s+6}_T}$ is not completely straightforward. This can be addressed using the theory of existence of solutions of integro-differential Volterra type equations (see \cite{GRIMMER1985189, Desch_Schappacher, Oka1995, MR0882711, Pruss_1983} and for a comprehensive review, \cite[Section 2]{10.1093/imamci/dnac004}).

Now we address the question of effective nuclear dynamics at higher orders. In light of part (b) of Theorem \ref{MAINTHEOREM_ABSTRACT-PREVIOUS}, this is now possible. Our next theorem deals with estimating the effective nuclear dynamics at second order using the expansion \eqref{wf(t)_expansion_Oka3}. In other words, the solution $\Psi(t)$ of the Schr{\"o}dinger IVP \eqref{abstractSE} with the initial condition $\Psi_0 = \psi_\circ f_0$ has the form 
\begin{equation}
    \Psi(t) = Q_P \tilde{f}(t) + O_{\cL_{7,0}}(\ka^2 e^{Ct}),
\end{equation}
for some constant $C > 0$, where $\tilde{f}(t)$ is the solution of the effective equation given below.

\begin{theorem} \label{TDBO_SECONDORDER_MAINTHEOREM}
(Effective nuclear dynamics to order $\ka^2$.) Let Assumptions \ref{assumption1}-\ref{assumption4} below hold with $H_{\ka}$ given by \eqref{Hkappa_def}. Let $\Psi(t)$ satisfy the Schr{\"o}dinger equation (\ref{abstractSE}) on $\cH = L^2(\R^m, \cH_{el})$ with $\Psi(0) = \psi_\circ f_0$ for some $f_0 \in H^{7}_{\ka ,y}$.
Then, there exists a $\ka_0 = \ka_0 (\delta, m, \norm{\nabla_{y_j} H_{bo}}, \dots, \norm{\nabla_{y_k} \psi_\circ}, \dots)$ such that for all $\ka \in (0, \ka_0)$, there exists a constant $C > 0$ such that
\begin{equation} \label{TDBOA_secondorder_estimate}
    \norm{\Psi(t) - Q_P \tilde{f}(t)} \leqslant \ka^2 C e^{C \tau} \norm{f_0}_{H^{7}_{\ka ,y}}, \qquad 0 \leqslant t \leq \tau,
\end{equation}
where $\tilde{f}(t)$ is a solution of the autonomous equation 
\begin{equation} \label{effeq_2ndorder}
    i \ka \p_t \tilde{f} = \heff^{(2)} \tilde{f}, \qquad \heff^{(2)} := T + E + \ka^2 v - \ka^2 w_1,
\end{equation}
with the initial condition $\tilde{f} |_{t=0} = f_0$, where $w_1$ (see \eqref{wf(t)_expansion_Oka3}) is given explicitly as 
\begin{equation} \label{tildeC_w2(0)_estimate}
    w_1 = -\frac{1}{\ka^2} \lan \psi_\circ, P T \oP \ \oR \ \oP T P \psi_\circ \ran_{\cH_{el}}, \qquad w_1 = O_{\cL_{s+2,s}}(1). 
\end{equation}
As $T + E = O_{\cL_{2,0}}(1)$, the operator $T + E$ is the time-dependent BOA to first order. For $\ka \in (0, \ka_0)$, the operator $\heff^{(2)}$ is self-adjoint on $L^2_y$ with domain $H^{2,\ka}_y$.
\end{theorem}

Theorems \ref{MAINTHEOREM_ABSTRACT-PREVIOUS} and \ref{TDBO_SECONDORDER_MAINTHEOREM} state that $\ka^2 v - \ka^2 w_1$ is the correct second-order time-dependent Born-Oppenheimer correction. By iterating the expansion \eqref{wf(t)_expansion_Oka3} to higher orders in $\ka$ (see \eqref{wkappa_expansion}), higher order corrections to the time-dependent BOA can be derived, and Theorem \ref{TDBO_SECONDORDER_MAINTHEOREM} can be extended to these higher orders.

Using simpler methods, in \eqref{effeq_2ndorder} we have obtained the same effective second-order correction as in \cite{PST, PST2003, Teufel}. In the physics literature, the correct second-order Born-Oppenheimer Hamiltonian was first obtained by Weigert and Littlejohn in \cite{WL1993} for matrix-valued Hamiltonians.

\begin{remark}
    The Sobolev order in the statement \eqref{TDBOA_secondorder_estimate} depends on the order of truncation. The corresponding statement for higher order truncations of the operator $w^\ka$ according to \eqref{wf(t)_expansion_Oka3} will require more regularity, note \eqref{wjf(t)_estimate} - \eqref{w3f(t)_estimate}. 
\end{remark}

\subsection{Assumptions in the abstract setting} \label{assumptions_section}

In order to track the dependence with respect to $\kappa$, we introduce the re-scaled differential operators (in nuclear variables)
\begin{equation}
    D_{y} := -i\kappa \partial_{y}, \qquad  D^\alpha = \prod_{j=1}^m D^{\alpha_j}_{y_j},
\end{equation}
and we denote $\partial_y^\alpha = \partial^{\alpha_1}_{y_1} \dots \partial^{\alpha_m}_{y_m}$ for any $\alpha = (\alpha_1, \dots, \alpha_m) \in \N^m$ with $\nrm{\alpha} = \sum_{j=1}^m \alpha_j$. Writing $L^2_y \equiv L^2(\R^m)$, we introduce the $\kappa$-scaled Sobolev spaces:
\begin{align}
    H_{\ka,y}^s &= \{\phi \in L^2(\R^m) \::\: \norm{\phi}_{H^{s, \kappa}_y}^2 = \sum_{\nrm{\alpha} = s} \norm{D_{y}^\alpha \phi}_{L^2_y}^2 + \norm{\phi}_{L^2_y}^2 < \infty\}. \label{Hsk_y_def}
\end{align} with the usual convention $H^{0}_{\ka, y} = L^2_y$. 


\begin{assumption}{[A1]} \label{assumption1}
The operators $H(y)$ are self-adjoint on $\cH_{el}$ with a dense domain $\cD_\circ$ independent of $y$, and there exist constants $\gamma$ and $k_A \geqslant 7$, independent of $y$, such that
\begin{equation} \label{H(y)bounded_below}
   \langle \phi, (H(y) + \gamma) \phi \rangle \geq  \gamma \Vert \phi \Vert^2, \qquad \phi \in \cH,
\end{equation} 
and
\begin{equation} \label{H(y)differentiable_assumption}
    \norm{\partial_y^{\alpha} H(y)}_{\cL(\cH_{el})} \leqslant C(\alpha), \quad \forall \alpha \text{ with } 1 \leqslant \nrm{\alpha} \leqslant k_A \text{ and } y \in \R^m,
\end{equation} 
where the constant $C(\alpha)$ depends only on $\alpha$.
\end{assumption}

\begin{assumption}{[A2]} \label{assumption2}
For all $\ka > 0$, the operators $H_{\kappa}$ are self-adjoint on $\cH$ with a dense domain $\cD(H_\ka)$ independent of $\ka$,
\begin{equation}
    \cD(H_{\ka}) = H^{2,\ka}_y \otimes  \cH_{el} \cap L^2_y \otimes \cD_\circ.
\end{equation}
Here, $\cD_\circ$ is the $y$-independent domain of the fibers $H(y)$ from Assumption \ref{assumption1}.
\end{assumption}

\begin{assumption}{[A3]} \label{assumption3}
The operators $H(y)$ have unique ground states, which we denote $\psi_\circ(y)$. The corresponding non-degenerate ground state energy, $E(y)$, satisfies
\begin{equation} \label{BO-ev}
 H(y)  \psi_\circ(y) = E(y) \psi_\circ(y), \qquad \norm{\psi_\circ(y)}_{\cH_{el}} = 1, \qquad \forall y \in \R^m.
\end{equation} 
Since $H(y)$ is real, we can assume $\psi_\circ(y)$ to be real. Let $P(y)$ denote the eigenprojection associated with $E(y)$, i.e., the orthogonal projection onto the ground state $\psi_\circ(y)$ of $H(y)$, 
\begin{equation} \label{abstract_defp1}
(P\Psi)(y) = \psi_\circ(y) \langle \psi_\circ(y), \Psi(y) \rangle_{\cH_{el}}.
\end{equation} It is easy to see that $P$ is an orthogonal projection on $\cH = L^2(\R^m, \cH_{el})$ of infinite rank. We write $\textrm{Ran }P$ as $\psi_\circ \otimes L^2_y$ and identify it with $L^2_y$. The projection $P$ defines a map from $\cH$ to $\psi_\circ \otimes L^2_y$. We further define $\overline{P}:=1-P$, the orthogonal complement to $P$.
\end{assumption}

\begin{assumption}{[A4]} \label{assumption4}
There is a gap between the ground state energy $E(y)$ of $H(y)$ and the rest of the spectrum of $H(y)$, which is positive uniformly for $y \in \R^m$, i.e., there exists a $\delta > 0$ such that
\begin{equation} \label{gapconditionuniformly}
    \inf_{y \in \R^m} \left\{ \vert \xi - E(y)   \vert;  \,  \xi \in \sigma(H(y)) \setminus \{E(y)\} \right\} \geqslant \delta > 0.
\end{equation}
\end{assumption}

\subsection{Main results in the molecular case}
\label{main-molecular-case}

Now, we specify the above construction to a molecular system as described by \eqref{Hmol}. We view the space $L^2(\R^{3(M+N)})$ as the space $L^2(\R^{3M}, \cH_{el}),$ where $\cH_{el} = L^2(\R^{3N})$. We re-write \eqref{Hmol} as
\begin{equation}
    H_{mol} = \Tn + \Hbo, \quad \textrm{ with } \quad H_{bo} = \int_{\mathbb{R}^{3M}}^\oplus H(y) dy, \label{Hmol=Hbo+Tnucl}
\end{equation}
where the operators $H(y)$ and $T$ acting on $\cH_{el} = L^2_x \equiv L^2(\R^{3N})$ and $L^2_y$ respectively are defined as 
\begin{align}
    &H(y) = - \sum_{j=1}^N \frac{1}{2} \Delta_{x_j} + V_e(x) + V_{en}(x,y) + V_n(y), \label{Hbo(y)def}\\
    &\Tn = - \sum_{j=1}^M \frac{1}{2 m_j} \Delta_{y_j} = -\ka^2 \sum_{j=1}^M \frac{1}{2 m_j'} \Delta_{y_j}, \qquad m_j' = \ka^2 m_j = \frac{m_j}{\min_{k} m_k}. \label{Tnucldef}
\end{align}
Note that, by introducing the inner product 
\begin{equation} \label{T=Nablay2_innerproduct}
    (y, \tilde{y}) = 2 \sum_{j=1}^M m_j' y_j \cdot \tilde{y}_j
\end{equation}
in $\R^{3M}$, the operator $\Tn$ can be written as 
\begin{equation}
    \Tn = - \ka^2 \Delta_y,
\end{equation}
where $\Delta_y$ is the Laplace-Beltrami operator in the metric \eqref{T=Nablay2_innerproduct}.

The electrons have charges $-e$, and the nuclei have charges $Z_j e$, where $j = 1, \dots, M$. The electrons are modeled as point charges, and the electronic repulsion is given by the Coulomb potential energy
\begin{equation} \label{Ve}
    V_e(x) = \sum_{i=1}^{N-1} \sum_{j=i+1}^N \frac{e^2}{\lvert{x_i - x_j}\rvert}.
\end{equation}

For technical and physical reasons, we shall model the nuclei as \textit{smeared} rigid charge distributions $\rho \in C^\infty_c(\mathbb{R}^3)$, $\rho \geqslant 0$, and $\lVert{\rho}\rVert_{L^1} = 1$, as opposed to the \textit{point} charges. These are referred to in the literature as \textit{form-factors} (see \cite{GS} and \cite{Teufel}). The potential for nuclear repulsion becomes:
\begin{equation} \label{Vn}
    V_n(y) = \sum_{i=1}^{M-1} \sum_{j=i+1}^M \int_{\mathbb{R}^6}  \frac{e^2 Z_i Z_j \rho(z - y_i) \rho(z' - y_j)}{\lvert{z - z'}\rvert} dz dz',
\end{equation}
and the attractive potential between electrons and nuclei becomes:
\begin{equation} \label{Ven}
    V_{en}(x,y) = - \sum_{i=1}^M \sum_{j=1}^N \int_{\mathbb{R}^3}  \frac{e^2 Z_i\rho(z - y_i)}{\lvert{z - x_j}\rvert} dz. 
\end{equation}
Observe that $V_n$ and $V_{en}$ are bounded and that $x \mapsto V_{en}(x,\cdot)$ is smooth, and $V_{en}(x,\cdot) \in L^\infty_y$ for all $x \in \R^{3N}$.

We apply Theorems \ref{TDBO_FIRSTORDER_MAINTHEOREM} - \ref{TDBO_SECONDORDER_MAINTHEOREM} to the molecular system given by \eqref{rescaledSE} with \eqref{Hmol=Hbo+Tnucl} - \eqref{Ven}, identifying $m = 3M$ and $\cH_{el} = L^2\left( \R^{3N} \right)$ to obtain the following theorems.

\begin{theorem} \label{TDBO_FIRSTORDER_MAINTHEOREM_MOL}
(Time-dependent BOA to order $\ka$ for molecules.)
Let Assumptions \ref{assumption3} and \ref{assumption4} hold. Then, for any $f \in H^{2,\ka}_y$, we have \eqref{H=Hbo+T_propagator_duhamel}, where $\heff = \Tn + E + \ka^2 v$, with $E$ and $v$ being multiplication operators by the functions $E(y)$ and $v(y):= \frac{1}{2} \norm{\n_{y} \psi_\circ}^2_{\cH_{el}}$. 
\end{theorem}


\begin{theorem} \label{maintheorem_MOL} 
(Effective nuclear dynamics for molecules.) Let Assumptions \ref{assumption3} and \ref{assumption4} hold. Let $H_{mol}$ be given by \eqref{Hmol=Hbo+Tnucl}. Let $\Psi_0 = \psi_\circ f_0$ for some $f_0 \in L^2_y$. We have the following:
\par 
\vspace{0.5em}   
(a) If $\Psi = \Psi(t)$ satisfies the Schr{\"o}dinger equation (\ref{SE}) on $L^2(\R^{3(N+M)})$ with $\Psi(0) = \Psi_0$, then $f(t) := \langle \psi_\circ, \Psi(t) \rangle_{L^2_x}$ (see \eqref{defoff}) obeys the Schr{\"o}dinger equation (\ref{effectiveeqforf}) with the effective (non-local) nuclear Hamiltonian \eqref{heff_maintheorem}-\eqref{rewritew}. Conversely, if $f(t)$ obeys (\ref{effectiveeqforf}) with the effective nuclear Hamiltonian $\heff^\ka$, then $Q_Pf(t)$ obeys (\ref{SE}) with initial conditions $Q_Pf(0) = \psi_\circ f_0$.
\par 
\vspace{0.5em}   
(b) Let $n \geqslant 1$ be an integer. Assume that $f_0 \in H^{s+2n}_{\ka ,y}$, for some integer $s \geqslant 0$, and let $f$ be the corresponding solution of \eqref{effectiveeqforf} in $B^{s+2n+2}_\tau$ for some $\tau>0$. Then, for every $0 \leq t \leq \tau$, the electronic feedback operator $w^\ka$, defined in \eqref{rewritew}, acting on $f$ admits the expansion \eqref{wf(t)_expansion_Oka3}-\eqref{w3f(t)_estimate}.
\end{theorem}


\begin{theorem} \label{TDBO_secondorder_MOL}
(Effective nuclear dynamics to order $\ka^2$ for molecules.) Let Assumptions \ref{assumption3} and \ref{assumption4} hold with $H_{mol}$ given by \eqref{Hmol=Hbo+Tnucl}. Let $\Psi(t)$ satisfy the Schr{\"o}dinger equation (\ref{SE}) on $L^2(\R^{3(N+M)})$ with $\Psi(0) = \psi_\circ f_0$ for some $f_0 \in H^{s+7}_{\ka ,y}$ for some $s > 0$, and let $\tilde{f}(t)$ be a solution of the equation \eqref{effeq_2ndorder}. Then, there exists a $\ka_0 = \ka_0(\delta, M, \norm{\nabla_{y_j} V}, \dots, \norm{\nabla_{y_k} \psi_\circ}, \dots) > 0$ such that for all $\ka \in (0, \ka_0)$, there exists a constant $C > 0$ such that \eqref{TDBOA_secondorder_estimate} holds.
\end{theorem}

To derive Theorems \ref{TDBO_FIRSTORDER_MAINTHEOREM_MOL} - \ref{TDBO_secondorder_MOL} from Theorems \ref{TDBO_FIRSTORDER_MAINTHEOREM} - \ref{TDBO_SECONDORDER_MAINTHEOREM}, we have to show that the family of operators in \eqref{Hbo(y)def} satisfies Assumptions \ref{assumption1} and \ref{assumption2}. \par 

The self-adjointness of $H_{mol}$ and $H_{bo}$ is standard (see for example \cite{GS, HS, RSv2, BSV4}). The differentiability follows from the fact that 
\begin{equation}
    \partial_y^\alpha H(y) = \partial_y^\alpha \left(V_n(y) + V_{en}(x,y) \right)
\end{equation}
where $V_n(y)$ and $V_{en}(x,y)$, given by \eqref{Vn} and \eqref{Ven}, are bounded and self-adjoint multiplication operators on $\cH_{el}$ by smooth and real functions of $y$. The semi-boundedness is also well-known (see \cite[Theorem 7.1.15]{BSV4} and the references above).



\subsection{Organization of the paper}

This paper is organised as follows. In Section \ref{nonabelian_ibp_Section} we develop the non-abelian integration by parts, NAIP, which is our central technique used to obtain Theorems \ref{TDBO_FIRSTORDER_MAINTHEOREM} - \ref{TDBO_SECONDORDER_MAINTHEOREM}. Then, in Section \ref{Section_proof of Thm-Propagators-Duhamel} we prove Theorem \ref{TDBO_FIRSTORDER_MAINTHEOREM}. After introducing the main ingredients of the proof with a minimum of technical points we proceed to close the argument and prove the theorem relying upon technical results that are proven in later Sections. In subsection \ref{Resolventestimates_Section} we show the gap estimate on the reduced resolvent $\oR = \oP\left\{\left(\Hbo - E\right)|_{\rRan \oP} \right\}^{-1} \oP$. 

In Section \ref{MainTheoremProof_Section} we prove Theorem \ref{MAINTHEOREM_ABSTRACT-PREVIOUS}. This first subsection deals with part a), which can be completed in a relatively straightforward way. In Subsections \ref{newibp_wf_Section} - \ref{proofIterative}, we focus on the proof of part (b) of Theorem \ref{MAINTHEOREM_ABSTRACT-PREVIOUS}, which also requires the NAIP developed in Section \ref{nonabelian_ibp_Section}. 

In Section \ref{TDBO_secondorder_Section} we prove Theorem \ref{TDBO_SECONDORDER_MAINTHEOREM} assuming Theorem \ref{MAINTHEOREM_ABSTRACT-PREVIOUS}. Again, we will make use of the NAIP of Section \ref{nonabelian_ibp_Section}. 


In Appendix \ref{BasicFiberEstimates_Section} we collect some basic results on fibered operators, while in Appendix \ref{SelfAdjointness_Section} we establish the self-adjointness of $P H_\ka P$ and $\oH = \oP H_\ka \oP$. Appendices \ref{commutatorestimates_Section} and \ref{Section_estimatesonPropagators} deal respectively with commutator estimates and propagators. 



\subsection{Notation} 
\label{notation}

In this section we gather some of the notation used throughout the paper. We reserve the symbols $f, g$ for elements of $C(\R, L^2(\R^m)$ and $u, v$ for functions of $L^2(\R^m)$ only. 

We denote by $\mathcal{L}(\mathcal{H})$ and $\cL(\cH_1, \cH_2)$ the spaces of bounded linear operators on a Hilbert space $\mathcal{H}$ (equipped with the standard operator norm $\lVert{\cdot}\rVert$) and of bounded linear operators from $\cH_1$ to $\cH_2$ (equipped with the operator norm).

In the rest of this document, we will write $\cH_1 \cH_2$ to denote the tensor product $\cH_1 \otimes \cH_2$ between two Hilbert spaces $\cH_1$ and $\cH_2$, equipped with the tensor product norm. We write $L^2_y = L^2(\R^m)$ and $\cH = L^2(\R^m, \cH_{el}) = L^2_y \cH_{el}$ and use the Sobolev spaces \eqref{Hsk_y_def} defined above and denote the tensor spaces
\begin{align}
    H^s_{\ka ,y} \cH_{el} &= H^s_{\ka ,y}(\R^{m}) \otimes \cH_{el} = H^s_{\ka ,y} \left(\R^{m}, \cH_{el} \right),
\end{align} 
with $\cL_{r,s} = \cL(H^{r,\ka}_y \cH_{el}, H^s_{\ka ,y} \cH_{el}).$ 

We shall make use of the time-dependent spaces $B^s_T = L^\infty(0, T; H^s_{\ka ,y})$, which are the Banach spaces equipped with norm 
\begin{equation} \label{BsT_norm_def}
    \norm{g}_{B^s_T} = \sup_{t \in [0,T]} \norm{g(t)}_{H^s_{\ka ,y}}.
\end{equation} 

The commutator of two unbounded operators on $\cH$, denoted $[\cdot,\cdot]$ and widely used throughout the article, is to be understood in the sense explained in Section \ref{commutatorestimates_Section} without explicit mention. 

Finally, $A \lesssim B$ stands for the inequality $A \le C B$, with some constant $C>0$ independent of $\kappa$. We also adopt the big $O$ notation in the sense that $A = O(\alpha)$ if there exists a constant $C > 0$ such that $\norm{A} \leqslant C \alpha$.

\subsection{Previous works and remarks}

The twiddle integration by parts (TIBP) method was introduced in \cite{Avron1987, JansenRuskaiSeiler} and applied to time-dependent BO by Teufel et al. \cite{Teufel, PST, ST_2001adiabatic, ST2002, PST2003} to derive first-order corrections. For higher order corrections, Teufel et al. made use of the pseudo-differential calculus with operator-valued symbols, which was used by other authors as well \cite{BrNo, ew96, ns04, ms02, s03}.

The novel contributions of this paper compared to earlier work are:

\begin{itemize}
    \item We develop a non-abelian integration by parts (NAIP) method, which iterates to higher orders, circumventing the need for complex pseudo-differential calculus used in prior approaches.
    \item By projecting out the electronic degrees of freedom from the full Schrödinger dynamics, we derive an effective equation for nuclear dynamics, from which approximations to all orders in $\ka$ can be obtained. The full molecular dynamics can be reconstructed or approximated from these solutions.
\end{itemize}

Our method also maintains dynamics on $\Ran P$ to higher orders in $\ka$, introducing non-local in time operators, which, though they complicate effective dynamics, avoid the expansion into larger subspaces (i.e. as required by Teufel et al.). In Theorem \ref{TDBO_SECONDORDER_MAINTHEOREM}, we match the effective second-order corrections derived by Teufel et al. (see for example \cite[Section 2.2]{PST2003}).

Hagedorn et al. follow a distinct approach, using semiclassical wavepackets to approximate solutions by combining semiclassical and adiabatic limits \cite{HagJoye_BO_overview, Ha-tdep-1, Ha-tdep-2, Ha-tdep-3, HaJ-tdep-1, HaJ-tdep-2}. Exploring the semiclassical limit of the effective nuclear dynamics (e.g., via semiclassical wavepackets or Wigner functions \cite{Lasser_Lubich_2020, Jin_Markowich_Sparber_2011}) remains an area for future work.

\begin{remark} 
Our results should extend naturally to higher-rank projections (e.g., $P$ projects onto $k$ energy bands, or the ground state is degenerate under magnetic fields). For applications of higher-order corrections and multi-band projections, see \cite{PST2003} on effective dynamics near conical intersections and in reactive scattering for $\ce{H2} + \ce{H} \rightarrow \ce{H} + \ce{H2}$. \end{remark}

\begin{remark} 
Equation \eqref{H=Hbo+T_propagator_duhamel} can be reformulated as \begin{equation} \label{TDBOA_firstorder_relation} U_t P = U^P_t P + O_{\cL_{2,0}}(\ka \lan t \ran^3), \end{equation} where $ U^P_t = e^{- i H^P t/\ka}$ and $H^P = P H P$, so $H^P (\phi f) = \phi (\heff f)$ (see Lemma \ref{lemma_PKP_and_UPt_estimate}). Alternatively, it can be written as
\begin{equation} U_t J = J \tilde{U}t + O{\cL_{2,0}}(\ka \lan t \ran^3), \end{equation} where $J : \psi_\circ \otimes f \mapsto \psi_\circ f$ and $\tilde{U}t = e^{-i \tilde H t/\ka}$ with $\tilde H (\psi\circ \otimes f) = \psi_\circ \otimes \heff f$, i.e., $\tilde H = 1 \otimes \heff$. 
\end{remark}

\begin{remark} 
The form of the potential $v$, also known as the Born-Huang potential, depends on the real-valued assumption of $\psi_\circ$ (justified by the non-degeneracy of $E(y)$). When $\psi_\circ$ is complex (e.g., under magnetic fields), we have \begin{align}\label{v} v = \norm{\n_{y} \psi_\circ}^2_{\cH_{el}} - \frac{i}{\ka} \sum_{j=1}^m (a_j D_{y_j} + D_{y_j} a_j), \end{align} where $a_j := \lan \psi_\circ, \n_{y_j} \psi_\circ \ran_{\cH_{el}}$. For a smooth, normalized family of eigenvectors $\psi_\circ$, $\textrm{Re }a_j = 0$ since
\begin{equation}
    \textrm{Re } \langle \psi_\circ, \nabla_y \psi_\circ \rangle_{\cH_{el}} = \frac{1}{2} \nabla_y \norm{\psi_\circ}^2_{\cH_{el}} = 0
\end{equation} 
for $\norm{\psi_\circ} = 1$. Thus, for real $\psi_\circ$, $a_j = 0$, simplifying $v$ to (\ref{v-expr1}). 
\end{remark}

\begin{remark} 
Assumption \ref{assumption3} holds for $H(y)$ as a Schr{\"o}dinger operator in an $L^2$ space. In many-body contexts (see \eqref{Hbo(y)def}), statistics are not included\footnote{Without statistics, Schr{\"o}dinger operators have unique, non-degenerate ground states due to the positivity of $e^{-\beta H}$ for $\beta > 0$ and the Perron-Frobenius theorem. When statistics are considered, ground state degeneracy may arise, but determining its multiplicity is complex.}. \end{remark}

\begin{remark}
Unless under very strict non-degeneracy assumptions (in the neutral or fermionic case), a global spectral gap for a generic molecule in the BO approximation was shown to exist by Gherghe \cite{Gherghe2024} only under bosonic statistics. Extending the results of this paper to accommodate a local spectral gap (within a spatial region) is another direction for future research. 
\end{remark}

\begin{remark} 
The spectral gap assumption, Assumption \ref{assumption4}, is essential for the adiabatic nature of the theory and is widely applied in similar contexts (e.g., eigenvalue perturbation theory \cite{RSv4, GD-IMS-BS, HS} or general adiabatic theories \cite{BornFock, Kato_adiabatic, Teufel}). It also influences the runtime bounds of the adiabatic theorem in quantum computing \cite{JansenRuskaiSeiler, ElgartHagedorn}.

This assumption offers two advantages. First, it allows the eigenprojection $P(y)$ to be expressed as a Riesz contour integral, linking the regularity of $P(y)$ to that of the fibers $H(y)$ (see Assumption \ref{assumption1}; this type of argument goes back to \cite{Kato_perturbation}). Second, in applying the NAIP, it enables the definition of the reduced resolvent $\oR = \oP\left\{\left(\Hbo - E\right)|_{\rRan \oP} \right\}^{-1} \oP$, which is bounded by the inverse of the spectral gap’s minimum size.

Adiabatic theories without a gap condition exist (e.g., \cite{AvronElgart, Bornemann} and in the quantum resonance context \cite{MM-IMS}), though they require direct assumptions on the regularity of $P(y)$. Additionally, $\oR$ can be defined by moving into the complex plane and using constructions such as $\lim_{\epsilon \downarrow 0} \oR(E + i \epsilon)$. 
\end{remark}


\section{Non-Abelian integration by parts} \label{nonabelian_ibp_Section}

In this section we derive an integration by parts formula that we will use in various forms to apply to operator-valued oscillatory integrals. This formula is different from the one used in \cite{Teufel, JansenRuskaiSeiler}, its main advantages being simplicity and iterability. 

\begin{lemma}[Exponent derivative representation] \label{expderrep_lemma}
Let $A$ and $B$ be anti-self-adjoint operators acting on an abstract Hilbert space $H$ such that $A-B$ is invertible and $[A,B]$ is well-defined. Then, for all $t \in \R$, the following identity holds:
\begin{equation} \label{expderrep_identity}
    e^{A t} = \partial_t \left(e^{A t}(A - B)^{-1}e^{- Bt} \right) e^{Bt} - e^{At} S,
\end{equation} 
where
\begin{equation} \label{expderrep_Sdef}
    S = (A-B)^{-1} [A, B] (A-B)^{-1}.
\end{equation}
\end{lemma}

\begin{remark}
If $[A, B] = 0$, with $D := A - B$ invertible, \eqref{expderrep_identity} yields the elementary formula $e^{Dt} = \partial_t \left( D^{-1} e^{Dt}\right)$, which is at the foundation of the pseudo-differential calculus and the stationary phase method. 
\end{remark}

\begin{proof}[Proof of Lemma \ref{expderrep_lemma}]
We compute:
\begin{align}
    \partial_t \left( e^{A t}(A-B)^{-1}e^{-Bt} \right) &= e^{A t} A (A-B)^{-1} e^{-Bt} - e^{A t} (A-B)^{-1} B e^{-Bt} \nonumber \\
    &= e^{At} e^{-Bt} + e^{At} [A, (A-B)^{-1}] e^{-Bt}. \label{expderrep_1}
\end{align}
Applying $e^{Bt}$ to both sides of \eqref{expderrep_1} from the right, we obtain \eqref{expderrep_identity} with $S = [(A-B)^{-1},A]$. To obtain the representation \eqref{expderrep_Sdef} for $S$, we use $1 = (A-B)(A-B)^{-1}$ and $1 = (A-B)^{-1}(A-B)$ to obtain
\begin{align}
    [(A-B)^{-1},A] &= (A-B)^{-1} [A, A-B] (A-B)^{-1} \nonumber \\
    &= - (A-B)^{-1} [A,B] (A-B)^{-1} =: -S.
\end{align}
This completes the proof.
\end{proof}

Identity \eqref{expderrep_identity} is applied in the situations where $[A,B]$ is of a \say{higher order} than either $A$ or $B$ and similarly for higher order commutators. Then, $S$ is small in some sense. Of course, the choice of a suitable operator $B$ is a crucial point.

\begin{lemma}[NAIP formula] \label{NAIP_lemma}
    Let $A$ and $B$ as in Lemma \ref{expderrep_lemma}. Let $t \mapsto F_t$ and $t \mapsto G_t$ be two smooth functions with values in $\cH$ or operators acting on $\cH$. Then, for all $t \geqslant 0$,
    \begin{align}
        \int_0^t G_s e^{As} F_s ds &= G_s e^{As} R F_s \bigg|_{s=0}^{s=t} + \int_0^t G_s e^{As} S F_s ds \nonumber \\ 
        &\hspace{15pt} - \int_0^t \left( G_s' e^{As} R F_s + G_s e^{As} R \left[B F_s + F_s' \right] \right) ds, \label{abstract_NAIP}
    \end{align}
    where we write $R = (A-B)^{-1}$ and recall $S = R [A, B] R$.
\end{lemma}

\begin{proof}[Proof of Lemma \ref{NAIP_lemma}]
    Using Lemma \ref{expderrep_lemma}, for all $s \geqslant 0$ we have 
    \begin{equation} \label{NAIP_1}
        G(s) e^{As} F(s) = G(s) \frac{d}{ds} \left[ e^{As} R e^{-Bs} \right] e^{Bs} F(s) + G(s) e^{As} S F(s).
    \end{equation}
    Then, we integrate over the interval $[0,t]$ and integrate by parts.
   Using $e^{-Bs} e^{Bs} = 1$ and
   \begin{align}
       e^{-Bs} \frac{d}{ds}\left[ e^{Bs} F_s \right] = e^{-Bs} \left[B e^{Bs} F_s + e^{Bs} F_s'\right] = B F_s + F_s',
   \end{align}
   (\ref{abstract_NAIP}) follows. %
\end{proof}

We will also make use of a left-handed version of the NAIP. The proof follows in the same way. 

\begin{lemma}[Left-handed NAIP]\label{NAIP_lemma_lefthanded}
    Let $A$ and $B$ as in Lemma \ref{expderrep_lemma} and $t \mapsto F_t$ and $t \mapsto G_t$ as in Lemma \ref{NAIP_lemma}. We have 
    \begin{equation} \label{expderrep_identity_lefthanded}
        e^{-A t} = e^{-Bt} \partial_t \left(e^{B t}(A - B)^{-1}e^{- At} \right) + S e^{-At},
    \end{equation} 
    where $S$ is as in Lemma \ref{expderrep_lemma}, and 
    \begin{align}
        \int_0^t F_s e^{-As} G_s ds &= F_s R e^{-As} G_s \bigg|_{s=0}^{s=t} - \int_0^t F_s S e^{-As} G_s ds \nonumber \\ 
        &\hspace{15pt} - \int_0^t \left( \left[F_s' - F_s B \right]  R e^{-As} G_s + F_s R e^{-As} G_s' \right) ds. \label{abstract_NAIP_lefthanded}
    \end{align}
\end{lemma}

\section{Proof of Theorem \ref{TDBO_FIRSTORDER_MAINTHEOREM}} \label{Section_proof of Thm-Propagators-Duhamel}

\subsection{Outline of the proof} \label{1storder_subsection} We now describe the approach and main steps of the proof of Theorem \ref{TDBO_FIRSTORDER_MAINTHEOREM} and outline the proof. 

Given a solution $f$ to the effective dynamics generated by $\heff$, we want to compare the evolution of $\psi_\circ f$ induced by the propagator $U_t$, acting in all variables, with the evolution of $f$ restricted to the nuclear variable. This amounts at considering the difference 
\begin{equation}
    D(t) := (U_t - U_t^P )P, \qquad \qquad t \geq 0, 
\end{equation} where we recall that $P$ stands for the (fibered) projections 
\begin{equation} \label{abstract_defp2}
    P = \int^{\oplus}_{\R^m} P(y) dy, \text{ with } P(y) = |\psi_\circ(y) \rangle \langle \psi_\circ(y) |,
\end{equation} or $P \Psi = \int^\oplus \psi_\circ(y) \langle \psi_\circ(y), \Psi(y) \rangle_{\cH_{el}} dy.$ Now, if $f = P\Psi$, we want to estimate the difference $D(t)f$ acting on solutions to the effective Hamiltonian and show that this difference is of order $\kappa$ in the adequate spaces. 

Thanks to Lemma \ref{lemma_PKP_and_UPt_estimate} and Lemma \ref{lemma_Ut-UPt_Xt}, we show that the difference $D(t)f$ can be recast as the integral in time of some operator-valued quantities. The next step is to prove that such integral terms are small using the NAIP of Lemma \ref{NAIP_lemma}, which relies crucially on identity \eqref{expderrep_identity}. On physical grounds suggested by the BO approximation, we shall apply \eqref{expderrep_identity} to the self-adjoint operators $A = \frac{i}{\ka} \oH$, $B = \frac{i}{\ka} \oK$ and the corresponding propagators\footnote{
The operators $\oK = \oP K \oP$ and $\oH$ are shown to be self-adjoint in Appendix \ref{SelfAdjointness_Section}.}
\begin{equation}
    e^{As} = e^{i\oH s/\ka} =: \oU_{-s}, \qquad \oH = \oP H_\ka  \oP, 
\end{equation} 
and 
\begin{equation}
    e^{Bs} = e^{i \oK s/\ka} =: \oV_{-s}, \qquad \oK = \oP K \oP, \qquad K = E + T. 
\end{equation} This choice, relying strongly on the physical grounds suggested by the BO approximation, is possible and effective since 
\begin{equation}
    (A-B)^{-1} = -i\ka \oP (\oH_{bo} - \oE)^{-1}\oP =: -i\ka \oR
\end{equation} is the reduced resolvent, which is well-defined (thanks to our assumptions and Lemma \ref{(Hbar - E)inverselemma} of Section \ref{Resolventestimates_Section}). Further, under the gap condition \eqref{gapconditionuniformly} and by the definition of $\oP$, we have 
\begin{equation}
    E(y) \in \rho(H(y)|_{\rRan \oP}), \qquad \forall y \in \R^m,
\end{equation} and the following result\footnote{This result is standard in adiabatic perturbation theory and we sketch the proof in Lemma \ref{smoothness_psicirc_lemma}.}.

\begin{lemma} \label{smoothness_proposition}
Let Assumptions \ref{assumption1}-\ref{assumption4} hold. Then, the ground state energy $E(y)$ and corresponding eigenprojection $P(y)$ are $k_A$-times differentiable in $y$ (with $k_A$ the same as in Assumption \ref{assumption1}):
\begin{equation} \label{smoothness_E_P_statement}
    E(\cdot) \in C_b^{k_A}(\R^m), \quad P(\cdot) \in C_b^{k_A}(\R^m, \cL(\cH_{el})).
\end{equation}
\end{lemma}

We further require the existence and smoothness of the reduced resolvent $\oR$ as in the following result\footnote{We defer the proof of this lemma to Subsection \ref{Resolventestimates_Section}.}.

\begin{lemma} \label{(Hbar - E)inverselemma}
Let Assumptions \ref{assumption1} - \ref{assumption4} hold. Then, the operator $\oH_{\rm bo} - \oE : \overline{\mathcal{D}} \to \rRan{\oP}$ is invertible and its inverse $\oR$
is bounded uniformly in $y \in \R^m$ and satisfies
\begin{equation} \label{oR_estimate}
    \lVert{\overline{R}}\rVert_{\cL(\rRan{\oP})} \leqslant \frac{1}{\delta}, \qquad \oR = O_{\cL_{s,s}}(1),
\end{equation}
where $\delta$ is given in (\ref{gapconditionuniformly}) and $0 \leqslant s \leqslant k_A$.
Moreover, 
\begin{equation} \label{Rbar(y)_fibers}
    \oR = \int^{\oplus}_{\R^m} \oR(y) dy,
\end{equation}
with the $\R^m$-fibers $\oR(y)$ satisfying $\oR(\cdot) \in C_b^{k_A}(\R^m, \cL(\cH_{el}))$. 
\end{lemma}

Thanks to the results mentioned above, which are consequence of the main assumptions in this work, and the application of the NAIP of Lemma \ref{NAIP_lemma} to the integral terms appearing in the difference $D(t)$ (as in Lemma \ref{lemma_Ut-UPt_Xt} below), it suffices to prove that some intermediate operators (see Lemma \ref{lemma_Xt_estimate} for details) are small in terms of $\kappa$ to conclude.  

In what follows, we present the main steps of the proof as a sequence of lemmas, postponing the proof of some of the technical results to appendices.

\subsection{Representation formulas}

Our first step is the following lemma, which allows us to relate the self-adjjoint operator\footnote{The self-adjointness of $H^P = P H_\ka P$ is proven in Appendix \ref{SelfAdjointness_Section}.} $H^P$ to the effective Hamiltonian $\heff$ and also rephrase $d$ in terms of $\left(U_t - U^P_t \right) P$.

\begin{lemma} \label{lemma_PKP_and_UPt_estimate}
Let Assumptions \ref{assumption1} - \ref{assumption4} hold. 
\par 
\vspace{0.5em}
a) For all $g \in H^{2,\ka}_y$, the following identity holds:  
\begin{equation} \label{PKP_identity}
    P K P \psi_\circ g = \psi_\circ (K + \ka^2 v) g,
\end{equation}
where $K = E + T$ and $v$ is the operator given by \eqref{v-expr1} when $\psi_\circ$ is real. In particular, if $t \mapsto \Psi(t) \in \cH$ and $f$ is given by (\ref{defoff}), the operator $H^P$ satisfies 
\begin{equation} \label{HPexpression}
    H^P \Psi = \psi_\circ \heff f,
\end{equation}
where $\heff = T + E + \ka^2 v$. 
\par 
\vspace{0.5em}
b) For all $\Psi \in \Ran P$, 
\begin{equation} \label{UPt_e-ihefft/ka_relation}
    U^P_t \Psi = \psi_\circ e^{- i \heff t/\ka} f, 
\end{equation}
where $f = \lan \psi_\circ, \Psi \ran_{\cH_{el}}$ and $\heff = T + E + \ka^2 v$.
\end{lemma}


\begin{proof}[Proof of Lemma \ref{lemma_PKP_and_UPt_estimate}]
\textit{Proof of a).} We begin by observing that, if  $g \in H^{2,\ka}_y$, then 
\begin{equation*}
P(\psi_\circ g) = \psi_\circ g. 
\end{equation*} Further, as $[T, \psi_\circ]g = T(\psi_\circ g) - \psi_\circ T g$, we may write  
\begin{align} 
K P \psi_\circ g = K \psi_\circ g = \psi_\circ (E + T) g + [T, \psi_\circ] g.
\end{align} Hence, taking the projection $P$ on the left yields
\begin{align}
P K P  \psi_\circ g & = P \psi_\circ (E + T) g + P [T, \psi_\circ] g = P \psi_\circ K g + P [T, \psi_\circ] g = \psi_\circ K g + \langle \psi_\circ, [T, \psi_\circ] \rangle_{\cH_{el}} g,
\end{align} 
where we have expanded the last term using the definition of $P$. Now, we need to compute the last term in the identity above. Expanding the commutator we find
\begin{align}
[T, \psi_\circ] g = - \ka^2  (\Delta_{y} \psi_\circ) g - 2i \ka (\nabla_{y} \psi_\circ) D_{y} g,
\end{align} 
and henceforth, using that $ g \in H^{2,\ka}_y$ and is independent of the electronic variables, 
\begin{align}
\langle \psi_\circ, [T, \psi_\circ] \rangle_{\cH_{el}} g & = - \ka^2 \langle \psi_\circ, \Delta_{y} \psi_\circ \rangle_{\cH_{el}} g - 2 i \ka \lan \psi_\circ, \nabla_{y} \psi_{y} \rangle_{\cH_{el}} D_{y} g \nonumber \\ 
& = \lan \psi_\circ, \n_{y} (\n_{y}\psi_\circ) \ran_{\cH_{el}} - 2 i \ka \lan \psi_\circ, \nabla_{y} \psi_{y} \rangle_{\cH_{el}} D_{y} g \nonumber \\ 
&= \n_{y} \lan \psi_\circ,  \n_{y} \psi_\circ \ran_{\cH_{el}} - \lan \n_{y} \psi_\circ, \n_{y} \psi_\circ \ran_{\cH_{el}} - 2 i \ka \lan \psi_\circ, \nabla_{y} \psi_{y} \rangle_{\cH_{el}} D_{y} g \nonumber \\
& = - \ka^2 (\nabla_{y} a) + \ka^2 \norm{\n_{y} \psi_\circ}^2_{\cH_{el}} - 2 i \ka a D_{y} g,
\end{align} 
where we have introduced the function $ y \mapsto a(y):=\lan \psi_{y}, \n_{y}\psi_{y}\ran_{\cH_{el}}$ and used the fact that $\Delta_{y} \psi_\circ = \nabla_{y} \cdot (\nabla_{y} \psi_\circ)$. Since $\psi_\circ(y)$ is real and $\norm{\psi_\circ}^2_{\cH_{el}} = 1$ for any $y \in \R^{m}$, we have 
\begin{align}
    a = \lan \psi_\circ, \n_{y} \psi_\circ \ran_{\cH_{el}} = \frac{1}{2} \n_{y} \norm{\psi_\circ}_{\cH_{el}}^2 = 0.
\end{align} Hence, 
\begin{equation}
\langle \psi_\circ, [T, \psi_\circ] \rangle_{\cH_{el}} g = \ka^2 \norm{\n_{y} \psi_\circ}^2_{\cH_{el}}g
\end{equation} and we deduce 
\begin{equation}
P K P  \psi_\circ g  = \psi_\circ K g + \ka^2 \norm{\n_{y} \psi_\circ}^2_{\cH_{el}} g
\end{equation} 
which implies (\ref{PKP_identity}) with $v$ of the form \eqref{v-expr1}. \par 

Finally, (\ref{HPexpression}) follows from (\ref{PKP_identity}). Indeed, given $t \mapsto \Psi(t) \in \cH$ and $f$ defined by (\ref{defoff}), we have that $P \Psi =  \psi_\circ f$. Then, using (\ref{PKP_identity}) with the choice $g:= f$, we get 
\begin{align}
 H^P \Psi & = P H P \Psi =  P (K - E + H_{bo}) \psi_\circ f = \psi_\circ (K + \ka^2 v) f + P (- E + H_{bo})  \psi_\circ f.
\end{align} 
But the last term vanishes as 
\begin{align}
P H_{bo} \psi_\circ f = P H_{bo}  \psi_\circ f = P (H_{bo}  \psi_\circ) f = P E \psi_\circ f, 
\end{align} whence $P (- E + H_{bo})  \psi_\circ f = 0$ and (\ref{HPexpression}) follows.
\par

\textit{Proof of b).} The operator $\heff$ is self-adjoint on $L^2_y$ with domain $H^2_{\ka,y}$ by the standard Kato-Rellich theory, since $T$ is self-adjoint on the same domain and $E +\ka^2 v$ are bounded perturbations. Hence the semigroup $e^{- i \heff t/\ka}$ exists. 
\par

All $\Psi \in \Ran P$ can be written in the form $\Psi = \psi_\circ f_0$, for some $f_0 \in L^2_y$ (see Lemma \ref{RanPcharacterizedlemma}). Denote $\Psi(t) = U^P_t \psi_0 f_0$ and using part a), 
\begin{equation}
    i \ka \p_t \Psi(t) = H^P \Psi(t) = \psi_\circ \heff f(t),
\end{equation}
where $f(t) := \lan \psi_\circ, \Psi(t) \ran_{\cH_{el}}$. We have $f(0) = f_0$ and 
\begin{equation}
    i \ka \p_t f(t) = \lan \psi_\circ, H^P \Psi(t) \ran = \heff f(t), 
\end{equation}
so the relation \eqref{UPt_e-ihefft/ka_relation} follows. 
\end{proof}

We write the difference $\left(U_t - U^P_t \right) P$ as an operator-valued integral in time in order to integrate by parts eventually. 


\begin{lemma}\label{lemma_Ut-UPt_Xt}
Let Assumptions \ref{assumption1} - \ref{assumption4} hold. Let $X = \frac{i}{\ka} \oP T P$. 
\par 
\vspace{0.5em}
a) We have 
\begin{equation} \label{Ut-Upt=UtXt_eq}
    \left(U_t - U^P_t \right) P = U_t X_t,
\end{equation}
where 
\begin{equation} \label{At_NAIP_1}
    X_t := - \int_0^t Y_s \oU_{-s} X U^P_s ds, \quad \text{ with } \quad Y_s := \one + \int_0^s U_{-a} X^* \oU_{a} da.
\end{equation}
\par 
\vspace{0.5em}
b) Applying Lemma \ref{NAIP_lemma} to $X_t$ in \eqref{At_NAIP_1} yields
    \begin{align}
        X_t &= i \ka U_{-s} \oR X U^P_s \bigg|_{0}^t - i \ka \int_0^t U_{-s} X^* \oR X P U^P_s ds - i \ka \int_0^t U_{-s} X_2 U^P_s ds, \label{At_NAIP_4}
    \end{align}
    where 
    \begin{equation} \label{X2_S_def}
        X_2 = \frac{i}{\ka} S X + \frac{i}{\ka} \oR \left( \oK X - X K^P \right), \qquad S = \oR [H_{bo} - E , T] \oR. 
    \end{equation}
    Here $K^P = P K P$.
\end{lemma}

\begin{proof}[Proof of Lemma \ref{lemma_Ut-UPt_Xt}]
\textit{Proof of a).} We write the difference $\left(U_t - U^P_t \right) P$ as the integral of the derivative (c.f. a Duhamel formula) to obtain
\begin{align}
    \left(U_t - U^P_t \right) P =  - U_t \left(U_{-t} U^P_t - \one \right) P &= - U_t \int_0^t \frac{d}{ds} \left(U_{-s} U^P_s \right) P ds  \nonumber \\
    &= - \frac{i}{\ka} \int_0^t U_{t-s} \left( H - PHP \right) P U^P_s ds, \label{Htilde_estimate_1}
\end{align}
where we used that $U^P_s P = P U^P_s$. Observe that $H_\ka = \Hbo + T$ and $[\Hbo, P] = 0$, so we have
\begin{equation}
    \left(H_\ka - H^P \right) P = \oP H_\ka P = \oP T P.
\end{equation}
We use this on the right-hand side of \eqref{Htilde_estimate_1} and use the relation $U_{t-s} = U_t U_{-s}$ to write \eqref{Ut-Upt=UtXt_eq} with 
\begin{equation}
   X_t = - \int_0^t U_{-s} X U^P_s ds, \qquad X := \frac{i}{\ka} \oP T P. \label{Htilde_estimate_3}
\end{equation}
In order to place $X_t$ in a form where we can integrate by parts, we use the Duhamel formula again to write 
\begin{equation}
    U_{-s} \oP = \oU_{-s} \oP + \frac{i}{\ka} \int_0^s U_{-s+r} P H \oP \: \oU_{-r} dr.
\end{equation}
Making the change of variables $r \mapsto a = s - r$ inside the integral above and using $- \frac{i}{\ka} P H \oP = - \frac{i}{\ka} P T \oP = X^*$, we write 
\begin{equation} \label{Ys_def}
    U_{-s} \oP = Y_s \oU_{-s} \oP, \qquad Y_s = \one + \int_0^s U_{-a} X^* \oU_{a} da.
\end{equation}
Introducing \eqref{Ys_def} into $X_t$, we obtain \eqref{At_NAIP_1}.
\par

\textit{Proof of b).} We apply the NAIP formula Lemma \ref{NAIP_lemma} to  $X_t$ in \eqref{At_NAIP_1} with $G_s = Y_s$, $A = \frac{i}{\ka} \oH$, $B = \frac{i}{\ka} \oK$ (where recall $K = E + T$ and $\oK = \oP K \oP$), and $F_s = X U^P_s$. Hence, 
\begin{align}
    X_t &= i \ka Y_s \oU_{-s} \oR X U^P_s \bigg|_{0}^t - \int_0^t Y_s \oU_{-s} S X P U^P_s ds \nonumber \\
    &\hspace{15pt} - i \ka \int_0^t \left[ \left(\partial_s Y_s\right) \oU_{-s} \oR X U^P_s + Y_s \oU_{-s} \oR \left( \frac{i}{\ka} \oK X U_s^P + \partial_s (X U_s^P) \right) \right] ds. \label{At_NAIP_3}
\end{align}
We re-write the terms on the right-hand side into a more suitable form. First, we compute, using that $H^P = K^P \equiv P K P$,
\begin{align}
    \frac{i}{\ka} \oK X U_s^P + \partial_s (X U_s^P) &= \frac{i}{\ka} \left( \oK X - X K^P \right) U^P_s, \label{WX_n=1case}
\end{align}
Define $X_2$ as in \eqref{X2_S_def}. Using \eqref{Ys_def}, we write $Y_s \oU_{-s} \oP = U_{-s} \oP$ and $\left(\partial_s Y_s\right) \oU_{-s} \oP = U_{-s} X^*$, where recall $X = \frac{i}{\ka} \oP T P$ and so $X^* = - \frac{i}{\ka} P T \oP$. Using these relations, we can re-write the right-hand side of \eqref{At_NAIP_3} as \eqref{At_NAIP_4}. 

Finally it remains to show the second part of \eqref{X2_S_def}. Using that $\oR \: \oP = \oR = \oP \: \oR$ and $[H_{bo} - E, \oP] = 0$, we can write
\begin{align}
    S = \oR \left((H_{bo}- E) \oP K - K \oP (H_{bo} - E) \right) \oR &= \oR \left( (H_{bo} - E) K - K (H_{bo} - E) \right) \oR \nonumber \\
    &= \oR [H_{bo} - E , K] \oR.
\end{align}
Using $K = E + T$ and the fact that $E$ commutes with $H_{bo}$ and itself, we find $[H_{bo} - E , K] = [H_{bo} - E , T]$. This concludes the proof. 

\end{proof}


\begin{lemma}\label{lemma_Xt_estimate}
Let Assumptions \ref{assumption1} - \ref{assumption4} hold. Then, we have
\begin{equation} \label{S_and_X2_estimates}
    S = O_{\cL_{s+1,s}}(\ka), \qquad X_2 =  O_{\cL_{2,0}}(1),
\end{equation}
and estimating the right-hand side of \eqref{At_NAIP_4} we obtain
\begin{equation}
    X_t = O_{\cL_{2,0}}(\ka \lan t \ran^3).
\end{equation}
\end{lemma}


\begin{proof}[Proof of Lemma \ref{lemma_Xt_estimate}]
Our first step is to show the estimate for $S$ in \eqref{S_and_X2_estimates}. In Lemma \ref{commutatorwithT_lemma} we show that $[H_{bo}, T]$, and $[E,T]$ are $O_{\cL_{s+1,s}}(\ka)$. Writing $S = \oR \left([H_{bo}, T] - [E , T] \right) \oR$ and using 
\begin{equation} \label{Rbar_bound}
    \oR = O_{\cL_{s,s}}(1),
\end{equation} 
from Lemma \ref{(Hbar - E)inverselemma}, the desired estimate follows.

Now, we estimate the terms on the right-hand side of \eqref{At_NAIP_3}. Using $\oP P = 0$, we have $X = \frac{i}{\ka} \oP [T, P] P$ and $X^* = \frac{i}{\ka} P [T, P] \oP$. Then by Lemma \ref{commutatorwithT_lemma} part a), $X$ and $X^*$ are both $O_{\cL_{1,0}}(1)$. Making use of the propagator estimates 
\begin{equation} \label{Ut_UPt_estimates}
    U_t = O_{\cL_{s,s}}(\lan t \ran^s), \quad U^P_t = O_{\cL_{s,s}}(\lan t \ran^s),
\end{equation}
(see Appendix \ref{Section_estimatesonPropagators}), and \eqref{oR_estimate}, this shows that the first and second terms on the right-hand side of \eqref{At_NAIP_4} are $O_{\cL_{2,0}}(\ka\lan t \ran^3)$. In order to estimate the third term, i.e. $X_2$, we use the orthogonality of $P$ and $\oP$ combined with Lemma \ref{lemma_recursivecommutator}, whose prove can be found in the Appendix. In particular, applying Lemma \ref{lemma_recursivecommutator} with $X^\circ = \frac{i}{\ka} [T,P]$, then $X = \oP X^\circ P$ and hence
\begin{align}
   X_2 &= \frac{i}{\ka} S \oP X^\circ P + \frac{i}{\ka} \oR \left[ X^\circ, [K,P] - K \right] P U^P_s = O_{\cL_{2,0}}(1), \label{[T,P]_commutator_oPKP}
\end{align}
where the estimate is shown in Lemma \ref{C1_C2_estimates_lemma}. In combination with the estimates \eqref{Ut_UPt_estimates} this shows that the third term on the right-hand side of \eqref{At_NAIP_4} is also $O_{\cL_{2,0}}(\ka \lan t \ran^3)$. This gives $X_t = O_{\cL_{2,0}}(\ka \lan t \ran^3)$.
\end{proof}

\subsection{Conclusion of the proof}

We now conclude the proof of Theorem \ref{TDBO_FIRSTORDER_MAINTHEOREM}, modulo Lemma \ref{(Hbar - E)inverselemma}, which we defer to the Appendix.

\begin{proof}[Proof of Theorem \ref{TDBO_FIRSTORDER_MAINTHEOREM}]

Let $f$ as in the statement. Then $f = P\Psi$ and thanks to identity (\ref{HPexpression}) in Lemma \ref{lemma_PKP_and_UPt_estimate} we may write
\begin{equation*}
    U_t(\psi_\circ f ) - \psi_\circ e^{i \heff t/k } f = (U_t - U_t^P )P\Psi.
\end{equation*} Next, using Lemma \ref{lemma_Ut-UPt_Xt} we have
\begin{equation*}
    (U_t - U_t^P )P\Psi = U_t X_t \Psi,
\end{equation*} for $X_t$ defined above. Using Lemma \ref{lemma_Xt_estimate} we deduce that there exists a constant $C>0$ such that  for every $\Psi \in H^2_{\kappa,y} \cH_{el}$
\begin{equation}
    \Vert X_t \Psi \Vert_{H^2_{\kappa,y} \cH_{el}} \leq C \kappa \lan t \ran^3  \Vert \Psi \Vert_{L^2_{\kappa,y} \cH_{el}}.
\end{equation} Then, using the propagator estimate (\ref{Ut_estimate}) of Lemma \ref{Ut_Ubart_estimate_theorem}, we get 
\begin{align*}
  \Vert  U_t(\psi_\circ f ) - \psi_\circ e^{i \heff t/k } f \Vert_{H^2_{\kappa,y} \cH_{el}} &= \Vert  U_t X_t \Psi \Vert_{H^2_{\kappa,y} \cH_{el}} \lesssim  \Vert  X_t \Psi \Vert_{H^2_{\kappa,y} \cH_{el}} \lesssim C \kappa \lan t \ran^3 \Vert \Psi \Vert_{L^2_{\kappa,y} \cH_{el}}.
\end{align*} This implies the result. 

\end{proof}

\begin{remark}
One can also show that $X_t = O_{\cL_{2,0}}(\ka \lan t \ran^3)$ by using the TIBP, see \cite[Theorem 2.10]{Teufel}.
\end{remark}

\section{Proof of Theorem \ref{MAINTHEOREM_ABSTRACT-PREVIOUS}} \label{MainTheoremProof_Section}

\subsection{Proof of Theorem \ref{MAINTHEOREM_ABSTRACT-PREVIOUS}, part (a)} 

We introduce some notation to be used below and the following lemma that will be useful in the proof. We will also make use of Lemma \ref{lemma_PKP_and_UPt_estimate}.

\begin{lemma}
Let $\Psi_0 \in \cH_{el}$ and let $\Psi$ satisfy (\ref{abstractSE}). The projections $\phi(t) := P \Psi(t)$ and $ \overline{\phi}(t) := \overline{P} \Psi(t)$ satisfy
\begin{equation}
\left\{\begin{array}{ll}
 i \kappa \partial_t \phi = H^P \phi + P \Tn \overline{\phi},  & \phi(0) = P\Psi_0,   \\
    i \kappa \partial_t \overline{\phi} = \overline{H}\: \overline{\phi} + \overline{P} \Tn \phi, & \overline{\phi}(0) = \overline{P}\Psi_0.
\end{array} \right.
\label{eqns phi phi bar}
\end{equation}
\end{lemma}

\begin{proof}
We apply the projection $P$ to (\ref{abstractSE}). As $P$ commutes with $\partial_t$ we get 
\begin{equation}
i \kappa \partial_t \phi = P H_{\ka} \Psi.
\end{equation} Next, we compute
\begin{align}
    P H_{\ka} \psi = P H_{\ka} (\phi + \overline{\phi} ) = P H_{\ka} \phi + P (T + H_{bo}) \overline{\phi}
	 & = (P H_{\ka} P) P \phi + PT \overline{\phi} + P H_{bo} \overline{P} \psi \nonumber \\ 
	 & = H^P \phi + PT \overline{\phi},
\end{align} 
where we have used that $P^2 = P $ and $P H_{bo} \oP = 0$ since $P$ commutes with $H_{bo}$. As a result, we get 
\begin{equation}
i \kappa \partial_t \phi = H^P \phi + PT \overline{\phi}
\end{equation} 
and the first equation in (\ref{eqns phi phi bar}) follows. Next, by applying the projection $\overline{P}$ to (\ref{abstractSE}) and using that $P H_{bo} \oP = 0 = \oP H_{bo} P$, so that $P H \oP = P \Tn \oP$ and $\oP H P = \oP \Tn P$, arguing as before yields the second equation in (\ref{eqns phi phi bar}). 
\end{proof}

We are now in position to prove the result. 

\begin{proof}[Proof of Theorem \ref{MAINTHEOREM_ABSTRACT-PREVIOUS} (a)]

\textit{Step 1: Showing that solutions of (\ref{abstractSE}) imply solutions of (\ref{effectiveeqforf}).} Let $f_0 \in L^2_y$ and $\Psi_0 = \psi_0 f_0 $ be as in the statement and let $\Psi$ be the solution to (\ref{abstractSE}) associated to the initial datum $\Psi_0$ given above. Then, thanks to Lemma \ref{eqns phi phi bar}, $\phi(t) = P \Psi(t)$ and $\overline{\phi}(t) = \overline{P} \Psi(t)$ satisfy (\ref{eqns phi phi bar}). Moreover, if we associate to $\Psi$ the path $t \mapsto f(t)$ defined by (\ref{defoff}), we may write further 
\begin{equation}
\phi(y,t) = \psi_\circ(y) f(y,t).
\label{phi path}
\end{equation} In particular, by construction, $\overline{\phi}(0) = \overline{P} \Psi(0) = \overline{P}( \psi_0 f_0 ) = 0$. Hence, $\overline{\phi}$ satisfies the inhomogeneous Cauchy problem
\begin{equation}
i \kappa \partial_t \overline{\phi} = \overline{H}\: \overline{\phi} + \overline{P} \Tn \phi, \qquad
\phi(0) = 0.
\end{equation} 
Using the semigroup $\oU_t = \e^{-i\overline{H} t/\ka}$, $t\in \R$, and the Duhamel principle we may write
\begin{equation} \label{duhamel1}
    \overline{\phi}(t) = - \frac{i}{ \kappa } \int_0^t \oU_{t-s} \overline{P} \Tn \phi(s) ds.
\end{equation} Next, substituting (\ref{duhamel1}) into ((\ref{eqns phi phi bar})), we obtain
\begin{equation} \label{eqforphi}
    i \kappa \partial_t \phi = H^P \phi + \cK \phi,
\end{equation}
where the operator $\cK$ is defined as
\begin{equation} \label{K2eq}
    (\cK \phi)(t) = - \frac{i}{ \kappa }  P \Tn \overline{P} \int_0^t  \oU_{t-s} \overline{P} \Tn P \phi(s) ds.
\end{equation} 
Now, we want to show that the $\cK$ can be rewritten as 
\begin{equation} \label{K_1}
    (\cK \phi)(x,y,t) = \psi_\circ(x,y) w^\ka [f](y,t)
\end{equation} where $f$ is given in (\ref{defoff}) and $w $ is given by (\ref{rewritew}). Recalling the direct integral representation of $P$, we find  
\begin{align}
    (\cK \phi)(y,t) &= - \frac{i}{\kappa} \psi_\circ(y) \langle \psi_\circ(y), P \Tn \oP \int_0^t \oU_{t-s} \overline{P} \Tn P \psi_\circ(y) f(s) ds \rangle_{\cH_{el}} \nonumber \\
    &= \psi_\circ(y) \left( - \frac{i}{\kappa} \int_0^t \langle \psi_\circ(y), P \Tn \oP \ \oU_{t-s} \overline{P} \Tn P \psi_\circ(y) \rangle_{\cH_{el}} f(s) ds \right),
\end{align} 
so (\ref{rewritew}) follows. Using (\ref{HPexpression}), (\ref{eqforphi}), (\ref{K_1}), and dropping $\psi_\circ$ from equation (\ref{eqforphi}), we obtain (\ref{effectiveeqforf}).

\textit{Step 2: Showing that solutions of (\ref{effectiveeqforf} imply solutions of (\ref{abstractSE}).} Let $f$ satisfy equation (\ref{effectiveeqforf}) with initial conditions $f_0$, and let $\Psi = Q_P f$ with $\overline{\phi}_0 = 0$. Then, equation (\ref{Qpexpr}) shows that $\Psi|_{t=0} = \psi_\circ(x,y) f_0(y)$. Now, we show that $\Psi$ satisfies equation (\ref{abstractSE}). Using (\ref{effectiveeqforf}), we compute 
\begin{align}
    i \kappa \partial_t \Psi &= \psi_\circ (i \kappa \partial_t f) + \overline{H} \Psi_1 + \overline{P} \Tn \psi_\circ f = \psi_\circ \left( E + \Tn + \kappa^2 v\right) f + \psi_\circ w^\ka [f] + \overline{H} \Psi_1 + \overline{P} \Tn \psi_\circ f, \label{equiveq1}
\end{align} where 
\begin{equation*}
\Psi_1 := - \frac{i}{ \kappa } \int_0^t \oU_{t-s} \overline{P} \Tn \psi_\circ f ds.
\end{equation*}
Now, we identify the terms on the right-hand side of (\ref{equiveq1}). By equation (\ref{HPexpression}) and the relation $P \Psi = \psi_\circ f$, we have
\begin{equation}
    \psi_\circ \left( E + \Tn + \kappa^2 v \right) f = H^P \Psi. \label{match1}
\end{equation}
Next, note that $P H \overline{P} = P \Tn \overline{P}$ and hence, using (\ref{rewritew}) and the definition of $P$ in (\ref{abstract_defp1}), we find that
\begin{equation}
    \psi_\circ w^\ka [f] =  P \Tn \Psi_1 =  P H \overline{P} \Psi. \label{match2}
\end{equation}
Similarly $\overline{P} H P = \overline{P} \Tn P$, and so, by $\psi_\circ f = P \Psi$, we have
\begin{equation}
    \overline{P} \Tn \psi_\circ f = \overline{P} H P \Psi. \label{match3}
\end{equation}
Using (\ref{match1}), (\ref{match2}), and (\ref{match3}) we see that the right-hand side of (\ref{equiveq1}) equals $(P HP + P H \overline{P} + \overline{P} H \overline{P} + \overline{P} H P)\Psi = H \Psi$, and hence $\Psi = Q_P f$ satisfies the Schr{\"o}dinger equation (\ref{rescaledSE}) with initial condition $\Psi_0(x,y) = \psi_\circ(x,y) f_0(y)$. 
\end{proof}


\subsection{Main ideas of the proof of Theorem \ref{MAINTHEOREM_ABSTRACT-PREVIOUS}, part (b)} \label{newibp_wf_Section}

In this subsection we explain the main idea behind the proof of part (b) of Theorem \ref{MAINTHEOREM_ABSTRACT-PREVIOUS}. In subsequent subsections, we will close the argument and prove Theorem \ref{MAINTHEOREM_ABSTRACT-PREVIOUS} (b) while relying on Lemmas \ref{smoothness_proposition} and \ref{(Hbar - E)inverselemma}, as well as results from Appendices \ref{commutatorestimates_Section} and \ref{Section_estimatesonPropagators}.

Our goal is to decompose the electronic feedback operator $w^{\ka}$ defined in \eqref{rewritew} into a local (Markov) term and a higher order (in $\ka$) non-local one. To this end, assume that $f$ obeys the Schr{\"o}dinger equation (\ref{effdynamics_abstract}) with the effective nuclear Hamiltonian \eqref{heff_maintheorem}. Recall in previous sections we used $X = \frac{i}{\ka} \oP T P$ and its adjoint $X^* = - \frac{i}{\ka} P T \oP$. We re-write \eqref{rewritew} as 
\begin{align} 
   w^\ka[f](t) &= - i \ka \langle \psi_\circ, X^* \left(A_t f\right)(t) \rangle_{\cH_{el}}, \label{new_wf_newexpr}
\end{align} 
where we have introduced the operator
\begin{equation} \label{newA_def}
    \left(A_tf\right)(t)= \int_0^t \oU_{t-r} X \psi_\circ f(r) dr.
\end{equation} 
Now, similar to the treatment of $X_t$ in Theorems \ref{TDBO_FIRSTORDER_MAINTHEOREM}, we expand $A_t f$ further by using the NAIP. Then, each integration by parts produces a number of local and non-local terms involving commutators, with each commutator contributing $O(\ka)$, as shown in Section \ref{commutatorestimates_Section}. 

\subsection{Expansion of the electronic feedback operator $w^{\ka}$}
\label{sec:Expansion Atf}

In this section we consider integrals of the form 
\begin{equation} \label{AtCf_def}
    \left(A_t^Y f \right)(t) = \int_0^t \oU_{t-r} Y \psi_\circ f(r) dr,
\end{equation} 
where $Y: \rRan P \to \rRan \oP$ is a given operator, and study the expansion of such $A_t^Y f$ in terms of other operators of the same kind. Proposition \ref{AYjtf_ibp_expansion_lemma} is one of the cornerstones of our proof of Theorem \ref{MAINTHEOREM_ABSTRACT-PREVIOUS}, part (b) and essentially, it is an induction step for the expansion of $A_t^X f$. 

\begin{lemma} \label{AYjtf_ibp_expansion_lemma}
Let $Y: \rRan P \to \rRan \oP$. For all $t \in \mathbb{R}$, and for all $f$ solutions of \eqref{effectiveeqforf}, $A_t^{Y}f$ as given in \eqref{AtCf_def} has the following expansion: 
\begin{align}
    \left(A_t^{Y} f\right)(t) &= - i \ka \left[ \oR Y \psi_\circ f(t) - \oU_t \oR Y \psi_\circ f_0 \right] - i \ka \left(A_t^{\oR Y X^*} A^X_{(\cdot)} f \right)(t) + i \ka \left(A^{W[Y]}_t f\right)(t), \label{AtCf_ibp_expansion}
\end{align}
where the operator $Y_{j+1}$ is defined recursively through the recursion relation \eqref{Xj_recursiverelation}, i.e.
\begin{equation} \label{W(X)_def}
    W[Y] = \frac{i}{\ka} S Y + \frac{i}{\ka} \oR \left( \oK Y - Y K^P \right)P,
\end{equation}
where recall $S$ is given by \eqref{X2_S_def}. Notice that $W[Y]:\rRan P \to \rRan \oP$.
\end{lemma}

Repeated applications of the operation $W[\cdot]$ on $X = \frac{i}{\ka} \oP T P$ generates the recursive relation
\begin{equation} \label{Xj_recursiverelation}
    \begin{cases}
    X_1 &= \frac{i}{\ka} \oP T P, \\
    X_j &= \frac{i}{\ka} S X_{j-1} + \frac{i}{\ka} \oR \left(K \oP X_{j-1} - X_{j-1} P K \right) P, \quad j \geqslant 2,
    \end{cases}
\end{equation}
defining the family $\{X_j\}_{j \geqslant 1}$ of differential operators.

\begin{proof}[Proof of Lemma \ref{AYjtf_ibp_expansion_lemma}]
We apply Lemma \ref{NAIP_lemma} to integrate $A^Y_t f$ by parts, with $G_s = 1$, $A = \frac{i}{\ka} \oH$, $B = \frac{i}{\ka} \oK$, and $F_s = Y \psi_\circ f(s)$. Then, we apply $\oU_t$ to the right-hand side of \eqref{abstract_NAIP} to obtain
\begin{align} \label{AtY_NAIP_1}
    \left(A_t^{Y} f\right)(t) &= \oU_{t-r} \oR Y \psi_\circ f(r) \bigg|_{r=0}^{r=t} + \int_0^t \oU_{t-r} S Y \psi_\circ f(r) dr - \int_0^t \oU_{t-r} \oR C_r dr, 
\end{align}
where as before $S = \oR [\oH_{bo} - \oE, \oK] \oR$ and $\oR = \oP (\oH_{bo} - \oE)^{-1}\oP$, and
\begin{align} 
    C_r &:= \frac{i}{\ka} \oK Y \psi_\circ f(r) + \frac{\p}{\p r} \left(Y \psi_\circ f(r) \right).
\end{align}
Computing the time-derivative in $C_r$ and using that $f(r)$ solves \eqref{effectiveeqforf}, we write
\begin{align}
    - i \ka C_r &= \oK Y \psi_\circ f(r) - Y \psi_\circ (i \ka \partial_{r} f(r)) = \oK Y \psi_\circ f(r) - Y \psi_\circ \left(K + \ka^2 v\right) f(r) - Y \psi_\circ w^\ka[f](r). \label{At_estimate_1.5}
\end{align}
Using the identity \eqref{PKP_identity} 
and writing 
\begin{equation}
    \psi_\circ w^\ka[f](r) = - i \ka X^* \left( A_r^X f \right)(r),
\end{equation}
see \eqref{new_wf_newexpr}, on the right-hand side of \eqref{At_estimate_1.5}, we obtain
\begin{align}
    - i \ka C_r  &= \left( \oK Y - Y K^P \right) \psi_\circ f(r) - i \ka Y X^* \left( A_r^X f \right)(r). \label{At_estimate_2}
\end{align}
Substituting \eqref{At_estimate_2} into \eqref{AtY_NAIP_1} and inserting $W[Y]$ using \eqref{W(X)_def}, we can identify all four terms in \eqref{AtCf_ibp_expansion}.
\end{proof}

Recall that $X$ and $X^*$ are both $O_{\cL_{s+1,s}}(1)$. Hence, a direct estimate on $\left(A^{X}_t f \right)(t)$ would be $O(1)$, but estimating after expanding using \eqref{AtCf_ibp_expansion} gives us an estimate of $O(\ka)$. In fact, we obtain
\begin{align}
    \left(A_t^{X} f \right)(t) &= - i \ka \left[ \oR X \psi_\circ f(t) - \oU_t \oR X \psi_\circ f_0 \right] - i \ka \left(A_t^{\oR X X^*} A^X_{(\cdot)} f \right)(t) + i \ka \left(A^{X_2}_t f\right)(t), \label{AtXf_expansion}
\end{align}
where we have two local-terms and two non-local terms. The non-local terms can be expanded further by iterating the expansion to obtain higher order local corrections and non-local remainders. This is in fact necessary in order to obtain a higher order remainder, as recall by Proposition \ref{Xj_recursive_estimates_proposition}, $X_2=O_{\cL_{s+2,s}}(1)$ in $\ka$ and hence both non-local terms in \eqref{AtXf_expansion} are of the same order as the local terms, $O(\ka)$. This is the content of the Lemma \ref{w3f_expansion_lemma}.

\begin{lemma} \label{w3f_expansion_lemma}
We can expand $w^\ka[f](t)$ into the following form
\begin{align} \label{wkappa_expansion_n=2}
    w^\ka[f](t) = (-i\ka)^{2} \left( w_1 f(t) - \tilde w_1(t) f_0 \right) + (-i\ka)^{3} w^{\ka}_{2}[f](t),
\end{align}
where
\begin{align}
    w_1 f(t) &= \lan \psi_\circ, X^* \oR X \psi_\circ f(t) \ran_{\cH_{el}}, \label{w1_def} \\
    \tilde w_1(t) f_0 &= \lan \psi_\circ, X^* \oU_t \oR X \psi_\circ f_0 \ran_{\cH_{el}}, \label{tildew1(t)_def} \\
    w^{\ka}_{2}[f](t) &= - \left(w_2 f(t) - \tilde w_2(t) f_0 \right) + \int_0^t \lan \psi_\circ, X^* \oU_{t-r} \left(X_3 + \oR X X^* \oR X \right) \psi_\circ f(r) \ran_{\cH_{el}} dr \nonumber \\
    &\hspace{15pt} + \int_0^t \int_0^r \lan \psi_\circ, X^* \oU_{t-r} \oR X_2 X^* \oU_{r-s} X \psi_\circ f(s) \ran_{\cH_{el}} ds dr \nonumber \\
    &\hspace{15pt} + \int_0^t \int_0^r \lan \psi_\circ, X^* \oU_{t-r} \oR X X^* \oU_{r-s} X_2 \psi_\circ f(s) \ran_{\cH_{el}} ds dr \nonumber \\
    &\hspace{15pt} + \int_0^t \int_0^r \int_0^s \lan \psi_\circ, X^* \oU_{t-r} \oR X X^* \oU_{r-s} \oR X X^* \oU_{s-q} X \psi_\circ f(q) \ran_{\cH_{el}} dq ds dr \label{wkappa2_def}
\end{align}
and 
\begin{align}
    w_2f(t) &= \lan \psi_\circ, X^* \oR X_2 \psi_\circ f(t) \ran_{\cH_{el}}, \label{w2_def} \\
    \tilde w_2(t) f_0 &= \lan \psi_\circ, X^* \oU_t \oR X_2 \psi_\circ f_0 \ran_{\cH_{el}} - \int_0^t \lan \psi_\circ, X^* \oU_{t-r} \oR X X^* \oU_{r} \oR X \psi_\circ f_0 \ran_{\cH_{el}} dr. \label{tildew2(t)_def}
\end{align}
In particular, recall that $X = \frac{i}{\ka} \oP [T, P] P$, $X^* = \frac{i}{\ka} P[T, P] \oP$, and, c.f. \eqref{Xj_recursiverelation}, 
\begin{align}
    X_2 &= -\frac{1}{\ka^2} S \oP [T, P] P - \frac{1}{\ka^2} \oR \left[[T, P], [T, P] - K \right] P, \label{X2_explicit} \\
    X_3 &= -\frac{i}{\ka^3} S^2 \oP [T, P] P - \frac{i}{\ka^3} \oR \left[S, [T, P] - K \right] [T, P] P - \frac{i}{\ka^3} \oR S \left[[T, P], [T, P] - K \right] P \nonumber \\
    &\hspace{15pt} - \frac{i}{\ka^3} S \oR \left[[T, P], [T, P] - K \right] P - \frac{i}{\ka^3} \oR \left[ \oR, [T, P] - K \right] \left[[T, P], [T, P] - K \right] P \nonumber \\
    &\hspace{15pt} - \frac{i}{\ka^3} \oR^2 \left[ \left[[T, P], [T, P] - K \right], [T, P] - K \right]. \label{X3_explicit}
\end{align}
\end{lemma}

\begin{proof}[Proof of Lemma \ref{w3f_expansion_lemma}]
We begin with the expansion (\ref{AtXf_expansion}) with $n=1$ and expand the two non-local terms. For the term $\left(A_t^{\oR X X^*} A^X_{(\cdot)} f \right)(t)$, we use Lemma \ref{AYjtf_ibp_expansion_lemma} with $Y = X$ to expand $A^X_{(\cdot)} f$ and obtain
\begin{align}
    \left(A_t^{\oR X X^*} A^X_{(\cdot)} f \right)(t) &= (-i\ka) \left( A_t^{\oR X X^* \oR X} f\right)(t) - (-i\ka) \left( A_t^{\oR X X^* } \oU_{(\cdot)}\right)(t) \oR X \psi_\circ f_0 \nonumber \\
    &\hspace{15pt} - (-i\ka) \left( A_t^{\oR X X^* } A_{(\cdot)}^{X_2} f \right)(t) + (-i\ka) \left( A_t^{\oR X X^*} A_{(\cdot)}^{\oR X X^*} A_{(\cdot)}^{X} f \right)(t).
\end{align}
Expanding $\left(A^{X_2}_t f\right)(t)$ using Lemma \ref{AYjtf_ibp_expansion_lemma} with $Y = X_2$ yields
\begin{align}
    \left(A^{X_2}_t f\right)(t) &= - (-i\ka) \left(\oR X_2 \psi_\circ f(t) - \oU_t \oR X_2 \psi_\circ f_0 \right) + (-i\ka) \left(A_t^{\oR X_2 X^*} A^X_{(\cdot)} f \right)(t) \nonumber \\
    &\hspace{15pt} - (-i\ka) \left(A_t^{X_3} f \right)(t).
\end{align}
Using these expansions and plugging them back into (\ref{AtXf_expansion}), we obtain an expansion of the form 
\begin{align}
    \left(A_t^{X} f \right)(t) &= - i \ka \left( \oR X \psi_\circ f(t) - \oU_t \oR X \psi_\circ f_0 \right) - (-i\ka)^2 \left(\oR X_2 \psi_\circ f(t) - \oU_t \oR X_2 \psi_\circ f_0 \right) \nonumber \\
    &\hspace{15pt} - (-i\ka)^2 \left( A_t^{\oR X X^* } \oU_{(\cdot)}\right)(t) \oR X \psi_\circ f_0 + (-i\ka)^2 \left(A_t^{X_3 + \oR X X^* \oR X} f \right)(t) \nonumber \\
    &\hspace{15pt} + (-i\ka)^2 \left( A_t^{\oR X X^* } A_{(\cdot)}^{X_2} f \right)(t) + (-i\ka)^2 \left(A_t^{\oR X_2 X^*} A^X_{(\cdot)} f \right)(t) \nonumber \\
    &\hspace{15pt} + (-i\ka)^2 \left( A_t^{\oR X X^*} A_{(\cdot)}^{\oR X X^*} A_{(\cdot)}^{X} f \right)(t). \label{AtXf_expansion_n=2}
\end{align}
Now, we can write the expansion for $w^\ka[f](t)$ by inserting \eqref{AtXf_expansion_n=2} into the expression \eqref{new_wf_newexpr},
\begin{align}
    w^\ka[f](t) = (-i\ka)^{2} \left( w_1 f(t) - \tilde w_1(t) f_0 \right) + (-i\ka)^{3} w^{\ka}_{2}[f](t),
\end{align}
where
\begin{align}
    w_1 f(t) &= \lan \psi_\circ, X^* \oR X \psi_\circ f(t) \ran_{\cH_{el}}, \\
    \tilde w_1(t) f_0 &= \lan \psi_\circ, X^* \oU_t \oR X \psi_\circ f_0 \ran_{\cH_{el}}, \\
    w^{\ka}_{2}[f](t) &= - \left(\lan \psi_\circ, X^* \oR X_2 \psi_\circ f(t) \ran_{\cH_{el}} - \lan \psi_\circ, X^* \oU_t \oR X_2 \psi_\circ f_0 \ran_{\cH_{el}} \right) \nonumber \\
    &\hspace{15pt}- \lan \psi_\circ, X^* \left(A^{\oR X X^*}_t \oU_{(\cdot)} \right)(t) \oR X \psi_\circ f_0 \ran_{\cH_{el}} \nonumber \\
    &\hspace{15pt}+ \lan \psi_\circ, X^* \left(A_t^{X_3 + \oR X X^* \oR X} f \right)(t) \ran_{\cH_{el}} + \lan \psi_\circ, X^* \left(A_t^{\oR X_2 X^*} A_{(\cdot)}^{X} f \right)(t) \ran_{\cH_{el}} \nonumber \\
    &\hspace{15pt}+ \lan \psi_\circ, X^* \left(A_t^{\oR X X^*} A_{(\cdot)}^{X_2} f \right)(t) \ran_{\cH_{el}} + \lan \psi_\circ, X^* \left(A_t^{\oR X X^*} A_{(\cdot)}^{\oR X X^*} A_{(\cdot)}^{X} f \right)(t) \ran_{\cH_{el}}.
\end{align}
Using the definition \eqref{AtCf_def} of the operators $A^{Y}_t f$ and identifying $w_2f(t)$ and $\tilde w_2(t) f_0$ via \eqref{w2_def} and \eqref{tildew2(t)_def}, we obtain \eqref{wkappa_expansion_n=2} - \eqref{tildew2(t)_def}. The expansions \eqref{X2_explicit} and \eqref{X3_explicit} follow directly from the recursive relation \eqref{Xj_recursiverelation}.
\end{proof}

Next, we would like to obtain a closed-form expression of $\left(A^{X}_t f \right)(t)$ expanded to arbitrary order. It is clear that one can always expand using Lemma \ref{AYjtf_ibp_expansion_lemma} to arbitrary order, and in the following proposition we aim to at least characterize this expansion. However, the recursion relations are too difficult to characterize explicitly. 

\begin{proposition} \label{AXtf_expansion_nthorder_prop}
Suppose Assumptions \ref{assumption1} - \ref{assumption4} hold. For all $t \in \mathbb{R}$, and for all $f$ solutions of \eqref{effectiveeqforf}, $A_t^X$ as given in \eqref{newA_def} admits an expansion of the following form:
\begin{equation}
    \left(A_t^{X} f\right)(t) = \sum_{j=1}^n - (-i\ka)^j \left[ L_j \psi_\circ f(t) - \tilde L_j(t) \psi_\circ f_0 \right] + (-i\ka)^n \cR_n [f](t), \label{AtCf_nfold_ibp_expansion}
\end{equation}
where $L_j$ and $L_j(t)$ are local operators satisfying the recursion relation below, and the remainder $\cR_n$ is given by
\begin{align}
    \cR_n [f](t) &= \left(A^{Y_n}_t f \right) (t) + \sum_{j=2}^n \sum_{l=1}^{N(n,j)} \left( A^{Y_{n, j}^{l,j}}_t A^{Y_{n, j}^{l,j-1}}_{(\cdot)} A^{Y_{n, j}^{l,j-2}}_{(\cdot)} \dots A^{Y_{n, j}^{l,1}}_{(\cdot)} f \right)(t) \nonumber \\
    &\hspace{15pt} + \left( A^{\oR X X^*}_t \left(\prod_{1}^{n-1} A^{\oR X X^*}_{(\cdot)}\right) A^{X}_{(\cdot)} f \right)(t), \label{AtCf_nfold_ibp_expansion_remainder}
\end{align}
for local time-independent differential operators $Y_{n, j}^{l,k}$. The $n=1$ case is given by \eqref{AtXf_expansion}, in which case we have $L_1 = \oR X$, $\tilde L_1(t) = \oU_t \oR X$, and $Y_1 = X_2$; and the $n=2$ case is given by \eqref{AtXf_expansion_n=2}. 

For $n \geqslant 2$, we describe the following recursive relations. $N(n,j)$ is a sequence of integers counting the number of respective terms, given by the recursive relation $N(n,1) = 1$, $N(n,n+1) = 1$, $N(n, j) = 0$ for $j \geqslant n+2$, and 
\begin{equation} \label{N(n,j)_recursionrelation}
    N(n+1,j) = N(n, j-1) + N(n,j) + N(n, j+1).
\end{equation}
For all values of $n$ and $j$ the operators $Y_n$ and $Y_{n,j}^{l,k}$ are differential operators belonging to the class $\cC^p$ for some $p \geqslant 1$. Furthermore, we have the following recursion relations:
\begin{equation} \label{Ln_recursionrelations}
     L_{n+1} = \oR Y_n, \qquad n \geqslant 0,
\end{equation}
where 
\begin{equation} \label{Yn_recursionrelations}
    Y_{n} = - W(Y_{n-1}) + \sum_{l=1}^{N(n-1,2)} Y_{n,2}^{l,2} \oR Y_{n,2}^{l,1}, \qquad Y_1 := X,
\end{equation}
with $W(Y_{n-1}) = \frac{i}{\ka} S Y_{n-1} + \frac{i}{\ka} \oR \left(\oK Y_{n-1} - Y_{n-1} K^P \right)P$. Also, $\tilde L_{1}(t) = \oU_t \oR X$, and for $n \geqslant 1$,
\begin{align} 
    \tilde L_{n+1}(t) &= \oU_t \oR Y_n - \sum_{j=2}^n \sum_{l=1}^{N(n,j)} \left( A^{Y_{n, j}^{l,j}}_t A^{Y_{n, j}^{l,j-1}}_{(\cdot)} A^{Y_{n, j}^{l,j-2}}_{(\cdot)} \dots \oU_{(\cdot)} \right)(t) Y_{n, j}^{l,1} \nonumber \\
    &\hspace{15pt} - \left( A^{\oR X X^*}_t \left(\prod_{1}^{n-1} A^{\oR X X^*}_{(\cdot)}\right) \oU_{(\cdot)} \right)(t) \oR X. \label{tildeLn(t)_recursionrelations}
\end{align} 
\end{proposition}

\begin{proof}[Proof of Proposition \ref{AXtf_expansion_nthorder_prop}]
For the inductive step, we look at expanding the remainder $(\cR_n f)(t)$ by expanding the innermost $A^{Y}_{(\cdot)} f$ term in every term of the form 
\begin{equation}
    \left( A^{Y_{n, j}^{l,j}}_t A^{Y_{n, j}^{l,j-1}}_{(\cdot)} \dots A^{Y_{n, j}^{l,2}}_{(\cdot)} A^{Y_{n, j}^{l,1}}_{(\cdot)} f \right)(t),
\end{equation}
which is a term of length $j$ in the sense that there are $j$ compositions. Expanding $A^{Y_{n, j}^{l,1}}_{(\cdot)} f$ using the expansion \eqref{AtCf_ibp_expansion} with $Y = Y_{n, j}^{l,1}$ yields 
\begin{align}
    \left( A^{Y_{n, j}^{l,j}}_t A^{Y_{n, j}^{l,j-1}}_{(\cdot)} \dots A^{Y_{n, j}^{l,2}}_{(\cdot)} A^{Y_{n, j}^{l,1}}_{(\cdot)} f \right)(t) &= (-i\ka) \left( A^{Y_{n, j}^{l,j}}_t A^{Y_{n, j}^{l,j-1}}_{(\cdot)} \dots A^{Y_{n, j}^{l,2} \oR Y_{n,j}^{l,1}}_{(\cdot)} f \right)(t) \nonumber \\
    &\hspace{15pt}- (-i\ka) \left( A^{Y_{n, j}^{l,j}}_t A^{Y_{n, j}^{l,j-1}}_{(\cdot)} \dots A^{Y_{n, j}^{l,2}}_{(\cdot)} \oU_{(\cdot)} \right)(t) \oR Y_{n,j}^{l,1} \psi_\circ f_0 \nonumber \\
    &\hspace{15pt}- (-i\ka) \left( A^{Y_{n, j}^{l,j}}_t A^{Y_{n, j}^{l,j-1}}_{(\cdot)} \dots A^{Y_{n, j}^{l,2}}_{(\cdot)} A^{W(Y_{n, j}^{l,1})}_{(\cdot)} f \right)(t) \nonumber \\
    &\hspace{15pt}+ (-i \ka) \left( A^{Y_{n, j}^{l,j}}_t A^{Y_{n, j}^{l,j-1}}_{(\cdot)} \dots A^{Y_{n, j}^{l,2}}_{(\cdot)} A^{\oR Y_{n, j}^{l,1} X^*}_{(\cdot)} A^{X}_{(\cdot)} f \right)(t). \label{AXt_expansion_crazy}
\end{align}
Here we have one non-local term of length $j-1$ acting on $\psi_\circ f(t)$, one local time-dependent term of length $j-1$ acting on $\psi_\circ f_0$, one non-local term of length $j$ acting on $\psi_\circ f(t)$, and one non-local term of length $j+1$ acting on $\psi_\circ f(t)$. At every value of $n$ there will always be one term of length $1$ and one term of length $n+1$. This yields the recursion relation \eqref{N(n,j)_recursionrelation} for $N(n,j)$. 

To obtain the recursion relations \eqref{Ln_recursionrelations}-\eqref{tildeLn(t)_recursionrelations}, we expand the terms of length $j=2$ and $j=1$ in the same manner as \eqref{AXt_expansion_crazy} and collect the like terms. 
\end{proof}

Inserting \eqref{AtCf_nfold_ibp_expansion} into the expression \eqref{new_wf_newexpr} for $w^{\ka}[f](t)$ yields
\begin{equation} \label{wkappa_expansion}
    w^\ka[f](t) = \sum_{j=1}^{n-1} - (-i\ka)^{j+1} \left( w_j f(t) - \tilde w_j(t) f_0 \right) + (-i\ka)^{n+1} w^{\ka}_{n}[f](t),
\end{equation}
with the local operators $w_j$ and $\tilde w_j(t)$ and remainder $w^{\ka}_{n}[f](t)$ given by
\begin{align} 
    w_j f(t) &= \langle \psi_\circ, X^* L_j \psi_\circ f(t) \rangle_{\cH_{el}}, \label{wj_def} \\
    \tilde w_j(t) f_0 &= \langle \psi_\circ, X^* \tilde L_j(t) \psi_\circ f_0 \rangle_{\cH_{el}}, \label{wtildej_def} \\
    w^{\ka}_{n}[f](t) &= - \left(w_n f(t) - \tilde w_n(t) f_0\right) + \langle \psi_\circ, X^* \cR_n [f](t) \rangle_{\cH_{el}}. \label{w_remainder_def}
\end{align} 

In the next subsection we estimate the right-hand sides of \eqref{wj_def}-\eqref{w_remainder_def}. 

\subsection{Completion of proof of Theorem \ref{MAINTHEOREM_ABSTRACT-PREVIOUS} (b)} \label{proofIterative}

In this section, we estimate \eqref{wkappa_expansion} completing the proof of Theorem \ref{MAINTHEOREM_ABSTRACT-PREVIOUS} (b). All terms $L_j$, $\tilde L_j(t)$, and $\cR_n[f](t)$ can be estimated using the following two Proposition \ref{Xj_recursive_estimates_proposition} and Proposition \ref{AXtf_expansion_nthorder_prop}.  

Finding an explicit expression for the recursive relation \eqref{Xj_recursiverelation} is too hard. Instead, we will obtain our estimate recursively by first showing
\begin{equation} \label{Xj_estimates}
    X_j = O_{\cL_{s+j,s}}(1).
\end{equation}
To do so, we introduce $\cC^{k}$, the class of differential operators of order $k$ with coefficients of a certain form. Consider an operator formed by linear combinations of compositions of the bounded operators 
\begin{equation} \label{Bcondition_1}
    (\partial_y^{\alpha_1} H_{bo}), \:(\partial_y^{\alpha_2} P), \:(\partial_y^{\alpha_3} E), \text{ and } \oR,
\end{equation}
for multi-indices $\alpha_1 \geqslant 1$, and $\alpha_2, \alpha_3 \geqslant 0$ as permitted by the regularity from Assumption \ref{assumption1}. Consider operators of the form 
\begin{equation} \label{element_of_class_Ck}
    A = \sum_{0 \leqslant \nrm{\alpha} \leqslant k} c_\alpha B_{\alpha} D^\alpha = O_{\cL_{s+k,s}}(1),
\end{equation}
where, for all multi-indices $\alpha$ and all $j$, $c_\alpha$ are possibly $\ka^{-1}$-dependent coefficients and $B_\alpha$ are linear combinations of operators in \eqref{Bcondition_1} labelled by the multi-index $\alpha$. We define the following class of operators,
\begin{equation} \label{Ck_class_def}
    \cC^{k} := \left\{ A \text{ operators of the form } \eqref{element_of_class_Ck} \right\}.
\end{equation}

\begin{lemma} \label{Ck_class_composition_lemma}
If $C_1 \in \cC^{k}$ and $C_2 \in \cC^{l}$, then the composition $C_1 C_2 \in \cC^{k+l}$. 
\end{lemma}

\begin{proof}[Proof of Lemma \ref{Ck_class_composition_lemma}]
We write 
\begin{align}
    C_1 C_2 &= \sum_{\substack{\nrm{\alpha} \leqslant k, \\ \nrm{\beta} \leqslant l}} B_\alpha D^\alpha B_\beta D^\beta = \sum_{\substack{\nrm{\alpha} \leqslant k, \\ \nrm{\beta} \leqslant l}} B_\alpha B_\beta D^{\alpha + \beta} + \sum_{\substack{\nrm{\alpha} \leqslant k, \\ \nrm{\beta} \leqslant l}} B_\alpha [D^\alpha, B_\beta] D^\beta.
\end{align}
Clearly the first term belongs to $\cC^{k+l}$, so it remains to show the same for the second term. This follows via a straightforward induction argument, by showing 
\begin{equation} \label{Ck_composition_inductionstep}
    [D^\alpha, B_\beta] \in \cC^{\nrm{\alpha}-1}.
\end{equation}
As $B_\beta$ is a linear combination of elements of \eqref{Bcondition_1}, \eqref{Ck_composition_inductionstep} follows using Lemma \ref{commutatorwithT_lemma} parts a) and b). 
\end{proof}

\begin{lemma}\label{Bj_commutator_lemma}
Let $B$ be one of the operators in \eqref{Bcondition_1}. Then $B = O_{\cL_{s,s}}(1)$, $\frac{i}{\ka} [B, [T,P] - T] \in \cC^{1}$, and 
\begin{equation} \label{Bj_commutator_estimate}
    \frac{i}{\ka} [B, [T,P] - T] = O_{\cL_{s+1,s}}(1).
\end{equation}
\end{lemma}

\begin{proof}[Proof of Lemma \ref{Bj_commutator_lemma}]
Writing $T = D^2$, then 
\begin{equation}
    [T,P] - T = -\ka^2 (\Delta_y P) - 2 i \ka (\nabla_y P) D - D^2. 
\end{equation}
Hence, using that $B$ is one of the operators in \eqref{Bcondition_1} and $B = O_{\cL_{s,s}}(1)$, we estimate the commutator $[B, [T,P] - T]$ in the manner of \eqref{Bj_commutator_estimate} using Lemma \ref{commutatorwithT_lemma} part a), and in particular, $\frac{i}{\ka} [B, [T,P] - T] \in \cC^{1}$.
\end{proof}

Under Assumption \ref{assumption1}, $P$ and $\oP$ are $k_A$-times differentiable (cf. Lemma \ref{smoothness_proposition}), and hence $X = \oP X^\circ P$, where  
\begin{equation} \label{PbarTP_Ckspace}
    X^\circ = \frac{i}{\ka} [T, P] \in \cC^{1} \text{ and } X^\circ = O_{\cL_{s+1,s}}(1) \text{ for } s \leqslant k_A - 1.
\end{equation}
In fact, by the recursion relation \eqref{Xj_recursiverelation}, it is clear that $X_j = \oP X_j P$ for each $j \geqslant 1$. Then, we seek to show that there exists a $X_j^\circ \in \cC^j$ such that $X_j = \oP X_j^\circ P$.

\begin{proposition} \label{Xj_recursive_estimates_proposition}
If $X_1 = \oP X^\circ_1 P$ with $X_1^\circ \in \cC^1$ and $X_j$ for $j \geqslant 2$ is given by the recursive relation \eqref{Xj_recursiverelation}, then there exists $X_j^\circ \in \cC^{j}(\R^m)$ such that $X_j = \oP X_j^\circ P$ for all $j \geqslant 2$. In particular, \eqref{Xj_estimates} holds. 
\end{proposition}

\begin{proof}[Proof of Proposition \ref{Xj_recursive_estimates_proposition}]
We argue by induction on $j$. As the base case is covered by the hypothesis, only the induction step remains. Assume that $X_n^\circ \in \cC^{n}$ with $X_n = \oP X_n^\circ P$. Then, by the recursion relation \eqref{Xj_recursiverelation}, 
\begin{equation}
    X_{n+1} = \frac{i}{\ka} S X_n + \frac{i}{\ka} \oR \left(K \oP X_n - X_n P K \right) P
\end{equation}
and using $X_n = \oP X_n^\circ P$, 
\begin{equation}
    X_{n+1} = \frac{i}{\ka} S X_n^\circ P + \frac{i}{\ka} \oR \left(K \oP X_n^\circ - X_n^\circ P K \right) P.
\end{equation}
Now, applying Lemma \ref{lemma_recursivecommutator} with $X^\circ = X_n^\circ$ yields
\begin{equation}
    X_{n+1} = \frac{i}{\ka} S X_n^\circ P + \frac{i}{\ka} \oR [X_n^\circ, [K, P] - K] P.
\end{equation}
Using $K = T + E$ and $[E, P] = 0$, then $[K, P] = [T, P]$ so we write $X_{n+1}$ as the sum of three terms, 
\begin{equation}
    X_{n+1} = \frac{i}{\ka} S X_n^\circ P + \frac{i}{\ka} \oR [X_n^\circ, [T, P] - T] P - \frac{i}{\ka} \oR [X_n^\circ, E] P.
\end{equation}
Recall that $S = \oR [H_{bo} - E, T] \oR$ and using Lemma \ref{commutatorwithT_lemma}, it is clear that $S \in \cC^{1}$ (c.f. \eqref{S_and_X2_estimates}). Hence, we define
\begin{equation}
    X_{n+1}^\circ = \frac{i}{\ka} S X_n^\circ + \frac{i}{\ka} \oR [X_n^\circ, [T, P] - T] - \frac{i}{\ka} \oR [X_n^\circ, E].
\end{equation}

Clearly the class $\cC^{n+1}$ is closed under linear combinations, so we examine each term individually. Then, by Lemma \ref{Ck_class_composition_lemma}, the composition $S X_n^\circ \in \cC^{n+1}$. 

Now consider the term $\oR [X_n^\circ, [T, P] - T]$. Since $X_n^\circ \in \cC^{n}$, we write
\begin{align}
    [X_n^\circ, [T, P] - T] &= \sum_{0 \leqslant \nrm{\alpha} \leqslant n} [B_{\alpha} D^\alpha, [T, P] - T] \nonumber \\
    &= \sum_{0 \leqslant \nrm{\alpha} \leqslant n} [B_{\alpha}, [T, P] - T] D^\alpha + B_{\alpha} [D^\alpha, [T, P] - T]. \label{Cj_recursive_estimate1}
\end{align}
By Lemma \ref{Bj_commutator_lemma}, $[B_{\alpha}, [T,P] - T] \in \cC^{1}$. By Lemma \ref{commutatorwithT_lemma} part d) it follows that 
\begin{equation}
    [D^\alpha, [T, P] - T]  = [D^\alpha, [T, P]] \in \cC^{\nrm{\alpha}}.
\end{equation}
Then by the above, Lemma \ref{Ck_class_composition_lemma}, and 
\eqref{Cj_recursive_estimate1}, 
\begin{equation}
    [X_j^\circ, [T, P] - T] \in \cC^{\nrm{\alpha}+1}.
\end{equation}

The last term is $\oR[X_n^\circ, E]$. Writing 
\begin{equation}
    X_n^\circ = \sum_{0 \leqslant \nrm{\alpha} \leqslant n} B_\alpha D^\alpha,
\end{equation}
we see that since $B_\alpha$ satisfy \eqref{Bcondition_1}, by Lemma \ref{[Fibered,E]_lemma}, we have $[B_\alpha, E] = 0$. Hence,
\begin{equation}
    [X_n^\circ, E] = \sum_{0 \leqslant \nrm{\alpha} \leqslant n} B_\alpha [D^\alpha, E]
\end{equation}
and by the Leibnitz Rule Lemma \ref{LeibnizRule_lemma}, we have $[D^\alpha, E] \in \cC^{\nrm{\alpha}-1}$, so $\oR [X_n^\circ, E] \in \cC^{n-1}$. This concludes the proof.
\end{proof}

\begin{proposition} \label{Lj_tildeLj(t)_cRn_estimates_prop}
Let Assumptions \ref{assumption1} - \ref{assumption4} hold. Then, the following estimates hold for $s \geqslant 0$ a positive integer and some constant $C > 0$ depending on $s$. For $n \geqslant 1$ and $t \in \R$, 
\begin{equation} \label{Ln_tildeLn(t)_estimates}
    L_n = O_{\cL_{s+n,s}}(1), \qquad \tilde L_n (t) = O_{\cL_{s+2n-1,s}}(e^{Ct}). 
\end{equation}
For $n \geqslant 1$ and $0 \leqslant t \leqslant \tau$,
\begin{equation} \label{cRn[f](t)_estimate}
    \norm{\cR_n[f](t)}_{H^s_{\ka,y}} \lesssim e^{C \tau} \norm{f}_{B_\tau^{s+2n+1}}.
\end{equation}
\end{proposition}

\begin{proof}[Proof of Proposition \ref{Lj_tildeLj(t)_cRn_estimates_prop}]
The $n=1$ case \eqref{Ln_tildeLn(t)_estimates} follows directly from estimating the terms in \eqref{AtXf_expansion}. In particular, we use $\oR = O_{\cL_{s,s}}(1)$, $X = O_{\cL_{s+1,s}}(1)$, and $\oU_t = O_{\cL_{s,s}}(e^{Ct})$ to see that 
\begin{align}
    L_1 = \oR X = O_{\cL_{s+1,s}}(1), \qquad  L_1 (t) = \oU_t \oR X = O_{\cL_{s+1,s}}(e^{Ct}). 
\end{align}
To estimate $\cR_1[f](t)$, we use in addition to the previous the estimates $X_2 = O_{\cL_{s+2,s}}(1)$ and $X^* = O_{\cL_{s+1,s}}(1)$. In particular, from the definition \eqref{AtCf_def} of $A_t^Y f$, we see that if $Y = O_{\cL_{s+j,s}}(1)$ for some $j$, then for $0\leqslant t \leqslant \tau$, 
\begin{equation} \label{AtY_estimate}
    \norm{(A_t^Y f)(t)}_{H^s_{\ka,y}} \leqslant \int_0^t e^{C(t-s)} \norm{Y \psi_\circ f(r)}_{H^s_{\ka,y}} dr \leqslant e^{C \tau} \norm{f}_{B_\tau^{s+j}}.
\end{equation}
Hence, since (c.f. \eqref{AtXf_expansion}) 
\begin{equation}
    \cR_1[f](t) = - i \ka \left(A_t^{\oR X X^*} A^X_{(\cdot)} f \right)(t) + i \ka \left(A^{X_2}_t f\right)(t),
\end{equation}
estimating this right-hand side gives us the estimate \eqref{cRn[f](t)_estimate} for $n=1$. 

The $n=2$ case can also be seen explicitly, using the expansion \eqref{AtXf_expansion_n=2}. In particular, we have 
\begin{equation}
    L_2 = \oR X_2 = O_{\cL_{s+2,s}}(1),
\end{equation}
and estimating 
\begin{equation}
    \norm{\left(A_t^{\oR X X^*} \oU_{(\cdot)}\right)(t) u}_{H^s_{\ka,y}} \leqslant \int_0^t e^{C(t-r)} \norm{\oR X X^* \oU_r u}_{H^s_{\ka,y}} dr \lesssim e^{Ct} \norm{u}_{H^{s+2}_{\ka,y}}
\end{equation}
for some constant $C > 0$, we obtain
\begin{equation}
    \tilde L_2(t) = \oU_t \oR X_2 - \left(A_t^{\oR X X^*} \oU_{(\cdot)}\right)(t) \oR X = O_{\cL_{s+3,s}}(e^{Ct}).
\end{equation}
We can also see that
\begin{align}
    \cR_2[f](t) &= \left(A_t^{X_3 + \oR X X^* \oR X} f \right)(t) + \left( A_t^{\oR X X^* } A_{(\cdot)}^{X_2} f \right)(t) + \left(A_t^{\oR X_2 X^*} A^X_{(\cdot)} f \right)(t) \nonumber \\
    &\hspace{15pt} + \left( A_t^{\oR X X^*} A_{(\cdot)}^{\oR X X^*} A_{(\cdot)}^{X} f \right)(t). \label{cR2[f](t)_def}
\end{align}
By Proposition \ref{Xj_recursive_estimates_proposition}, $X_3 = O_{\cL_{s+3,s}}(1)$. Recall also that for terms of the form $\left( A^{Y_1}_t \dots A^{Y_n}_{(\cdot)} u \right)(t),$ by their length we mean the number of compositions of $A_t^{Y_j}$ that appear (for this generic term, the length is $n$). Then, the total differential order of such a term will be the sum of the differential orders of the $Y_1, \dots, Y_n$ operators. 

On the right-hand side of \eqref{cR2[f](t)_def}, we see that the term of length $1$, $\left(A_t^{X_3 + \oR X X^* \oR X} f \right)(t)$, is of differential order $3$. The terms of length $2$, $\left( A_t^{\oR X X^* } A_{(\cdot)}^{X_2} f \right)(t) + \left(A_t^{\oR X_2 X^*} A^X_{(\cdot)} f \right)(t)$, are of differential order $4$, and finally the term of length $3$, $\left( A_t^{\oR X X^*} A_{(\cdot)}^{\oR X X^*} A_{(\cdot)}^{X} f \right)(t)$, is of differential order $5$. The estimate \eqref{cRn[f](t)_estimate} for $n=2$ follows. 

We now proceed to argue by induction, using the previous estimates as the base case. Following the estimate for $\cR_2[f](t)$, our intuition is that the length of a term dictates the differential order. Hence, assume that in the expansion of order $n$, for all $1 \leqslant j \leqslant n+1$ we have
\begin{equation} \label{lengthjterm_differentialorder}
    \sum_{l=1}^{N(n,j)} \left( A^{Y_{n, j}^{l,j}}_t A^{Y_{n, j}^{l,j-1}}_{(\cdot)} \dots A^{Y_{n, j}^{l,2}}_{(\cdot)} A^{Y_{n, j}^{l,1}}_{(\cdot)} f \right)(t) = O_{\cL_{s+n+j+1}}(e^{Ct}). 
\end{equation}
Expanding $A^{Y_{n, j}^{l,1}}_{(\cdot)} f$ using the expansion \eqref{AtCf_ibp_expansion} with $Y = Y_{n, j}^{l,1}$ for all $1 \leqslant l \leqslant N(n,j)$ yields 
\begin{align}
    &\sum_{l=1}^{N(n,j)} \left( A^{Y_{n, j}^{l,j}}_t A^{Y_{n, j}^{l,j-1}}_{(\cdot)} \dots A^{Y_{n, j}^{l,2}}_{(\cdot)} A^{Y_{n, j}^{l,1}}_{(\cdot)} f \right)(t) \nonumber \\
    &= (-i\ka) \sum_{l=1}^{N(n,j)} \left( A^{Y_{n, j}^{l,j}}_t A^{Y_{n, j}^{l,j-1}}_{(\cdot)} \dots A^{Y_{n, j}^{l,2} \oR Y_{n,j}^{l,1}}_{(\cdot)} f \right)(t) \label{length_j-1} \\
    &- (-i\ka) \sum_{l=1}^{N(n,j)} \left( A^{Y_{n, j}^{l,j}}_t A^{Y_{n, j}^{l,j-1}}_{(\cdot)} \dots A^{Y_{n, j}^{l,2}}_{(\cdot)} \oU_{(\cdot)} \right)(t) \oR Y_{n,j}^{l,1} \psi_\circ f_0 \label{length_j-1_tildeL} \\
    &- (-i\ka) \sum_{l=1}^{N(n,j)} \left( A^{Y_{n, j}^{l,j}}_t A^{Y_{n, j}^{l,j-1}}_{(\cdot)} \dots A^{Y_{n, j}^{l,2}}_{(\cdot)} A^{W(Y_{n, j}^{l,1})}_{(\cdot)} f \right)(t) \label{length_j} \\
    &+ (-i \ka) \sum_{l=1}^{N(n,j)} \left( A^{Y_{n, j}^{l,j}}_t A^{Y_{n, j}^{l,j-1}}_{(\cdot)} \dots A^{Y_{n, j}^{l,2}}_{(\cdot)} A^{\oR Y_{n, j}^{l,1} X^*}_{(\cdot)} A^{X}_{(\cdot)} f \right)(t). \label{length_j+1}
\end{align}
The term \eqref{length_j-1} is of length $j-1$ and is of the same differential order as the term before expansion, $s+n+j+1$. The term \eqref{length_j-1_tildeL}, by \eqref{tildeLn(t)_recursionrelations}, contributes to $\tilde L_{n+1}(t)$. We remark that it is also of differential order $s+n+j+1$ and return to it later. Next, the terms \eqref{length_j} are of length $j$ and are of differential order $s+(n+1)+j+1$, since by Proposition \ref{Xj_recursive_estimates_proposition}, $W(Y^{l,1}_{n,j})$ will be of one differential order higher that $Y_{n,j}^{l,1}$. Lastly, the terms \eqref{length_j+1} are of length $j+1$ and of differential order $s+(n+1)+(j+2)+1$, since $\oR Y_{n,j}^{l,1} X^*$ is of one differential order higher than $Y_{n,j}^{l,1}$ and $X$ is of differential order $1$. Hence, for all $n \geqslant 2$ and $1 \leqslant j \leqslant n+1$, \eqref{lengthjterm_differentialorder} follows.

Using \eqref{lengthjterm_differentialorder}, we can now estimate $L_n$, $\tilde L_n(t)$, and $\cR_n[f](t)$ using the recursive relations \eqref{Ln_recursionrelations} - \eqref{tildeLn(t)_recursionrelations}. We have $Y_{n-1} = O_{\cL_{s+n,s}}(1)$, so \eqref{Ln_recursionrelations} yields the first estimate in \eqref{Ln_tildeLn(t)_estimates}. Next, in \eqref{tildeLn(t)_recursionrelations}, we see that the term of highest differential order is 
\begin{equation}
    \left( A^{\oR X X^*}_t \left(\prod_{1}^{n-2} A^{\oR X X^*}_{(\cdot)}\right) \oU_{(\cdot)} \right)(t) \oR X = O_{\cL_{s+2n-1}}(e^{Ct}), 
\end{equation}
so the second estimate in \eqref{Ln_tildeLn(t)_estimates} follows. Finally, in \eqref{AtCf_nfold_ibp_expansion}, see that the term of highest differential order is 
\begin{equation}
    \norm{\left( A^{\oR X X^*}_t \left(\prod_{1}^{n-1} A^{\oR X X^*}_{(\cdot)}\right) A^{X}_{(\cdot)} f \right)(t)}_{H^s_{\ka,y}} \lesssim e^{Ct} \norm{f}_{B^{s+2n+1}_t}, 
\end{equation}
and the estimate \eqref{cRn[f](t)_estimate} follows.
\end{proof}

Now we can estimate the operators \eqref{wkappa_expansion} - \eqref{w_remainder_def}.

\begin{proposition} \label{wkappa_expansions_estimate_proposition}
Let Assumptions \ref{assumption1} - \ref{assumption4} hold. Then, the following estimates hold for $s \geqslant 0$ a positive integer and some constant $C > 0$ depending on $s$ and independent of $\ka$. For $n \geqslant 1$ and $t \in \R$, 
\begin{equation} \label{wtildej_estimate}
    \norm{w_n f(t)}_{H^s_{\ka ,y}} \lesssim \norm{f(t)}_{H^{s+n+1}_{\ka ,y}}, \qquad \norm{\tilde w_n(t) f_0}_{H^s_{\ka,y}} \lesssim e^{Ct} \norm{f_0}_{H^{s+2n}_{\ka,y}}. 
\end{equation}
For $n \geqslant 1$ and $0 \leqslant t \leqslant \tau$, 
\begin{equation} \label{wkappa_remainder_estimate1}
    \norm{w^\ka_n[f](t)}_{H^s_{\ka,y}} \lesssim e^{C \tau} \left( \norm{f_0}_{H^{s+2n}_{\ka,y}} + \norm{f}_{B_\tau^{s+2n+2}} \right).
\end{equation}
\end{proposition}

\begin{proof}[Proof of Proposition \ref{wkappa_expansions_estimate_proposition}]
The estimates follow directly from the definitions \eqref{wkappa_expansion} - \eqref{w_remainder_def} and the estimates in Proposition \ref{Lj_tildeLj(t)_cRn_estimates_prop}. We illustrate the estimate of $w_n f(t)$. Using Cauchy-Schwartz and the normalization $\norm{\psi_\circ}_{\cH_{el}} = 1$, it follows from \eqref{wj_def} that
\begin{equation}
    \norm{w_n f(t)}_{H^s_{\ka ,y}} \lesssim \norm{X^* L_n \psi_\circ f}_{H^s_{\ka ,y} \cH_{el}}.
\end{equation}
Recall that $X^* = \frac{i}{\ka} P [T, P] \oP = O_{\cL_{s+1,s}}(1)$ (see Lemma \ref{commutatorwithT_lemma} part a)). Combining this with $L_n = O_{\cL_{s+n,s}}(1)$ (from Proposition \ref{Lj_tildeLj(t)_cRn_estimates_prop}), we obtain the first estimate in \eqref{wtildej_estimate}. The remaining estimates follow similarly. In particular, estimating the right-hand side of \eqref{w_remainder_def} yields 
\begin{equation}
    \norm{w^\ka_n[f](t)}_{H^s_{\ka,y}} \lesssim \norm{f(t)}_{H^{s+n+1}_{\ka ,y}} + e^{C \tau} \left( \norm{f_0}_{H^{s+2n}_{\ka,y}} + \norm{f}_{B_\tau^{s+2n+2}} \right).
\end{equation}
\eqref{wkappa_remainder_estimate1} follows by using $\norm{f(t)}_{H^{s+n+1}_{\ka ,y}} \leqslant \norm{f}_{B^{s+n+1}_t} \lesssim \norm{f}_{B^{s+2n+2}_t}$. 
\end{proof}

The proof of Theorem \ref{MAINTHEOREM_ABSTRACT-PREVIOUS} (b) now follows immediately by combining the expansion \eqref{wkappa_expansion} - \eqref{w_remainder_def} with the estimates in Proposition \ref{wkappa_expansions_estimate_proposition}. \qed

\section{Proof of Theorem \ref{TDBO_SECONDORDER_MAINTHEOREM} assuming Theorem \ref{MAINTHEOREM_ABSTRACT-PREVIOUS}} \label{TDBO_secondorder_Section}

\subsection{Outline of the proof}

We first describe the approach and main steps of the proof of Theorem \ref{TDBO_SECONDORDER_MAINTHEOREM} and outline the proof. The approach is very similar to the proof of Theorem \ref{TDBO_FIRSTORDER_MAINTHEOREM}: we seek to re-write the difference $\Psi(t) - (Q_P\tilde{f})(t)$ as the integral in time of some operator-valued quantities, and we show these quantities are small by using the NAIP (the left-handed version, Lemma \ref{NAIP_lemma_lefthanded}). 

As in the theorem, let $\Psi(t)$ satisfy the Schr{\"o}dinger equation (\ref{abstractSE}) with $\Psi(0) = \psi_\circ f_0$. Then, using Theorem \ref{MAINTHEOREM_ABSTRACT-PREVIOUS} part (a) and the definition of $Q_P$, see \eqref{Qpexpr}, we can write 
\begin{equation}
    \Psi(t) = (Q_P f)(t) = \psi_\circ f(t) + \int_0^t \oU_{t-s} X^* \psi_\circ f(s) ds,
\end{equation}
where recall $X^* = -\frac{i}{\ka} \oP H_\ka P$, and $f(t)$ solves the effective nuclear dynamics \eqref{effectiveeqforf}, i.e. the dynamics under the full non-local effective nuclear Hamiltonian $\heff^\ka$ given in \eqref{heff_maintheorem} with initial condition $f(0) = f_0$. This representation implies 
\begin{align}
    \Psi(t) - (Q_P\tilde{f})(t) &= \psi_\circ \left(f(t) - \tilde{f}(t) \right)  + \int_0^t \oU_{t-s} X^* \psi_\circ \left(f(s) - \tilde{f}(s) \right) ds, \label{Psi-Qp_secondorder_1}
\end{align}
where recall $\tilde{f}(t)$ is the solution of the truncated effective nuclear equation \eqref{effeq_2ndorder}, under the truncated local effective nuclear Hamiltonian $\heff^{(2)}$. In order to proceed, we prove $\psi_\circ\left(f(t) - \tilde{f}(t) \right) = O_{\cL_{s+6,s}}(\ka^2 e^{C \tau})$ for $0 \leqslant t \leqslant \tau$ by recasting the difference as the integral in time of some operator-valued quantities. We then prove that such integral terms are small using the left-handed NAIP of Lemma \ref{NAIP_lemma_lefthanded}, \eqref{abstract_NAIP_lefthanded}. We apply the NAIP with the same choices as before, i.e. $A = \frac{i}{\ka} \oH$, $B = \frac{i}{\ka} \oK$ (where recall $K = E + T$ and $\oK = \oP K \oP$), and the corresponding propagators \begin{equation}
    e^{As} = e^{i\oH s/\ka} =: \oU_s, \qquad \oH = \oP H_\ka \oP, \qquad e^{Bs} = e^{i \oK s/\ka} =: \oV_s, \qquad \oK = \oP K \oP,    
\end{equation} 
and the reduced resolvent
\begin{equation}
    (A-B)^{-1} = -i \ka \oP (\oH_{bo} - \oE)^{-1}\oP =: - i \ka \oR. 
\end{equation}
As in the previous proofs, we make use of Lemmas \ref{smoothness_proposition} and \ref{(Hbar - E)inverselemma}, the commutator estimates of Appendix \ref{commutatorestimates_Section}, and the propagator estimates of Appendix \ref{Section_estimatesonPropagators}.



\subsection{Representation formulas}

As a first step in the proof of Theorem \ref{TDBO_SECONDORDER_MAINTHEOREM}, we prove the self-adjointness of $\heff^{(2)}$ in the following lemma.

\begin{lemma} \label{heff(2)_selfadjoint_prop}
Let Assumptions \ref{assumption1}-\ref{assumption4} hold. a) For $\ka$ sufficiently small, 
\begin{equation} \label{k0_def}
    0 < \ka < \sqrt{\frac{2}{\tilde C \sqrt{m(m+1)}}}, \qquad \tilde C = \tilde C (\delta, \norm{\nabla_{y_j} H_{bo}}, \dots),
\end{equation}
where $\tilde C$ is the smallest constant such that $\norm{w_1 u} \leqslant \tilde C \norm{u}_{H^{2,\ka}_y}$, the operator $\heff^{(2)}$ is self-adjoint on $L^2_y$ with domain $H^{2}_{\ka,y}$. b) The operator $H^{(2)}_\ka : \rRan P \to \rRan P$ acting via $H^{(2)}_\ka (\psi_\circ f) = \psi_\circ (\heff^{(2)} f )$ is also self-adjoint. It generates the evolution semigroup 
\begin{equation}
    U^{(2)}_t = e^{- i H^{(2)}_\ka t/\ka},
\end{equation}
which for all $\Psi \in \Ran P$ acts via 
\begin{equation} \label{U2t_action}
    U^{(2)}_t \Psi = \psi_\circ e^{- i \heff^{(2)} t/\ka} g,
\end{equation}
where $g = \lan \psi_\circ, \Psi \ran_{\cHel}$.
\end{lemma}

\begin{proof}[Proof of Lemma \ref{heff(2)_selfadjoint_prop}]
\textit{Proof of a).} Using the Fourier transform on $\R^m$, we can show 
\begin{equation}
    \norm{u}_{H^{2,\ka}_y} \leqslant \frac{\sqrt{m(m+1)}}{2} \norm{T u} + \norm{u}.
\end{equation}
Combining this with the estimate $\norm{w_1 u} \leqslant \tilde C \norm{u}_{H^{2,\ka}_y}$, implies 
\begin{equation}
    \norm{w_1 u} \leqslant \tilde C \frac{\sqrt{m(m+1)}}{2} \norm{T u} + \tilde C \norm{u},
\end{equation}
and hence for $\ka$ satisfying \eqref{k0_def}, the operator $- \ka^2 w_1$ is a relatively bounded perturbation of $T$ with relative bound less than $1$. It follows easily as well that $w_1$ is a symmetric operator. As $E + \ka^2 v$ are bounded symmetric operators, it follows by the Kato-Rellich theorem that $\heff^{(2)}$ is self-adjoint on $L^2_y$ with domain $\cD(T) = H^{2,\ka}_y$.

\textit{Proof of b).} The self-adjointness of $H^{(2)}_\ka$ follows immediately. The proof of \eqref{U2t_action} follows analogous to that of Lemma \ref{lemma_PKP_and_UPt_estimate} b). 
\end{proof}

\begin{remark}
It was show in Lemma 8 of \cite{JansenRuskaiSeiler} that 
\begin{equation}
    \norm{\nabla_{y_j} P} \leqslant \frac{1}{\delta} \norm{\nabla_{y_j} H_{bo}}.
\end{equation}
Generalizing these sorts of estimates would allow us to explicitly compute $\tilde C$ in terms of the spectral gap $\delta$ and the operator norms of derivatives of $H_{bo}$ (which are all bounded by Assumption \ref{assumption1}).
\end{remark}

We write the difference $\psi_\circ\left(f(t) - \tilde f(t)\right)$ as an operator-valued integral in time in order to integrate by parts eventually. 

\begin{lemma}\label{lemma_f(t)-tildef(t)}
Let Assumptions \ref{assumption1} - \ref{assumption4} hold. Recall $X = \frac{i}{\ka} \oP T P$ and its adjoint $X^* = -\frac{i}{\ka} P T \oP$. a) We have 
\begin{equation} \label{f-f2_secondorder_2}
    \psi_\circ\left(f(t) - \tilde f(t)\right) = U^{(2)}_{t} Z_t \oR X \psi_\circ f_0 - \ka^2 \int_0^t U^{(2)}_{t-r} \psi_\circ w_2^\ka [f](r) dr, 
\end{equation}
where 
\begin{equation} \label{Bt_def_f-f2}
    Z_t = - i \ka \int_0^t U^{(2)}_{-r} X^* \oU_r dr. 
\end{equation}
b) Applying Lemma \ref{NAIP_lemma_lefthanded} to $Z_t$ in \eqref{Bt_def_f-f2} yields
\begin{align} \label{Zt_NAIP_1}
    Z_t &= -\ka^2 U_{-r}^{(2)} \oR X \oU_r \bigg|_{r=0}^{r=t} - \ka^2 \int_0^t U^{(2)}_{-r} X_{(2)} \oU_r dr, 
\end{align}
where 
\begin{equation} \label{X(2)_def}
    X_{(2)} = \left(X_2\right)^* - i \ka \psi_\circ w_1 \lan \psi_\circ, X^* \left(\cdot\right) \ran_{\cHel}.
\end{equation}
Here by $*$ we denote the adjoint and recall $X_2$ is defined in \eqref{X2_S_def}.
\end{lemma}

It is easy to see that the right-hand side of \eqref{Bt_def_f-f2} is $O(\ka)$, so the integration by parts is necessary in order to estimate $Z_t = O(\ka^2)$.

\begin{proof}[Proof of Lemma \ref{lemma_f(t)-tildef(t)}]
\textit{Proof of a).} Using the expansion \eqref{wf(t)_expansion_Oka3} from Theorem \ref{MAINTHEOREM_ABSTRACT-PREVIOUS} part (b) and introducing $\heff^{(2)}$ as defined in \eqref{effeq_2ndorder}, we can write
\begin{equation} \label{heff_expansion}
    \heff f(t) = \heff^{(2)} f(t) + \ka^2 \tilde w_1(t) f_0 + i \ka^3 w_2^\ka [f](t). 
\end{equation}
Using the semigroup $U^{(2)}_t$ and \eqref{heff_expansion} we can re-write equation \eqref{effeq_2ndorder} for $\tilde{f}(t)$ and equation \eqref{effectiveeqforf} for $f(t)$ in Duhamel form, and taking the difference yields
\begin{align} \label{f-f2_secondorder_1}
    \psi_\circ \left(f(t) - \tilde{f}(t) \right) &= - i \ka \int_0^t U^{(2)}_{t-r} \psi_\circ \tilde w_1(r) f_0 dr - \ka^2 \int_0^t U^{(2)}_{t-r} \psi_\circ w_2^\ka [f](r) dr. 
\end{align}
By the definition of $\tilde w_1(r)$, see \eqref{tildew1(t)_def}, we have 
\begin{equation}
    U^{(2)}_{t-r} \psi_\circ \tilde w_1(r) f_0 = U^{(2)}_{t-r} X^* \oU_t \oR X \psi_\circ f_0,
\end{equation}
so we write \eqref{f-f2_secondorder_1} in the form \eqref{f-f2_secondorder_2} - \eqref{Bt_def_f-f2}.

\textit{Proof of b).} We now integrate $Z_t$ by parts using the left-handed NAIP as described above. We apply the integration by parts expansion \eqref{abstract_NAIP_lefthanded} with $F_r = U^{(2)}_{t-r} X^*$ and $G_r = 1$ (then $R = -i\ka \oR$), to the right-hand side of \eqref{Bt_def_f-f2}. In this case $F_r' = \frac{i}{\ka} U^{(2)}_{t-r} H^{(2)}_\ka X^*$, so we obtain
\begin{align}
    Z_t &= - \ka^2 U^{(2)}_{t-r} X^* \oR \: \oU_r \bigg|_{r=0}^{r=t} + i \ka \int_0^t U^{(2)}_{t-r} X^* S \oU_r dr  - i \ka \int_0^t U^{(2)}_{t-r} \left( H^{(2)}_\ka X^* -  X^* \oK \right)  \oR \: \oU_{r} dr. \label{Bt_NAIP_f-f2}
\end{align}
Combining the integrals, we obtain \eqref{Zt_NAIP_1} with
\begin{equation}
    X_{(2)} = -\frac{i}{\ka} X^* S - \frac{i}{\ka} \left( H^{(2)}_\ka X^* -  X^* \oK \right) \oR,
\end{equation}
so it remains to show \eqref{X(2)_def}. Using the definition of $H^{(2)}_\ka$ and the identity \eqref{PKP_identity}, we have that for any $g \in L^2_y$, $H^{(2)}_\ka \psi_\circ g = K^P \psi_\circ g - \ka^2 \psi_\circ w_1 g$. Since $X^* = P X^*$, \eqref{X(2)_def} follows.
\end{proof}

We estimate the right-hand side of \eqref{f-f2_secondorder_2} using \eqref{Zt_NAIP_1} and \eqref{X(2)_def}. To do so we will need the following two additional estimates.

\begin{lemma} \label{Bstau_norm_estimate_lemma}
Let Assumptions \ref{assumption1}-\ref{assumption4} hold. a) Estimating the right-hand side of \eqref{Zt_NAIP_1} yields
\begin{equation} \label{Zt_estimate}
    Z_t = O_{\cL_{s+2,s}}(\ka^2 e^{Ct}),
\end{equation}
for some constant $C > 0$ independent of $t$ and $\ka$. b) For solutions $f(t)$ of \eqref{effectiveeqforf} with initial conditions $f(0) = f_0$, we have
\begin{equation} \label{Bstau_norm_estimate_IC}
    \norm{f}_{B^s_\tau} \leqslant C \lan \tau \ran^s \norm{f_0}_{H^s_{\ka ,y}}.
\end{equation}
for some constant $C > 0$ independent of $\tau$ and $\ka$. 
\end{lemma}

\begin{proof}[Proof of Lemma \ref{Bstau_norm_estimate_lemma}]
\textit{Proof of a).} To estimate $X_{(2)}$, we use the estimate of $X_2 =  O_{\cL_{s+2,s}}(1)$ which follows from \eqref{S_and_X2_estimates} in Lemma \ref{lemma_Xt_estimate}, and from the estimate \eqref{wtildej_estimate} we have that $\psi_\circ w_1 \lan \psi_\circ, X^* \left(\cdot\right) \ran_{\cH_{el}} = O_{\cL_{s+2,s}}(1)$. These two estimates together imply 
\begin{equation}
    X_{(2)} = O_{\cL{s+2,s}}(1).
\end{equation}
Combining this with the estimates for the propagator $\oU_t = O_{\cL_{s,s}}(e^{Ct})$ (see Theorem \ref{Ut_Ubart_estimate_theorem}) and using $X = O_{\cL_{s+1,s}}(1)$ and $\oR = O_{\cL_{s,s}}(1)$, we obtain \eqref{Zt_estimate}. 

\textit{Proof of b).} From the definition of the operator $Q_P$ given in (\ref{Qpexpr}), we observe that $\psi_\circ f(t) = P (Q_Pf)(t)$. By Theorem \ref{MAINTHEOREM_ABSTRACT-PREVIOUS} part (a), $Q_Pf$ is a solution of the full Schr{\"o}dinger equation $i \ka \p_t Q_Pf(t) = H_\ka Q_Pf(t)$ with initial conditions $(Q_Pf)(0) = \psi_\circ f_0$. Then,
\begin{equation} \label{PQPf_0}
    \psi_\circ f(t) = P (Q_Pf)(t) = P U_t \psi_\circ f_0.
\end{equation}
Using this and the fact that $P = O_{\cL_{s,s}}(1)$ (see Lemma \ref{smoothness_proposition}), the definition of the $B^s_\tau$ norm \eqref{BsT_norm_def} implies
\begin{align}
    \lVert{f(t)}\rVert_{B^s_\tau} = \sup_{0 \leqslant t \leqslant \tau} \norm{\psi_\circ f(t)}_{\cH_{el} H^s_{\ka ,y}} \lesssim \sup_{0 \leqslant t \leqslant \tau} \norm{U_t \psi_\circ f_0}_{\cH_{el} H^s_{\ka ,y}}.
    \label{HsboundofQP}
\end{align}
The last step is to use the estimate $U_t = O_{\cL_{s,s}}(\lan t \ran^s)$, which is proven in Theorem \ref{Ut_Ubart_estimate_theorem}, equation \eqref{Ut_estimate}, on the right-hand side of \eqref{HsboundofQP}. Then the claim follows.
\end{proof}

\subsection{Conclusion of proof} We now complete the proof of Theorem \ref{TDBO_SECONDORDER_MAINTHEOREM} using the lemmas we have established.

\begin{proof}[Proof of Theorem \ref{TDBO_SECONDORDER_MAINTHEOREM}]

Let $\Psi(t)$ satisfy the Schr{\"o}dinger equation (\ref{abstractSE}) with $\Psi(0) = \psi_\circ f_0$ and let $\tilde{f}(t)$ be the solution of the truncated effective nuclear equation \eqref{effeq_2ndorder} with $\tilde{f}(0) = f_0$. We set $\ka_0$ to the right-hand side of the inequality in \eqref{k0_def} and consider $\ka \in (0, \ka_0)$. Then, using the definition of $Q_P$, see \eqref{Qpexpr}, we can write 
\begin{align} \label{Psi-Qptildef_2}
    \Psi(t) - (Q_P\tilde{f})(t) &= \psi_\circ \left(f(t) - \tilde{f}(t) \right)  + \int_0^t \oU_{t-s} X^* \psi_\circ \left(f(s) - \tilde{f}(s) \right) ds,
\end{align}
where $f(t)$ solves the effective nuclear dynamics \eqref{effectiveeqforf} with $f(0) = f_0$. Using the self-adjointness of $\heff^{(2)}$ (see Lemma \ref{heff(2)_selfadjoint_prop}), we can further write
\begin{equation} \label{f-f2_secondorder_3}
    \psi_\circ\left(f(t) - \tilde f(t)\right) = U^{(2)}_{t} Z_t \oR X \psi_\circ f_0 - \ka^2 \int_0^t U^{(2)}_{t-r} \psi_\circ w_2^\ka [f](r) dr, 
\end{equation}
where $Z_t$, after using the NAIP, admits the expansion \eqref{Zt_NAIP_1} - \eqref{X(2)_def} (see Lemma \ref{lemma_f(t)-tildef(t)}). 

To estimate the first term on the right-hand side of \eqref{f-f2_secondorder_3}, we use Lemma \ref{Bstau_norm_estimate_lemma} part a) to estimate $Z_t$ and the estimate $U^{(2)}_t = O_{\cL_{s,s}} ( \lan t \ran^s)$, which can be obtained by arguing analogously as in the case of $U^P_t$ (see Theorem \ref{theorem_UPt_estimate}). The second term can be estimated by using Lemma \ref{Bstau_norm_estimate_lemma} part b) together with the estimate \eqref{w3f(t)_estimate} to conclude
\begin{equation} \label{w3f(t)_estimate_neq}
    \norm{w^{\ka}_2 [f](t)} \lesssim e^{C \tau} \norm{f_0}_{H^{s+6}_{\ka ,y}}, \qquad 0 \leq t \leq \tau.
\end{equation} 
Using this on the right-hand side of \eqref{f-f2_secondorder_3}, we conclude that
\begin{align}
    \norm{\psi_\circ \left(f(t) - \tilde{f}(t) \right)}_{H^s_{\ka ,y}} &\lesssim \ka^2 \lan t \ran^s \norm{f_0}_{H^{s+3}_{\ka ,y}} + \ka^2 e^{C \tau} \norm{f_0}_{H^{s+6}_{\ka ,y}} \lesssim \ka^2 e^{C \tau} \norm{f_0}_{H^{s+6}_{\ka ,y}}, \qquad 0 \leqslant t \leqslant \tau.
\end{align}
Using this estimate on the right-hand side of \eqref{Psi-Qptildef_2}, we obtain \eqref{TDBOA_secondorder_estimate} and the proof concludes. 
\end{proof}

\appendix

\section{Basic fiber smoothness estimates} \label{BasicFiberEstimates_Section}

In this section, we collect some preliminary lemmas characterizing the range of $P$ and the smoothness of the eigenfunction $\psi_\circ(y)$ as a function of $y \in \R^m$. The proof of Lemma \ref{RanPcharacterizedlemma} is trivial and we omit it. 

\begin{lemma} \label{RanPcharacterizedlemma}
Let Assumptions \ref{assumption1} - \ref{assumption4} hold. The subspace $\textrm{Ran }P \subset \cH = L^2(\R^m, \cH_{el})$ can be characterized in the following way:
\begin{equation}
    \textrm{Ran }P = \{g(y) \psi_\circ(y) \::\: \forall g(y) \in L^2(\R^m)\}, \label{RanPcharacterized}
\end{equation}
where $\psi_\circ = \int^{\oplus}_{\R^m} \psi_\circ(y) dy$.
\end{lemma}

\subsection{Fiber smoothness}

\begin{lemma} \label{smoothness_psicirc_lemma}
Let Assumptions \ref{assumption1}-\ref{assumption4} hold. Using the theory of Riesz integrals, 
\begin{equation} \label{P(y)_Riesz_ytilde}
    P(y) = \frac{1}{2 \pi i} \oint_{\Gamma(\Tilde{y})} R(z,y) dz, \qquad \forall y \in U(\Tilde{y}),
\end{equation} 
where $\Gamma(\tilde{y}) \subset \C$ is a positively oriented closed curve encircling $E(\tilde{y})$ once and $U(\Tilde{y})$ is an open neighbourhood of $\Tilde{y}$ such that 
\begin{equation} \label{Utilde_nbhd}
    \inf_{y \in U(\Tilde{y})} \textrm{dist}(\Gamma(\tilde{y}), \sigma(H(y))) \geqslant \frac{\delta}{4}.
\end{equation}

Then, the smoothness of $P(\cdot)$ and $E(\cdot)$ as in \eqref{smoothness_E_P_statement} holds, and the ground state $\psi_\circ(y)$ is a $k_A$-times differentiable as a vector-function of $y$, in the sense that
\begin{equation} \label{psiy_regularity}
    \sup_y \lVert{\partial_y^\alpha \psi_\circ(y)}\rVert_{\cH_{el}} \leqslant C \quad\text{ for all multi-index } \lvert{\alpha}\rvert \leqslant k_A.
\end{equation}
\end{lemma}

\begin{proof}[Proof of Lemma \ref{smoothness_psicirc_lemma}] The smoothness of the eigenprojection $P(\cdot)$ follows from \eqref{P(y)_Riesz_ytilde}-\eqref{Utilde_nbhd} and the first resolvent identity (see for instance \cite[Theorem 2.2]{Teufel}), making use of the spectral gap and smoothness assumption on the fibers $H(\cdot)$. To prove the differentiability of the ground state $\psi_\circ(\cdot)$ and ground state energy $E(\cdot)$ we use that the eigenvalues are simple and argue as follows. Let $P(y)$, $y, \tilde{y} \in \R^m$, and $U(\tilde{y})$ be as in \eqref{P(y)_Riesz_ytilde}-\eqref{Utilde_nbhd}, i.e. taking $U(\tilde{y})$ sufficiently small and using that $P(\tilde{y}) \psi_\circ(\tilde{y}) = \psi_\circ(\tilde{y})$, we ensure that $P(y) \psi_\circ(\tilde{y}) \neq 0$ for all $y \in U(\tilde{y})$. Hence, by Assumption \ref{assumption1} and $P(\cdot) \in C_b^k(\R^m, \cL(\cH_{el}))$, the following functions 
\begin{equation}
    \nu(y) = \langle \psi_\circ(\tilde{y}), P(y) \psi_\circ(\tilde{y}) \rangle_{\cH_{el}}, \qquad \mu(y) = \langle H(y) \psi_\circ(\tilde{y}), P(y) \psi_\circ(\tilde{y}) \rangle_{\cH_{el}},
\end{equation}
are $k_A$-times differentiable in $y$ and $\inf_{y \in \R^m} \nu(y) > 0$. Then, we conclude \eqref{smoothness_E_P_statement} and \eqref{psiy_regularity} using 
\begin{equation} \label{eqfor_E(y)}
    E(y) = \nu(y)^{-1} \mu(y), \qquad \psi_\circ(y) = \nu(y)^{-\frac{1}{2}} P(y) \psi_\circ(\tilde{y}).
\end{equation}
\end{proof}

\subsection{Resolvent (gap) estimates: Proof of Lemma \ref{(Hbar - E)inverselemma}} \label{Resolventestimates_Section}

We mention first that Assumption \ref{assumption1} implies the boundedness of the operators $\partial_y^\alpha H_{bo}$, where 
\begin{equation}
    \partial_y^\alpha H_{bo} = \int_{\R^m}^{\oplus} \partial_y^\alpha H(y) dy.
\end{equation}



\begin{proof}[Proof of Lemma \ref{(Hbar - E)inverselemma}]
Since $E$ is a multiplication operator by the bounded real function $E(y)$, $\oH_{\rm bo} - \oE$ is a self-adjoint operator with domain $\overline{\mathcal{D}}$. They by \eqref{Hkappa_def} and (\ref{gapconditionuniformly}), we have 
\begin{equation}
\oH_{\rm bo} = \int^\oplus_{\R^m} \overline{H}(y) dy \geqslant \int^\oplus_{\R^m} E_1(y) \overline{P}(y) dy,
\end{equation}
where $E_1(y) := \inf \left(\sigma(H(y)) \setminus \{E(y)\} \right)$. Let $\Psi := \int^\oplus_{\R^m} \Psi(y) dy$ be any element of $\textrm{Ran }\overline{P}$. Then, by the gap condition (\ref{gapconditionuniformly}), 
\begin{align}
    \langle \Psi, (\oH_{\rm bo}-\oE) \Psi \rangle \geqslant \int^\oplus_{\R^m} \langle \Psi(y), (\overline{H}(y) - \oE(y)) \Psi(y) \rangle dy &\geqslant \int^\oplus_{\R^m} \langle \Psi(y), (E_1(y)-E(y)) \Psi(y) \rangle dy \nonumber \\
    &\geqslant \int^\oplus_{\R^m} \delta \lVert{\Psi(y)}\rVert^2_{\cH_{el}} dy.
\end{align}
Since $\int^\oplus_{\R^m} \lVert{\Psi(y)}\rVert^2_{\cH_{el}} dy = \lVert{\Psi}\rVert^2$, this gives
\begin{equation}
    \langle \Psi, (\oH_{\rm bo} - \oE) \Psi \rangle \geqslant \delta \lVert{\Psi}\rVert^2. \label{lowerboundHbar-E}
\end{equation}
Since $\oH_{\rm bo} - \oE$ is self-adjoint, equation (\ref{lowerboundHbar-E}) yields $\sigma(\oH_{\rm bo} - \oE) \subset [\delta, \infty)$ and therefore the operator $\oH_{\rm bo} - \oE$ is invertible. Furthermore, we have by a standard spectral estimate, 
\begin{equation}
    \lVert{(\oH_{\rm bo} - \oE)^{-1}}\rVert \leqslant \frac{1}{\textrm{dist}(0, \sigma(\oH_{\rm bo} - \oE))},  
\end{equation}
which implies the first estimate in (\ref{oR_estimate}). To prove that the fibers 
\begin{equation}
    \oR(y) = \oP(y) (\oH(y) - \oE(y))^{-1} \oP(y)
\end{equation}
satisfy $\oR(\cdot) \in C_b^k(\R^m, \cL(\cH_{el}))$, we use that, by Lemma \ref{smoothness_proposition} and the definition $\oP(y) = \one - P(y)$, we have $\oP(\cdot) \in C_b^k(\R^m, \cL(\cH_{el}))$ with $\partial_y^\alpha \oP(y) = - \partial_y^\alpha P(y)$. Since $[P(y), H(y)] = 0$, we have that $\oH(y) = \oP(y) H(y) \oP(y) = H(y)\oP(y)$. By the Liebnitz rule, for a multi-index $\alpha = (\alpha_1, \dots, \alpha_m)$,  
\begin{align}
    \partial_y^\alpha \oH(y) = \partial_y^\alpha (H(y) \oP(y)) = - \sum_{0 \leqslant \beta \leqslant \alpha} {{\alpha}\choose{\beta}} \left(\partial_y^\beta H(y) \right) \left( \partial_y^{\alpha-\beta} P(y) \right), \label{H(y)_partialyP(y)_0}
\end{align}
where recall for multi-indices, $\beta \leqslant \alpha$ if $\beta_j \leqslant \alpha_j$ for all $j = 1, \dots, m$ and ${{\alpha}\choose{\beta}} := {{\alpha_1}\choose{\beta_1}} \dots {{\alpha_m}\choose{\beta_m}}$. By Assumption \ref{assumption1} it suffices to consider the term $H(y) \partial_y^\alpha P(y)$. We choose $\tilde{y} \in \R^m$ and a neighbourhood $U(\tilde{y})$ such that 
\begin{equation} \label{Utildey_condition}
    \inf_{y \in U(\Tilde{y})} \textrm{dist}(\Gamma(\tilde{y}), \sigma(H(y))) \geqslant \frac{\delta}{4}.
\end{equation}
Then, for $y \in U(\Tilde{y})$, 
\begin{equation}
    H(y) \partial_y^\alpha P(y) = \frac{1}{2 \pi i} \oint_{\Gamma(\tilde{y})} H(y) \partial_y^\alpha R(z,y) dz. \label{H(y)_partialyP(y)_1}
\end{equation}
Using the first resolvent identity to differentiate $R(z,y)$ and using the identity $H(y) R(z,y) = 1 + z R(z,y)$, it follows from \eqref{H(y)_partialyP(y)_1} that $H(\cdot) \partial_y^\alpha P(\cdot) \in C^{k - \nrm{\alpha}}_b(\R^m, \cL(\cH_{el}))$. Using this, \eqref{H(y)_partialyP(y)_0}, and Assumption \ref{assumption1}, we obtain 
\begin{equation}
    \partial_y^\alpha \oH_{bo}(\cdot) \in C^{k-\nrm{\alpha}}_b(\R^m, \cL(\cH_{el})).
\end{equation}
With this, proceeding as in the proof of Lemma \ref{smoothness_proposition}, we find that $\oR(\cdot) \in C_b^k(\R^m, \cL(\cH_{el}))$ giving \eqref{Rbar(y)_fibers}. The estimate \eqref{oR_estimate} follows from 
\begin{equation}
    D^s \oR(y) = [D^s, \oR(y)] + \oR(y) D^s
\end{equation}  
combined with the regularity $\oR(\cdot) \in C_b^k(\R^m, \cL(\rRan{\oP}))$ and Lemma \ref{LeibnizRule_lemma}.
\end{proof}

\section{Self-adjointness} \label{SelfAdjointness_Section}

In this section, we establish the self-adjointness of the reduced Hamiltonians $H^P$ and $\oH$ in Lemma \ref{Hbarselfadjointlemma}. Thanks to this lemma, $\overline{H}$ generates the one-parameter unitary group $e^{-i \overline{H} t} $. For the reader's convenience, we recall that 
\begin{equation}
D_{y_j} := -i\kappa \partial_{y_j}, \qquad  D^\alpha = \prod_{j=1}^m D^{\alpha_j}_{y_j}. 
\label{eq:differentials in kappa}
\end{equation}
First, we prove a preliminary lemma that we will use in the proof of self-adjointness.

\begin{lemma}\label{T_relativelybounded_Hkappa_lemma}
Let Assumptions \ref{assumption1} - \ref{assumption4} hold. Then, the re-scaled Laplacian $T = - \ka^2 \Delta_y$ is a relatively bounded perturbation of $H_\ka$, i.e. for some constant $C > 0$,
\begin{equation} \label{TrelboundedbyH}
\lVert{-\ka^2 \Delta_y \phi}\rVert \leqslant \lVert{H_\ka \phi}\rVert + C \lVert{\phi}\rVert, \quad \phi \in \cD(H_\ka).
\end{equation}
\end{lemma}

\begin{proof}[Proof of Lemma \ref{T_relativelybounded_Hkappa_lemma}]
To begin, consider the quantity $2 \rRe \langle -\ka^2 \Delta_y \phi, \tilde{H}_{bo} \phi \rangle$. Here, $\tilde{H}_{bo} = H_{bo} + \gamma$ for some constant $0 < \gamma < \infty$  so that $\tilde{H}_{bo} \geqslant 0$ (see Assumption \ref{assumption1}). Since $- \ka^2 \Delta_y = (i \ka \nabla_y)^2$, we have using $\tilde{H}_{bo} \geqslant 0$ that 
\begin{align}
    \langle -\ka^2 \Delta_y \phi, \tilde{H}_{bo} \phi \rangle &= \langle i \ka \nabla_y \phi, \tilde{H}_{bo} (i \ka \nabla_y \phi) \rangle + \langle  i \ka \nabla_y \phi, (i \ka \nabla_y H_{bo} ) \phi \rangle \geqslant \langle i \ka \nabla_y \phi, i \ka (\nabla_y H_{bo} ) \phi \rangle,
\end{align}
which implies 
\begin{equation} \label{Hdiagselfadjoint_Repart_boundedbelow1}
    2 \rRe \langle -\ka^2 \Delta_y \phi, \tilde{H}_{bo} \phi \rangle \geqslant 2 \rRe \langle i \ka \nabla_y \phi, i \ka (\nabla_y H_{bo} ) \phi \rangle.
\end{equation}
Using the product rule and Assumption \ref{assumption1}, $i \ka (\nabla_y H_{bo} )$ is antisymmetric,
\begin{align}
    \langle \phi, i \ka (\nabla_y H_{bo} ) \psi \rangle = \langle \phi, i \ka \nabla_y (H_{bo} \psi) - H_{bo} (i \ka \nabla_y \psi) \rangle &= \langle H_{bo} (i \ka \nabla_y \phi) - i \ka \nabla_y (H_{bo} \phi), \psi \rangle dy \nonumber \\
    &= \langle - i \ka (\nabla_y H_{bo}) \phi, \psi \rangle.
\end{align}
This implies
\begin{align}
    \overline{\langle i \ka \nabla_y \phi, i \ka (\nabla_y H_{bo}) \phi \rangle} = \langle i \ka (\nabla_y H_{bo}) \phi, i \ka \nabla_y \phi \rangle 
    &= \langle (-\ka^2 \Delta_y H_{bo}) \phi, \phi \rangle + \langle i \ka (\nabla_y H_{bo}) (i \ka \nabla_y \phi), \phi \rangle \nonumber \\
    &= \langle (-\ka^2 \Delta_y H_{bo}) \phi, \phi \rangle - \langle i \ka \nabla_y \phi, i \ka (\nabla_y H_{bo}) \phi \rangle,
\end{align}
so that by re-arranging we obtain 
\begin{equation} \label{Hdiagselfadjoint_Repart_boundedbelow2}
    2 \rRe \langle i \ka \nabla_y \phi, i \ka (\nabla_y H_{bo} ) \phi \rangle = 2 \rRe \langle (-\ka^2 \Delta_y H_{bo}) \phi, \phi \rangle.
\end{equation}
By Assumption \ref{assumption1}, $(-\Delta_y H_{bo})$ is a bounded operator. Hence \eqref{Hdiagselfadjoint_Repart_boundedbelow1} and \eqref{Hdiagselfadjoint_Repart_boundedbelow2} together imply that
\begin{equation}
    2 \rRe \langle -\ka^2 \Delta_y \phi, \tilde{H}_{bo} \phi \rangle \geqslant - C \ka^2 \norm{\phi}^2
\end{equation}
for some constant $C > 0$. As $\ka < 1$, we obtain
\begin{equation}
    2 \rRe \langle -\ka^2 \Delta_y \phi, \tilde{H}_{bo} \phi \rangle + C \norm{\phi}^2 \geqslant 0.
\end{equation}
Add to both sides of the inequality above, we obtain
\begin{align}
    \norm{- \ka^2 \Delta_y \phi}^2 &\leqslant \norm{- \ka^2 \Delta_y \phi}^2 + \norm{\tilde{H}_{bo} \phi}^2 + 2 \rRe \langle -\ka^2 \Delta_y \phi, \tilde{H}_{bo} \phi \rangle + C \norm{\phi}^2 \nonumber \\
    &= \norm{(-\ka^2 \Delta_y + \tilde{H}_{bo}) \phi}^2 + C \norm{\phi}^2 \nonumber \\
    &\leqslant \left(\norm{(-\ka^2 \Delta_y + H_{bo}) \phi} + (\sqrt{C} + \nrm{\gamma}) \norm{\phi} \right)^2
\end{align}
and so (\ref{TrelboundedbyH}) follows. 
\end{proof}

\begin{proposition} \label{Hbarselfadjointlemma} Let Assumptions \ref{assumption1} - \ref{assumption4} hold. The operators $H^P$ and $\overline{H}$ defined by
\begin{equation}
    H^P = P H_{\kappa} P, \quad \overline{H} = \oP H_{\kappa} \oP,
\end{equation}
are self-adjoint on the orthogonal subspaces $\textrm{Ran }P$ and $\textrm{Ran }\overline{P}$ with domains $\mathcal{D}(H^P) = H^{2,\ka}_y \otimes \psi_\circ$ and $\mathcal{D}(\overline{P}) = \overline{P} \mathcal{D}(H_{\ka})$, respectively.
\end{proposition}

\begin{proof}[Proof of Proposition \ref{Hbarselfadjointlemma}]

\textit{Step 1. The diagonalized Hamiltonian $H_{diag} := H^P + \overline{H}$ is self-adjoint on $\mathcal{D}(H)$.} We claim it is self-adjoint on $\cH = L^2(\R^m, \cH_{el})$ with domain $\mathcal{D}(H_{diag}) = \mathcal{D}(H_{\ka})$. Writing
\begin{align}
    H_{diag} &= H - \Lambda, \textrm{ where } \Lambda = P H_{\ka} \overline{P} + \overline{P} H_{\ka} P, \label{Hdiag}
\end{align}
we argue that the operator $\Lambda$ is a first-order differential operator. To see this, first consider $P H_{\ka} \overline{P}$. Since $H_{\ka} = -\ka^2 \Delta_y + H_{bo}$ and $P H_{bo} \oP = H_{bo} P \oP = 0$, then
\begin{equation}
    P H_{\ka} \overline{P} = - \ka^2 P \Delta_y \oP = \sum_{j=1}^m P D_{y_j}^2 \oP. \label{Hdiag_selfadjoint1}
\end{equation}
Commuting $D_{y_j}$ to the right, we have by the orthogonality of $P$ and $\oP$ that $P D_{y_j}^2 \oP = P [D_{y_j}^2, \oP]$ and 
\begin{equation}
    P [D_{y_j}^2, \oP] = P \left( D_{y_j} (D_{y_j} \oP) + (D_{y_j} \oP) D_{y_j} \right) = P (D^2_{y_j} \oP) + 2 P (D_{y_j} \oP) D_{y_j}. \label{Hdiag_selfadjoint2}
\end{equation}
The operators $(D^2_{y_j} \oP)$ and $(D_{y_j} \oP)$ are bounded operators by Lemma \ref{smoothness_proposition}. Combining \eqref{Hdiag_selfadjoint2} with \eqref{Hdiag_selfadjoint1}, it follows that
\begin{equation}
    \lVert{\overline{P} H_{\ka} P \phi}\rVert^2 \leqslant C_1 \sum_1^m \lVert{-i \ka \partial_{y_j} \phi}\rVert^2 + C_2 \lVert{\phi}\rVert^2.  \label{PbarTPrelativebound0}
\end{equation}
It is standard that $-i\ka \partial_{y_j}$ is relatively bounded with respect to $-\ka^2 \Delta_y$ with relative bound $0$ (cf. \cite{GS}, \cite{RSv2}, \cite{BSV4}), so \eqref{PbarTPrelativebound0} implies that $P H_{\ka} \oP$ is relatively bounded by $-\ka^2 \Delta_y$ with relative bound $0$. Arguing similarly for the adjoint $\oP H_{\ka} P = (P H_{\ka} \oP)^*$, we conclude that $\Lambda = \overline{P} H_{\ka} P + P H_{\ka} \overline{P}$ is a relatively-bounded perturbation of $-\ka^2 \Delta_y$ with relative bound $0$. Now, $-\ka^2 \Delta_y$ itself is a relatively-bounded perturbation of $H_\ka$ thanks to Lemma \ref{T_relativelybounded_Hkappa_lemma} and \eqref{TrelboundedbyH}, so we conclude that $\Lambda$ is a relatively bounded perturbation of $H_\ka$ with relative bound $0$. Hence the operator $H_{diag}$ is self-adjoint on $\mathcal{D}(H)$. 

\textit{Step 2. The operators $\overline{H}$ and $H^P$ are self-adjoint.} Clearly they are both symmetric, so it remains to show they are densely-defined, closed, and satisfy the fundamental criterion of self-adjointness. To see that they are densely defined, we use (\ref{RanPcharacterized}) from Lemma \ref{RanPcharacterizedlemma}. Now, as $H^{2,\ka}_y$ is dense in $L^2_y$, (\ref{RanPcharacterized}) implies that $\mathcal{D}(H^P) = H^{2,\ka}_y \otimes \psi_\circ$ is dense in $\textrm{Ran }P$ and $\mathcal{D}(\overline{H})$ is dense in $\textrm{Ran }\overline{P}$, respectively.

$\overline{P}$ is a bounded operator and so $H \overline{P}$ is closed since $H$ is closed. This implies that $(H\overline{P})^* = \overline{P} H$ is closed as it is densely defined (cf Theorem 7.1.1 in \cite{BSV4}). Thus, $\overline{P} H \overline{P} = \overline{H}$ is also closed and analogously $H^P$ is closed. 

It remains to show the fundamental criterion of self-adjointness: that for any $f \in \textrm{Ran } \overline{P}$ and $g \in \textrm{Ran }P$ there exist $u_\pm \in \mathcal{D}(\overline{H})$ and $v_\pm \in \mathcal{D}(H^P)$ such that
\begin{equation} \label{HbarHPself-adjoint}
    (\overline{H} \pm i)u_\pm = f, \quad (H^P \pm i)v_\pm = g.
\end{equation}
In what follows we will consider $\overline{H} + i$ and $H^P + i$, as the argument follows analogously for the other case. As $H_{diag}$ is self-adjoint it must satisfy the fundamental criterion of self-adjointness, so for any $h \in L^2$ there exists a $w \in \mathcal{D}(H)$ such that 
\begin{equation} \label{Hdiagselfadjoint}
    (H_{diag} + i) w = h.
\end{equation}
Writing $H_{diag} = \overline{H} + H^P$ and applying the projections $P$ and $\overline{P}$ to (\ref{Hdiagselfadjoint}) yields the following two equations:
\begin{equation}
    H^P u + i u = P h, \quad \overline{H} v + i v = \overline{P} h, \label{Hdiagselfadjoint1} 
\end{equation}
where $u = P w$ and $v = \overline{P} w$. Hence, for any $f \in \textrm{Ran } \overline{P}$ and $g \in \textrm{Ran }P$, solving (\ref{Hdiagselfadjoint1}) for $h = f + g$ will yield the solutions $u = \overline{P} w$ and $v = P w$ to (\ref{HbarHPself-adjoint}), so $\overline{H} + i$ and $H^P + i$ are surjective on $\textrm{Ran }\overline{P}$ and $\textrm{Ran }P$ respectively. Arguing analogously yields that $\overline{H} - i$ and $H^P - i$ are surjective as well.
\end{proof}

\section{Estimates of commutators} \label{commutatorestimates_Section}

In this section we collect all estimates on commutators used in this paper. According to e.g. \cite[Section 12.4]{Grubb} or \cite[Section 6.2]{AMG}, if $A$ and $B$ are two given differential operators acting on Sobolev spaces of the form $H^s_{\ka ,y} \cH_{el}$, their commutator $[A,B]$ is the operator defined by
\begin{equation}
    \langle f, [A,T] g \rangle = \langle A^* f, T g \rangle - \langle T^* f, A g \rangle,
\label{eq:commutator definition}
\end{equation} for $f,g$ in a some dense subspace of the domains ob both $A$ and $B$. We shall use this definition throughout this section.

Recall that the operators $D_{y_j}$ and $D^\alpha$ are given in (\ref{eq:differentials in kappa}).

We use the following elementary lemma (presented without proof) when dealing with commutators of fibered operators.



\begin{lemma} \label{[Fibered,E]_lemma}
Let $A = \int_{\R^m}^{\oplus} A(y) dy$ and $F = \int_{\R^m}^{\oplus} F(y) dy$ be two fibered operators on $\cH$. Then, 
\begin{equation}
    [A, F] = \int_{\R^m}^{\oplus} [A(y), F(y)] dy.
\end{equation}
If, in particular, $F(y) = f(y)$ where $f: \R^m \to \C$, then
\begin{equation} 
    [A, F] = 0.
\end{equation}
Moreover, 
\begin{equation} 
    [D_{y_j}, A] = \int_{\R^m}^{\oplus} (D_{y_j} A(y)) dy \qquad \forall \: j.
\end{equation}
\end{lemma}

We will also make extensive use of the Leibniz (generalized product) rule. 

\begin{lemma} \label{LeibnizRule_lemma}
(Leibniz Rule) Let $F(y) \in C_b^k \left(\R^m, \cL_{s_1, s_2} \right)$ for some integers $s_1, s_2 \geqslant 0$ and $k \geqslant 1$. Then, for any $y_i \in \R$,
\begin{equation} \label{[Dyk,F(y)]_expansion}
    [D_{y_i}^k, F(y)] = \sum_{j=0}^{k-1} {{k}\choose{k-j}} (-i\ka)^{k-j} (\partial_{y_i}^{k-j} F(y)) D_{y_i}^j.
\end{equation}
For a multi-index $\alpha = (\alpha_1, \dots, \alpha_m)$ with $\nrm{\alpha} \leqslant k$, 
\begin{equation} \label{[Dyalpha,F(y)]_expansion_multiindex}
    [D^\alpha, F(y)] = \sum_{\substack{0 \leqslant \beta \leqslant \alpha \\ \beta \neq \alpha}} {{\alpha}\choose{\alpha - \beta}} (-i\ka)^{\nrm{\alpha - \beta}} (\partial_y^{\alpha - \beta} F(y)) D_{y}^\beta,
\end{equation}
where recall for multi-indices, $\beta \leqslant \alpha$ if $\beta_j \leqslant \alpha_j$ for all $j = 1, \dots, m$ and ${{\alpha}\choose{\beta}} := {{\alpha_1}\choose{\beta_1}} \dots {{\alpha_m}\choose{\beta_m}}$.

In particular, if $F = \int_{\R^m}^{\oplus} F(y) dy$ is a decomposable operator, then \eqref{[Dyalpha,F(y)]_expansion_multiindex} implies
\begin{align}
    [D^\alpha, F] &= \sum_{\substack{0 \leqslant \beta \leqslant \alpha \\ \beta \neq \alpha}} {{\alpha}\choose{\alpha - \beta}} (-i\ka)^{\nrm{\alpha - \beta}} (\partial_y^{\alpha - \beta} F) D_{y}^\beta,
\end{align}
where $\partial_y^{\alpha - \beta} F = \int_{\R^m}^{\oplus} (\partial_y^{\alpha - \beta} F(y)) dy.$
\end{lemma}

We shall also need the following general result.

\begin{lemma} \label{lemma_recursivecommutator}
Let $X^\circ$ be a differential operator. Then,
\begin{equation} \label{C_recursivecommutator}
    \oP \left( K \oP X^\circ - X^\circ P K \right) P = \oP \left[X^\circ, [K, P] - K \right] P.
\end{equation}
\end{lemma}

\begin{proof}[Proof of Lemma \ref{lemma_recursivecommutator}]
Commuting the $K$ operators through the projections $\oP$ and $P$, we obtain
\begin{align}
    \oP \left( K \oP X^\circ -  X^\circ P K \right) P = \oP \left(K X^\circ + [K, \oP] X^\circ - X^\circ K - X^\circ [P, K] \right) P. \label{Cj_recursive_0}
\end{align}
We re-arrange the terms on the right-hand side of the above as follows. First, we notice that $K X^\circ - X^\circ K = - [X^\circ, K]$. Furthermore, due to $[K, \oP] = - [K, P]$ and $-[P, K] = [K, P]$, we have $[K, \oP] X^\circ - X^\circ [P, K] = [X^\circ, [K, P]]$. Using these two identities on the right-hand side of \eqref{Cj_recursive_0} yields \eqref{C_recursivecommutator}.
\end{proof}

\subsection{Commutators involving $H_\ka$, $\oH$ and derivatives}

\begin{lemma} \label{UtUbartcommutator_extend_bounded_lemma}
Let assumption \ref{assumption1} hold for some $k \geqslant 1$ and let $\alpha$ be a multi-index with $1 \leqslant \nrm{\alpha} = s \leqslant k$. Then, 
\begin{align}
    [i H_\ka, D^\alpha] &= O_{\cL_{s-1, 0}}(\ka), \label{HkaDyalpha_form_bounded} \\
    [i \oH, \overline{D^\alpha}] &= O_{\cL_{s, 0}}(\ka). \label{HbarDyalphabar_form_bounded}
\end{align}
\end{lemma}

\begin{proof}[Proof of Lemma \ref{UtUbartcommutator_extend_bounded_lemma}] 
We define $[i H_\ka, D^\alpha] = -[D^\alpha, i H_\ka]$ via the sesquilinear form 
\begin{equation} \label{HkaDyalpha_form_definition}
    \langle f, [i H_\ka, D^\alpha] g \rangle := \langle D^\alpha f, i H_\ka g \rangle - \langle i H_\ka f, D^\alpha g \rangle
\end{equation}
for $f, g \in \mathcal{D}(H_\ka) \cap H^s_{\ka, y} \cH_{el}$. Similarly, we define $[i \oH, \overline{D^\alpha}] = -[\overline{D^\alpha}, i \oH]$ via the sesquilinear form 
\begin{equation} \label{HbarDyalphabar_form_definition}
    \langle f, [i \oH, \overline{D^\alpha}] g \rangle := \langle \overline{D^\alpha} f, i \oH g \rangle - \langle i \oH f, \overline{D^\alpha} g \rangle
\end{equation}
for $f, g \in \mathcal{D}(\oH) \cap \oP H^s_{\ka, y} \cH_{el}$. 

It remains to show that the forms $[i H_\ka, D^\alpha]$ and $[i \oH, \overline{D^\alpha}]$ defined in \eqref{HkaDyalpha_form_definition} and \eqref{HbarDyalphabar_form_definition} respectively extend to the bounded operators on the domains $H^s_{\ka, y} \cH_{el}$ and  $\oP H^s_{\ka, y} \cH_{el}$ respectively. We use the Leibniz Rule Lemma \ref{LeibnizRule_lemma} to compute the commutators on a suitable dense subset. 

\textit{Proof of \eqref{HkaDyalpha_form_bounded}:} For $f, g \in H^s_{\ka ,y} \cH_{el} \cap \cD(H_\ka) = H^{s+2}_{\ka ,y} \cH_{el} \cap L^2_y \cD$, 
\begin{equation}
    \langle D^\alpha f, i H_\ka g \rangle - \langle i H_\ka f, D^\alpha g \rangle = \langle f, [D^\alpha, i H_\ka] g \rangle.
\end{equation}
Recall that $H_\ka = T + H_{bo}$, where $T = -\ka^2 \Delta_y$. For any multi-index $\alpha$, $[D^\alpha, T] = [D^\alpha, -\ka^2 \Delta_y] = 0$. Then,
\begin{equation}
    [D^\alpha, H_\ka] = [D^\alpha, H_{bo}],
\end{equation}
and using the Leibniz Rule Lemma \ref{LeibnizRule_lemma} we obtain 
\begin{equation} \label{[Dyalpha,Hbo]_1}
    [D^\alpha, H_{bo}] = \sum_{\substack{0 \leqslant \beta \leqslant \alpha \\ \beta \neq \alpha}} {{\alpha}\choose{\alpha - \beta}} (-i\ka)^{\nrm{\alpha - \beta}} (\partial_y^{\alpha - \beta} H_{bo}) D_{y}^\beta,
\end{equation}
where $\partial_y^{\alpha - \beta} H_{bo} = \int^{\oplus}_{\R^m} (\partial_y^{\alpha - \beta} H(y)) dy.$ For all terms in the sum it is true that $\nrm{\alpha - \beta} \leqslant k$, and the furthermore the condition $\beta \neq \alpha$ implies $\nrm{\alpha - \beta} \geqslant 1$ for all $\beta$. Hence by assumption \ref{assumption1}, $(\partial_y^{\alpha - \beta} H(y)) = O(1)$ for all $\beta$, and thus $[D^\alpha, H_\ka]$ is of the form 
\begin{equation}
    [D^\alpha, H_\ka] = \sum_{j=0}^{\nrm{\alpha}-1} O_{\cL_{j,0}}(\ka^{\nrm{\alpha} - j}) = O_{\cL_{\nrm{\alpha} - 1,0}}(\ka), \label{[Dyalpha,Hbo]_2}
\end{equation}
using $\norm{\phi}_{H^{j,\ka}_y \cH_{el}} \lesssim \norm{\phi}_{H^{n,\ka}_y \cH_{el}}$ for all $j < n$. Summing over all $\nrm{\alpha} = s$ yields the desired estimate \eqref{HkaDyalpha_form_bounded}.

\textit{Proof of \eqref{HbarDyalphabar_form_bounded}:}  For $f, g \in \oP H^s_{\ka ,y} \cH_{el} \cap \cD(\oH) = \oP H^{s+ 2, \ka}_y \cH_{el} \cap L^2_y \cD_{el}$, 
\begin{equation}
    \langle \overline{D^\alpha} f, i \oH g \rangle - \langle i \oH f, \overline{D^\alpha} g \rangle = \langle f, [\overline{D^\alpha}, i \oH] g \rangle.
\end{equation}
Using $\oH = \oH_{bo} + \overline{T}$, we have 
\begin{equation} \label{[oDalpha,oH]_1}
    [\overline{D^\alpha}, i \oH] = [\overline{D^\alpha}, i \oH_{bo}] + [\overline{D^\alpha}, i \overline{T}].
\end{equation}
Consider the first commutator on the right-hand side of \eqref{[oDalpha,oH]_1}. Since $[\oP, H_{bo}] = 0$, we have $\oH_{bo} = \oP H_{bo} \oP = H_{bo} \oP = \oP H_{bo}$, and thus
\begin{align}
    [\overline{D^\alpha}, i \oH_{bo}] = \oP D^\alpha \oP H_{bo} \oP - \oP H_{bo} \oP D^\alpha \oP = i \oP \left( D^\alpha H_{bo} - H_{bo} D^\alpha \right) \oP = i \oP [D^\alpha, H_{bo}] \oP \label{[oDalpha,oH]_1.5}
\end{align}
Using \eqref{[Dyalpha,Hbo]_1} and \eqref{[Dyalpha,Hbo]_2} it is clear that this term is $O_{\cL_{\nrm{\alpha}-1, 0}}(\ka)$, hence 
\begin{equation}
    [\overline{D^\alpha}, i \oH_{bo}] = O_{\cL_{\nrm{\alpha}-1, 0}}(\ka). \label{[oDalpha,oH]_2}
\end{equation}
Now consider the second commutator on the right-hand side of \eqref{[oDalpha,oH]_1}. Writing $T = \sum_{j=1}^m D_{y_j}^2$, we have 
\begin{equation} \label{[oDalpha,oH]_2.5}
    [\overline{D^\alpha}, i \overline{T}] = i \sum_{j=1}^m [\overline{D^\alpha}, \overline{D_{y_j}^2}], 
\end{equation}
and we compute the commutator $[\overline{D^\alpha}, \overline{D_{y_j}^2}]$ to obtain 
\begin{equation} \label{[oDalpha,oH]_6}
    [\overline{D^\alpha}, i \overline{T}] = i \sum_{j=1}^m \oP [[D^\alpha, P],[D_{y_j}^2,P] ] \oP. 
\end{equation} 
We postpone the computation of \eqref{[oDalpha,oH]_6} to the end. Using the Leibniz Rule Lemma \ref{LeibnizRule_lemma}, we compute
\begin{align*}
    [D^\alpha, P] &= \sum_{\substack{0 \leqslant \beta \leqslant \alpha \\ \beta \neq \alpha}} {{\alpha}\choose{\alpha - \beta}} (-i\ka)^{\nrm{\alpha - \beta}} (\partial_y^{\alpha - \beta} P) D_{y}^\beta, \\
    [D_{y_j}^2,P] &= -\ka^2 (\partial_{y_j}^2 P) - 2 i \ka (\partial_{y_j} P) D_{y_j},
\end{align*}
where by Lemma \ref{smoothness_proposition}, $\partial_y^{\gamma} P = O(1)$ for all $\gamma$ with $\nrm{\gamma} \leqslant k$, $k$ given in assumption \ref{assumption1}. Thus, the commutator $[\overline{D^\alpha}, i \overline{T}]$ is of the form
\begin{align}
    [\overline{D^\alpha}, i \overline{T}] &= \sum_{\substack{0 \leqslant \beta \leqslant \alpha, \: \beta \neq \alpha \\ j = 1, \dots, m}} O(\ka^3) \oP [(\partial_y^{\alpha - \beta} P) D_{y}^\beta, (\partial_{y_j}^2 P)] \oP + O(\ka^2) \oP [(\partial_y^{\alpha - \beta} P) D_{y}^\beta, (\partial_{y_j} P) D_{y_j}] \oP.
\end{align}
As $[(\partial_y^{\alpha - \beta} P), (\partial_{y_j} P)] \neq 0$ and $\nrm{\beta} = \nrm{\alpha} - 1$, it follows that
\begin{equation} \label{[oDalpha,oH]_7}
    [\overline{D^\alpha}, i \overline{T}] =  O_{\cL_{\nrm{\alpha}, 0}}(\ka^2).
\end{equation}
Apply the estimates \eqref{[oDalpha,oH]_2} and \eqref{[oDalpha,oH]_7} together to the right-hand side of \eqref{[oDalpha,oH]_1}, we obtain 
\begin{equation}
    [\overline{D^\alpha}, i \oH] = O_{\cL_{\nrm{\alpha}-1, 0}}(\ka) + O_{\cL_{\nrm{\alpha}, 0}}(\ka^2),
\end{equation}
from which we obtain \eqref{HbarDyalphabar_form_bounded}.

It remains to show \eqref{[oDalpha,oH]_6}. Analogous to \eqref{[oDalpha,oH]_1.5}, 
\begin{equation}
    [\overline{D^\alpha}, \overline{D_{y_j}^2}] = \oP \left( D^\alpha \oP D_{y_j}^2 - D_{y_j}^2 \oP D^\alpha \right) \oP.
\end{equation}
Using $\oP = 1 - P$, 
\begin{equation}
    [\overline{D^\alpha}, \overline{D_{y_j}^2}] = \oP \left( [D^\alpha, D_{y_j}^2] - D^\alpha P D_{y_j}^2 + D_{y_j}^2 P D^\alpha \right) \oP,
\end{equation}
and since $[D^\alpha, D_{y_j}^2] = 0$ we obtain
\begin{equation} \label{[oDalpha,oH]_3}
    [\overline{D^\alpha}, \overline{D_{y_j}^2}] = \oP \left(D_{y_j}^2 P D^\alpha - D^\alpha P D_{y_j}^2 \right) \oP.
\end{equation}
Writing $P^2 = P$, we use that $\oP D_{y_j}^2 P = \oP [D_{y_j}^2, P]$ and $P D^\alpha \oP = [P, D^\alpha] \oP = - [D^\alpha, P] \oP$ to write
\begin{equation} \label{[oDalpha,oH]_4}
    \oP D_{y_j}^2 P D^\alpha \oP = - \oP [D_{y_j}^2, P] [D^\alpha,P] \oP,
\end{equation}
and analogously for the adjoint terms, 
\begin{equation} \label{[oDalpha,oH]_5}
    - \oP D^\alpha P D_{y_j}^2 \oP = \oP [D^\alpha, P] [D_{y_j}^2, P] \oP.
\end{equation}
Then combining \eqref{[oDalpha,oH]_4} and \eqref{[oDalpha,oH]_5} with \eqref{[oDalpha,oH]_3}, we obtain
\begin{align}
    [\overline{D^\alpha}, \overline{D_{y_j}^2}] &= \oP \left( - [D_{y_j}^2, P] [D^\alpha,P] + [D^\alpha, P]  [D_{y_j}^2,P] \right) = \oP [[D^\alpha, P],[D_{y_j}^2,P] ] \oP,
\end{align}
so that returning to \eqref{[oDalpha,oH]_2.5} we obtain \eqref{[oDalpha,oH]_6}.
\end{proof}

\subsection{Commutators involving $H_{bo}$, $E$, $P$, $T$ and derivatives}

\begin{lemma} \label{commutatorwithT_lemma}
Let Assumptions \ref{assumption1}-\ref{assumption4} hold. Then, for any multi-index $\nrm{\alpha} \leqslant k$, where $k$ is from \ref{assumption1}, \\
a) the commutators $[D^\alpha, H_{bo}]$, $[D^\alpha, E]$, and $[D^\alpha, P]$ are $O_{\cL_{s+\nrm{\alpha}-1,s}}(\ka)$. The same holds if we replace $H_{bo}$, $E$ and $P$ with higher-order derivatives $\partial^{\alpha_1} H_{bo}$, $\partial^{\alpha_2} E$, and $\partial^{\alpha_3} P$ respectively. \\
b) $[D^\alpha, \oR] = O_{\cL_{s+\nrm{\alpha}-1,s}}(\ka)$, \\
c) $[H_{bo}, [T, P]] = O_{\cL_{s+1,s}}(\ka)$, \\
d) $[D^\alpha, [T, P]] = O_{\cL_{s+\nrm{\alpha},s}}(\ka^2)$.
\end{lemma}

\begin{proof}[Proof of Lemma \ref{commutatorwithT_lemma}]
a) Recall from Lemma \ref{smoothness_proposition} that $E(y)$ and $P(y)$ are regular, and $H(y)$ is regular by assumption \ref{assumption1}. Consider first the commutator $[D^\alpha, H_{bo}]$. We write 
\begin{equation} \label{[Hbo,T]_0}
    \norm{[D^\alpha, H_{bo}] \phi}_{H^s_{\ka ,y} \cH_{el}}^2 = \norm{[D^\alpha, H_{bo}] \phi}^2 + \sum_{\nrm{\beta} = s} \norm{D^\beta [D^\alpha, H_{bo}] \phi}^2.
\end{equation}
We compute using the Leibniz Rule Lemma \ref{LeibnizRule_lemma},
\begin{align}
[D^\alpha, H_{bo}] &= \sum_{\substack{0 \leqslant \gamma \leqslant \alpha \\ \gamma \neq \alpha}} {{\alpha}\choose{\alpha - \gamma}} (-i\ka)^{\nrm{\alpha - \gamma}} \int^{\oplus}_{\R^m}  (\partial_y^{\alpha - \gamma} H(y)) dy D_{y}^\gamma = O_{\cL_{\nrm{\alpha}-1,0}}(\ka), \label{[Hbo,T]_1}
\end{align}
which estimates the first term on the right-hand side of \eqref{[Hbo,T]_0}. For the second term, consider 
\begin{equation}
    D^\beta [D^\alpha, H_{bo}] = [D^\beta, [D^\alpha, H_{bo}]] + [D^\alpha, H_{bo}] D^\beta 
\end{equation}
and use \eqref{[Hbo,T]_1} to obtain
\begin{equation} \label{[Hbo,T]_2}
    D^\beta [D^\alpha, H_{bo}] = [D^\beta, [D^\alpha, H_{bo}]] + O_{\cL_{s+ \nrm{\alpha} - 1,0}}(\ka). 
\end{equation}
For the remaining term on the right-hand side of \eqref{[Hbo,T]_2}, we have by \eqref{[Hbo,T]_1} and $[D^\beta, D^\gamma] = 0$ that
\begin{equation}
    [D^\beta, [D^\alpha, H_{bo}]] = \sum_{\substack{0 \leqslant \gamma \leqslant \alpha \\ \gamma \neq \alpha}} {{\alpha}\choose{\alpha - \gamma}} (-i\ka)^{\nrm{\alpha - \gamma}} [D^\beta, (\partial_y^{\alpha - \gamma} H_{bo})] D^\gamma.
\end{equation}
Applying Lemma \ref{LeibnizRule_lemma} to expand $[D^\beta, (\partial_y^{\alpha - \gamma} H_{bo})]$, we see that
\begin{align}
    [D^\beta, (\partial_y^{\alpha - \gamma} H_{bo})] = \sum_{\substack{0 \leqslant \sigma \leqslant \beta \\ \sigma \neq \beta}} {{\beta}\choose{\beta - \sigma}} (-i\ka)^{\nrm{\beta - \sigma}} (\partial_y^{\alpha + \beta - \gamma - \sigma} H_{bo}) D^\sigma
\end{align}
which implies
\begin{align}
    [D^\beta, [D^\alpha, H_{bo}]] &= \sum_{\substack{0 \leqslant \gamma \leqslant \alpha \\ \gamma \neq \alpha}} \sum_{\substack{0 \leqslant \sigma \leqslant \beta \\ \sigma \neq \beta}} {{\alpha}\choose{\alpha - \gamma}} {{\beta}\choose{\beta - \sigma}} (-i\ka)^{\nrm{\alpha - \gamma} + \nrm{\beta - \sigma}} (\partial_y^{\alpha + \beta - \gamma - \sigma} H_{bo}) D^{\gamma + \sigma} \nonumber \\
    &= O_{\cL_{s+\nrm{\alpha} - 2,0}}(\ka^2) \label{[Hbo,T]_3}
\end{align}
Using the bound $\norm{f}_{H^{s+\nrm{\alpha} - 2,\ka}_y} \lesssim \norm{f}_{H^{s+ \nrm{\alpha} - 1,\ka}_y}$, and combining \eqref{[Hbo,T]_2} and \eqref{[Hbo,T]_3}, we obtain
\begin{equation} \label{[Hbo,T]_4}
    D^s [H_{bo}, T] = O_{\cL_{s+\nrm{\alpha} - 2,0}}(\ka^2) + O_{\cL_{s+ \nrm{\alpha} - 1,0}}(\ka) = O_{\cL_{s+ \nrm{\alpha} - 1,0}}(\ka). 
\end{equation}
Using both estimates \eqref{[Hbo,T]_1} and \eqref{[Hbo,T]_4} on the right-hand side of \eqref{[Hbo,T]_0} we see that 
\begin{equation} 
    \norm{[H_{bo}, T] \phi}_{H^s_{\ka ,y}}^2 \lesssim \ka^2 \norm{\phi}_{H^{\nrm{\alpha} - 1,\ka}_y}^2 + \ka^2 \norm{\phi}_{H^{s+ \nrm{\alpha} - 1,\ka}_y}^2 \lesssim \ka^2 \norm{\phi}_{H^{s+ \nrm{\alpha} - 1,\ka}_y}^2,
\end{equation}
proving the claim. The estimates for the commutators $[D^\alpha, E]$, $[D^\alpha, P]$, and the higher order derivatives follow analogously, replacing $H(y)$ with $E(y)$ or $P(y)$ or higher order derivatives where appropriate.

b) We consider $[D^\alpha, \oR]$. Using $1 = P + \oP,$ we expand the commutator $[D^\alpha, \oR]$ to write
\begin{align}
    [D^\alpha, \oR] = D^\alpha \oR - \oR D^\alpha = [\overline{D^\alpha}, \oR] + P D^\alpha \oR - \oR D^\alpha P, \label{[Dyalpha,oR]_1}
\end{align}
where $\overline{D^\alpha} := \oP D^\alpha \oP$. By part a), $P D^\alpha \oP = [P,D^\alpha]\oP = O_{\cL_{s+\nrm{\alpha} - 1,s}}(\ka)$, and from Lemma \ref{(Hbar - E)inverselemma} we have $\oR = O_{\cL_{s,s}}(\ka)$. Then, 
\begin{equation}
    P D^\alpha \oR = O_{\cL_{s+\nrm{\alpha} - 1,s}}(\ka) = \oR D^\alpha P. \label{[Dyalpha,oR]_2}
\end{equation}
Next, we note that, since $ \oP (H_{bo} - E)  \oR = \oP = \oR (H_{bo} - E) \oP$, we have
\begin{equation}
    [\overline{D^\alpha}, \oR] = \oR [\oH_{bo} - \oE, \overline{D^\alpha}] \oR.
\end{equation}
Now, using that $\oR \: \oP = \oR = \oP \: \oR$ and $[H_{bo} - E, \oP] = 0$, we obtain
\begin{align}
    [\overline{D^\alpha}, \oR] = \oR \left((H_{bo}- E) \oP D^\alpha - D^\alpha \oP (H_{bo} - E) \right) \oR
    &= \oR \left((H_{bo}- E) D^\alpha - D^\alpha (H_{bo} - E) \right) \oR \nonumber \\
    &= \oR \left( [H_{bo} - E, D^\alpha] \right) \oR.
\end{align}
By part a), $[H_{bo}, D^\alpha]$ and $[E,D^\alpha]$ are $O_{\cL_{s+\nrm{\alpha} - 1,s}}(\ka)$. Hence,
\begin{equation}
    [\overline{D^\alpha}, \oR] = O_{\cL_{s+\nrm{\alpha} - 1,s}}(\ka). \label{[Dyalpha,oR]_3}
\end{equation}
Combining \eqref{[Dyalpha,oR]_1}, \eqref{[Dyalpha,oR]_2}, and \eqref{[Dyalpha,oR]_3} we obtain $[D^\alpha, \oR]$ is $O_{\cL_{s+\nrm{\alpha}-1,s}}(\ka)$, as desired.

c) Now consider $[[T, P], H_{bo}]$. While the operators $[T,P]$ and $[T, H_{bo}]$ are $O_{\cL_{s+1,s}}(\ka)$, it is not immediately clear whether the commutator of the coefficients of the differential operators in $[T, P]$ and $H_{bo}$ are bounded. Writing $T = D^2$, then 
\begin{equation} \label{[T,P]_1}
    [T, P] = [D^2, P] = -i\ka D (\nabla_y P) - i \ka (\nabla_y P) D
\end{equation}
so that
\begin{equation} \label{Dy(nablaP)_Hbo_commutator0}
    [[T, P], H_{bo}] = -i\ka [D (\nabla_y P), H_{bo}] - i \ka [(\nabla_y P) D, H_{bo}].
\end{equation}
Let us consider the first commutator on the right-hand side of \eqref{Dy(nablaP)_Hbo_commutator0}, proceeding analogously for the second. We have
\begin{equation} \label{Dy(nablaP)_Hbo_commutator1}
    [D (\nabla_y P), H_{bo}] = [D, H_{bo}] (\nabla_y P) + D [(\nabla_y P), H_{bo}]. 
\end{equation}
Note that $\nabla_y P$ is no longer an eigenprojection of $H_{bo}$, and $[(\nabla_y P), H_{bo}] \neq 0$. However, using the product rule we have
\begin{align}
    [(\nabla_y P), H_{bo}] &= (\nabla_y P)H_{bo} - H_{bo} (\nabla_y P) \nonumber \\
    &= \nabla_y (P H_{bo}) - P(\nabla_y H_{bo}) - \nabla_y (H_{bo} P) + (\nabla_y H_{bo}) P.
\end{align}
Since $H_{bo} P = P H_{bo}$, then $\nabla_y (P H_{bo}) - \nabla_y (H_{bo} P) = 0$, and thus
\begin{equation}
    [(\nabla_y P), H_{bo}] = [(\nabla_y H_{bo}), P],
\end{equation}
which is the commutator of two bounded operators and thus is bounded. Moreover, it is easy to see that it is in fact $O_{\cL_{s,s}}(1)$, so that the right-hand side of \eqref{Dy(nablaP)_Hbo_commutator1} is $O_{\cL_{s+1,s}}(\ka)$. Arguing analogously for the second commutator on the right-hand side of \eqref{Dy(nablaP)_Hbo_commutator0}, we conclude that $[[T,P], H_{bo}]$ is $O_{\cL_{s+1,s}}(\ka)$.

d) Now we consider $[D^\alpha, [T,P]]$. Writing $T = D^2$ and using \eqref{[T,P]_1} and $[D^\alpha, D] = 0$ we write
\begin{equation} \label{Dyalpha[T,P]_1}
    [D^\alpha, [T,P]] = -i\ka D [D^\alpha, (\nabla_y P)] - i\ka [D^\alpha, (\nabla_y P)] D.
\end{equation}
Arguing analogously to part a), $[D^\alpha, (\nabla_y P)] = O_{\cL_{s + \nrm{\alpha}-1,s}}(\ka)$, so the right-hand side of \eqref{Dyalpha[T,P]_1} implies $[D^\alpha, [T,P]] = O_{\cL_{s + \nrm{\alpha},s}}(\ka^2)$, as desired.
\end{proof}

The next result is crucial for the proof of Theorem \ref{MAINTHEOREM_ABSTRACT-PREVIOUS} part (b), and in particular Proposition \ref{Xj_recursive_estimates_proposition} in which we estimate the operators $X_j$ defined in the recursive relations \eqref{Xj_recursiverelation}. Recall that $K = E + T$.

\begin{lemma} \label{C1_C2_estimates_lemma}
Let Assumptions \ref{assumption1} - \ref{assumption4} hold. Then, 
\begin{align}
    &S \oP [T, P] P + \oR [[T, P], [K, P] - K] P = O_{\cL_{s+2,s}}(\ka^2). \label{C2_estimate_inlemma}
\end{align}
\end{lemma}

\begin{proof}[Proof of Lemma \ref{C1_C2_estimates_lemma}]
We first note that by Lemma \ref{commutatorwithT_lemma} part a), 
\begin{equation}
    \oP [T, P] P = O_{\cL_{s+1,s}}(\ka). \label{C1_estimate_inlemma}
\end{equation}
To show \eqref{C2_estimate_inlemma}, we study the two terms on the right-hand side individually. For the first term, $S \oP [T, P] P$, we recall that in \eqref{S_and_X2_estimates}, $S = O_{\cL_{s+1,s}}(\ka)$, so that in combination with \eqref{C1_estimate_inlemma} we obtain
\begin{equation}
    S \oP [T, P] P = O_{\cL_{s+2,s}}(\ka^2).
\end{equation}
For the second term, $\oR [[T, P], [K, P] - K] P$, using $\oR = O_{\cL_{s,s}}(1)$ from Lemma \ref{(Hbar - E)inverselemma} it suffices to estimate the commutator $[[T, P], [K, P] - K]$. Using $K = T + E$ and $[E, P] = 0$, then $[K, P] = [T, P]$ and we write 
\begin{align}
    \left[[T, P], [K, P] - K\right] &= \left[[T,P], [T, P] - T - E\right] = - \left[[T, P], T\right] - \left[[T, P], E\right]. \label{[K,[T,P]]_1}
\end{align}
$[[T, P], T] = O_{\cL_{s+2,s}}(\ka^2)$ follows from Lemma \ref{commutatorwithT_lemma} part d). A similar argument demonstrates $\left[[T, P], E\right] = O_{\cL_{s+2,s}}(\ka^2)$, as follows. Writing $T = D^2$, then
\begin{equation}
    [T,P] = [D^2, P] = [D,P] D + D[D,P] \label{[K,[T,P]]_2}
\end{equation}
where note $[D,P] = O(\ka)$. Using furthermore that $[E, [D,P]] = -i\ka [E, (\nabla_y P)] = i \ka [P, (\nabla_y E)] = 0$, as $\nabla_y E$ is multiplication by the function $(\nabla_y E)(y)$ (see Lemma \ref{[Fibered,E]_lemma}), we have
\begin{align}
    [E,[T,P]] &= [E, [D,P]] D + [D,P][E, D] + [E,D] [D,P] + D [E,[D,P]] \nonumber \\
    &= [D,P] [E, D] + [E, D] [D,P].
\end{align}
Since also $[E, D] = i \ka (\nabla_y E) = O(\ka),$ then it follows that 
\begin{equation}
    [E,[T,P]] = O(\ka^2). \label{[K,[T,P]]_3}
\end{equation}

Thus the right-hand side of \eqref{[K,[T,P]]_1} is $O_{\cL_{s+2,s}}(\ka^2)$, which implies that $\oR [[T, P], [K, P] - K] P$ is $O_{\cL_{s+2,s}}(\ka^2)$, so the estimate \eqref{C2_estimate_inlemma} follows.
\end{proof}

\section{Estimates on propagators} 
\label{Section_estimatesonPropagators}

In this section we collect the estimates on the propagators
\begin{equation}
    U_t = e^{- i H t/\ka}, \quad \oU_t = e^{-i \oH t/\ka}, \quad U^P_t = e^{-i P H_\ka P t/\ka}. 
\end{equation}
By standard results and Proposition \ref{Hbarselfadjointlemma}, the generators are all self-adjoint on the spaces $\cH = L^2_y \cH_{el}, \  \Ran \oP,$ and $\Ran P$ respectively. 

We establish the existence of the semigroup $U_t$ on the Hilbert spaces $H^s_{\ka ,y} \cH_{el}$, equipped with the tensor product norm of the Sobolev norm given in \eqref{Hsk_y_def} and the norm $\norm{\cdot}_{\cH_{el}}$ of $\cH_{el}$. In a similar manner we consider $\oU_t$ on the closed subspace $\oP H^s_{\ka ,y} \cH_{el}$. 

\begin{theorem} \label{Ut_Ubart_estimate_theorem}
Let Assumptions \ref{assumption1} - \ref{assumption4} hold. Then, $U_t$ is a  $C_0$-semigroup on $H^s_{\ka ,y} \cH_{el}$. Furthermore, for all $t \in \mathbb{R}$ and integers $0 \leqslant s \leqslant k$, where $k$ is given in \ref{assumption1}, there exists a constant $C > 0$ such that 
\begin{align} 
    \norm{U_t \phi}_{H^s_{\ka ,y} \cH_{el}} \leqslant C \lan t \ran^s \norm{\phi}_{H^s_{\ka ,y} \cH_{el}}. \label{Ut_estimate}
\end{align}
Furthermore, for $\phi \in \rRan \oP$, then for every $s$ there exists a constant $C_s > 0$ such that 
\begin{equation} \label{Ubart_estimate}
    \norm{\oU_t \phi}_{H^s_{\ka ,y} \cH_{el}} \leqslant C_s e^{C_s \nrm{t}} \norm{\phi}_{H^s_{\ka ,y} \cH_{el}}.
\end{equation}
\end{theorem}




\begin{proof}[Proof of Theorem \ref{Ut_Ubart_estimate_theorem}]

To establish Theorem \ref{Ut_Ubart_estimate_theorem}, we 
will make use of the Lumer-Phillips theorem (see e.g. Theorem 4.3 \cite{Pazy}), which amounts at finding some constants $C, \overline{C} > 0$ such that the operators 
\begin{align}
H_{\ka,C} := \frac{i}{\ka} H_{\ka} - C, &\qquad \cD(H_{\ka,C}) = \cD(H_{\ka}), \\
H_{\overline{C}} := \frac{i}{\ka} \oH - \overline{C}, &\qquad \cD(H_{\overline{C}}) = \cD(\oH),
\end{align} are dissipative, in the sense that
\begin{equation} \label{Hka_oH_dissipative}
    \rRe \langle (\frac{i}{\ka} H_{\ka} - C) \phi, \phi \rangle_{H^s_{\ka ,y} \cH_{el}} \leqslant 0 \:\text{ and }\: \rRe \langle (\frac{i}{\ka} \oH - \overline{C}) \phi, \phi \rangle_{H^s_{\ka ,y} \cH_{el}} \leqslant 0, \qquad \forall \phi \in H^s_{\ka ,y} \cH_{el}.
\end{equation}

\textit{Step 1. $H_{\ka,C} $ is dissipative for some constant $C> 0$.} We aim to establish the first inequality in \eqref{Hka_oH_dissipative} for a suitable constant $C>0$ to be chosen later on. By the definition of the tensor inner product and using that $D_y^{\alpha} (\frac{i}{\ka} H_\ka) = [D_y^{\alpha}, \frac{i}{\ka} H_\ka] + \frac{i}{\ka} H_\ka D_y^\alpha$ we find  
\begin{align}
    \langle \frac{i}{\ka} H_\ka \phi, \phi \rangle_{H^s_{\ka ,y} \cH_{el}} & = \sum_{\nrm{\alpha} = s} \langle D_y^{\alpha} (\frac{i}{\ka} H_\ka) \phi, D_y^{\alpha} \phi \rangle + \langle \frac{i}{\ka} H_{\ka} \phi, \phi \rangle \nonumber \\  
&= \sum_{\nrm{\alpha} = s} \langle [D_y^{\alpha}, \frac{i}{\ka} H_\ka] \phi, D_y^{\alpha} \phi \rangle
    + \langle (\frac{i}{\ka} H_\ka) D_y^{\alpha} \phi, D_y^{\alpha} \phi \rangle  + \langle \frac{i}{\ka} H_{\ka} \phi, \phi \rangle,
\end{align}  for every $\phi \in H^s_{\ka,y}\cH_{el}$. Using that $H_\ka$ is self-adjoint, taking the real part yields 
\begin{equation}
    \rRe \langle \frac{i}{\ka} H_\ka \phi, \phi \rangle_{H^s_{\ka ,y} \cH_{el}} = \sum_{\nrm{\alpha} = s} \rRe \langle [D_y^{\alpha}, \frac{i}{\ka} H_\ka] \phi, D_y^{\alpha} \phi \rangle.
\end{equation}
Now, by Lemma \ref{UtUbartcommutator_extend_bounded_lemma},     $[D_y^{\alpha}, \frac{i}{\ka} H_\ka] = O_{\cL_{s-1, 0}}$, and hence from Cauchy-Schwarz, 
\begin{equation} \label{Hka_dissipative_1}
    \rRe \langle \frac{i}{\ka} H_\ka \phi, \phi \rangle_{H^s_{\ka ,y} \cH_{el}} \leqslant C \norm{\phi}^2_{H^s_{\ka ,y} \cH_{el}},
\end{equation}
for some constant $C > 0$. Thus, $H_{\ka,C} = \frac{i}{\ka} H_{\ka} - C$ is dissipative and by the Lumer-Phillips it generates a $C_0$-semigroup on $H^s_{\ka ,y} \cH_{el}$. By classical arguments, this semigroup must coincide with $U_t \big|_{H^s_{\ka ,y} \cH_{el}}$ and in particular, we have the estimate
\begin{equation} \label{Ut_estimate_preliminary}
    \norm{U_t \phi}_{H^s_{\ka ,y} \cH_{el}} \lesssim e^{Ct} \norm{\phi}_{H^s_{\ka ,y} \cH_{el}}.
\end{equation}


\vspace{0.5em}
\par
\textit{Step 2. Proof of the polynomial estimate (\ref{Ut_estimate}).} To show \eqref{Ut_estimate}, we use the strong $C^1$ property of the map $t \mapsto D(t) := U_{-t} D^{\alpha} U_t$, for $|\alpha| = s$. Arguing as in \cite[Proposition 3.1]{Arbunich2021}, we may write 
\begin{equation} \label{Ut_C1_FTC}
    U_{-t} D^\alpha U_t \phi = D^\alpha \phi + U_{-t} [D^\alpha, U_t] \phi =  D^\alpha \phi + \int_0^t U_{-s} [\frac{i}{\ka} H_\ka, D^\alpha] U_s \phi ds.
\end{equation} Then, we shall use Lemma \ref{UtUbartcommutator_extend_bounded_lemma} to show that the unbounded commutator in the integrand of the right-hand side is $O_{\cL_{s-1, 0}}(1)$, which gives \eqref{Ut_estimate}.

We prove it by induction on $s \geqslant 1$ (when $s=0$ the estimate follows immediately by the unitary property of $U_t$). For the case $s=1$, commuting $D_{y_j} U_t = U_t D_{y_j} + [D_{y_j}, U_t]$, we have
\begin{align}
    \norm{U_t \phi}_{H^{1,\ka}_y \cH_{el}} &\leqslant \sum_{j=1}^m \norm{D_{y_j} U_t \phi} + \norm{U_t \phi} \nonumber \\
    &\leqslant \sum_{j=1}^m \norm{D_{y_j} \phi} + \norm{[D_{y_j}, U_t] \phi} + \norm{\phi}  \lesssim \norm{\phi}_{H^{1,\ka}_y \cH_{el}} + \sum_{j=1}^m \norm{[D_{y_j}, U_t] \phi}^2. \label{Dyj_e-iHt_1}
\end{align}
By Lemma \ref{UtUbartcommutator_extend_bounded_lemma},
\begin{equation} \label{Dyj_e-iHt_2}
    [D_{y_j}, U_t] = \int_0^t U_{t-r} [D_{y_j}, \frac{-i}{\ka} H] U_r dr = O(\nrm{t}),
\end{equation}
which combined with \eqref{Dyj_e-iHt_1} completes the proof of the base case. Now assume the estimate \eqref{Ut_estimate} holds for $s = n$ and we seek to show the estimate for $s = n+1$. Note that by the assumption \ref{assumption2}, it must be that $n+1 \leqslant k$. Analogous to \eqref{Dyj_e-iHt_1}, we have 
\begin{align}
    \norm{U_t \phi}_{H^{n+1,\ka}_y \cH_{el}} &\leqslant \sum_{\nrm{\alpha} = n+1} \norm{D_{y}^{\alpha} U_t \phi} + \norm{U_t \phi} \nonumber  \\
    &\leqslant \sum_{\nrm{\alpha} = n+1} \norm{D_{y}^{\alpha} \phi} + \norm{[D_{y}^{\alpha}, U_t] \phi} + \norm{\phi} \lesssim \norm{\phi}_{H^{n+1,\ka}_y \cH_{el}} + \sum_{\nrm{\alpha} = n+1} \norm{[D_{y}^\alpha, U_t] \phi}. \label{Dyj_e-iHt_3} 
\end{align}
By Lemma \ref{UtUbartcommutator_extend_bounded_lemma},
\begin{equation} 
    [D^\alpha, U_t] = \int_0^t U_{t-r} [D^\alpha, \frac{-i}{\ka} H] U_r dr =  \int_0^t U_{t-r} O_{\cL_{n, 0}}(1) U_r dr.
\end{equation}
Using the inductive assumption, $O_{\cL_{n, 0}}(1) U_r = O_{\cL_{n, 0}}((1+\nrm{r})^n)$, and integrating this estimate in $r$ results in
\begin{equation} \label{Dyj_e-iHt_4}
    [D^\alpha, U_t] = O_{\cL_{n, 0}}((1+\nrm{t})^{n+1}).
\end{equation}
Combining \eqref{Dyj_e-iHt_3} with \eqref{Dyj_e-iHt_4} and the bound $\norm{\phi}_{H^{n,\ka}_y \cH_{el}} \lesssim \norm{\phi}_{H^{n+1,\ka}_y \cH_{el}}$ yields the desired estimate for $s=n+1$, so the proof by induction is complete.

\vspace{0.5em}
\par
\textit{Step 3. $H_{\overline{C}} $ is dissipative for some constant $\overline{C}> 0$.} The proof is identical to Step 1, so we underline only the significant modifications if $\phi \in \oP H^s_{\ka ,y} \cH_{el}$. Using the self-adjointness of $\oH$ (see Lemma \ref{Hbarselfadjointlemma}),  
\begin{equation}
    \rRe \langle \frac{i}{\ka} \oH \phi, \phi \rangle_{H^s_{\ka ,y} \cH_{el}} = \sum_{\nrm{\alpha} = s} \rRe \langle D_y^{\alpha} (\frac{i}{\ka}\oH) \phi, D_y^{\alpha} \phi \rangle, \label{Hbar_dissipative_0}
\end{equation} as $ \rRe \langle \frac{i}{\ka} \oH \phi, \phi \rangle = 0$.  Now, as $\phi \in \rRan \oP$ and $1 = P + \oP$, using that $ P \oH = 0 $ and $P \phi = 0$ we may write 
\begin{align}
    \langle D_y^{\alpha} (\frac{i}{\ka}\oH) \phi, D_y^{\alpha} \phi \rangle & = \langle D_y^{\alpha} (P + \oP) (\frac{i}{\ka}\oH) \phi, D_y^{\alpha} (P + \oP) \phi \rangle = \langle D_y^{\alpha} \oP (\frac{i}{\ka}\oH) \phi, D_y^{\alpha} \oP \phi \rangle.
\end{align} Next, expanding the commutator  
\begin{equation}
\overline{D_y^{\alpha}} (\frac{i}{\ka}\oH) = [\overline{D_y^{\alpha}}, (\frac{i}{\ka}\oH)] + (\frac{i}{\ka}\oH) \overline{D_y^{\alpha}},
\end{equation} and using that $\rRe \langle (\frac{i}{\ka}\oH) \overline{D_y^{\alpha}} \phi, \overline{D_y^{\alpha}} \phi \rangle = 0$, taking the real part we have  
\begin{align}
   \rRe \langle D_y^{\alpha} (\frac{i}{\ka}\oH) \phi, D_y^{\alpha} \phi \rangle  
    &= \rRe \langle \overline{D_y^{\alpha}} (\frac{i}{\ka}\oH) \phi, \overline{D_y^{\alpha}} \phi \rangle + \rRe \langle P D_y^{\alpha} \oP (\frac{i}{\ka}\oH) \phi, P D_y^{\alpha} \oP \phi \rangle.
\end{align} Hence, summing over $\alpha$ in the range $\vert \alpha \vert = s$, (\ref{Hbar_dissipative_0}) can be rephrased as 
\begin{equation}
    \rRe \langle \frac{i}{\ka} \oH \phi, \phi \rangle_{H^s_{\ka ,y} \cH_{el}} = \sum_{\nrm{\alpha} = s} \rRe \left( I_1^\alpha + I_2^\alpha \right) , \label{Hbar_dissipative_decomposition}
\end{equation} where for every $\vert \alpha \vert = s$,
\begin{equation}
    I_1^\alpha := \langle \overline{D_y^{\alpha}} (\frac{i}{\ka}\oH) \phi, \overline{D_y^{\alpha}} \phi \rangle, \qquad I_2^\alpha :=   \langle P D_y^{\alpha} \oP (\frac{i}{\ka}\oH) \phi, P D_y^{\alpha} \oP \phi \rangle. 
\end{equation} We shall next fix $\alpha$ with $\vert \alpha \vert = s$ and examine each term separately. 

\par
\vspace{0.3em}

Consider the term $I_1^\alpha$. By Lemma \ref{UtUbartcommutator_extend_bounded_lemma}, the commutator $[\overline{D_y^\alpha}, \frac{i}{\ka} \oH]$ is $O_{\cL_{s, 0}}(1)$, and thus by Cauchy-Schwarz and the bound $\norm{\phi}_{H^{s-1, \ka}_y \cH_{el}} \lesssim \norm{\phi}_{H^s_{\ka, y} \cH_{el}}$, 
\begin{equation}
    \rRe I_1^\alpha = \rRe \langle [\overline{D_y^{\alpha}}, (\frac{i}{\ka}\oH)] \phi, \overline{D_y^{\alpha}} \phi \rangle \leqslant C \norm{\phi}_{H^{s-1, \ka}_y \cH_{el}} \norm{\phi}_{H^s_{\ka, y} \cH_{el}} \leqslant C \norm{\phi}_{H^s_{\ka, y} \cH_{el}}^2. \label{Hbar_dissipative_2}
\end{equation}

\par
\vspace{0.3em} Now we consider the term $I_2^\alpha$. Owing to the particular structure of this term, we aim to manipulate both sides of the scalar product in $I_2^\alpha$ in order to make appear a suitable commutator that we know how to bound.  
Writing $\oH = \oH_{bo} + \oT$, where $T = - \ka^2 \Delta_y$, and using the commutator $D_y^\alpha H_{bo} = [D_y^\alpha, H_{bo}] + H_{bo} D_y^\alpha$ we have 
\begin{align}
    P D_y^{\alpha} \oP (\frac{i}{\ka}\oH) & = \frac{i}{\ka} P D_y^\alpha \oP H_{bo} \oP + \frac{i}{\ka} P D_y^\alpha \oP T \oP =  \frac{i}{\ka} P [D_y^\alpha, H_{bo}] \oP + \frac{i}{\ka} PH_{bo} D_y^\alpha \oP + P D_y^\alpha \oP T \oP,
\end{align} as $[H_{bo}, \oP] = 0$. Using also $[H_{bo}, P] = 0$, this becomes
\begin{equation}
    P D_y^\alpha \oP H_{bo} \oP =  P D_y^\alpha H_{bo} \oP =  P [D_y^\alpha, H_{bo}] \oP + H_{bo} P D_y^\alpha \oP,
\end{equation} which combined with the previous identity implies 
\begin{equation}
    P D_y^{\alpha} \oP (\frac{i}{\ka}\oH) = \frac{i}{\ka} H_{bo} P D_y^\alpha \oP + \frac{i}{\ka} P [D_y^\alpha, H_{bo}] \oP + \frac{i}{\ka} P D_y^\alpha \oP T \oP. 
\end{equation} Now, injecting this in the definition of $I_2^\alpha$ we find 
\begin{equation}
    \rRe I_2^\alpha = \rRe \langle \frac{i}{\ka} P [D_y^{\alpha}, H_{bo}] \oP \phi, P D_y^{\alpha} \oP \phi \rangle  + \rRe \langle \frac{i}{\ka} P D_y^{\alpha} \oP T \oP \phi, P D_y^{\alpha} \oP \phi \rangle, \label{Hbar_dissipative_6}
\end{equation} where we have used that $ \rRe \langle \frac{i}{\ka} H_{bo} P D_y^\alpha \oP \phi, P D_y^\alpha \oP \phi \rangle = 0$ is purely imaginary by the self-adjointness of $H_{bo}$.

We expand next the commutator in the first term of (\ref{Hbar_dissipative_6}). By the Leibniz Rule (cf. Lemma \ref{LeibnizRule_lemma}), 
\begin{align}
    P [D_y^{\alpha}, H_{bo}] \oP &= \sum_{\substack{0 \leqslant \beta \leqslant \alpha \\ \beta \neq \alpha}} {{\alpha}\choose{\alpha - \beta}} (-i\ka)^{\nrm{\alpha - \beta}} P (\partial_y^{\alpha - \beta} H_{bo}) D_{y}^\beta \oP = O_{\cL_{s-1, 0}}(\ka)
\end{align}
and
\begin{align}
    P D_y^\alpha \oP = [P, D_y^\alpha] \oP &= - \sum_{\substack{0 \leqslant \beta \leqslant \alpha \\ \beta \neq \alpha}} {{\alpha}\choose{\alpha - \beta}} (-i\ka)^{\nrm{\alpha - \beta}} (\partial_y^{\alpha - \beta} P) D_{y}^\beta \oP = O_{\cL_{s-1, 0}}(\ka).
\end{align}
Using these estimates, by Cauchy-Schwarz we estimate the first term on the right-hand side of \eqref{Hbar_dissipative_6} by 
\begin{equation}
    \rRe \langle \frac{i}{\ka} P [D_y^{\alpha}, H_{bo}] \oP \phi, P D_y^{\alpha} \oP \phi \rangle \leqslant C \ka \norm{\phi}_{H^{s-1,\ka}_y \cH_{el}}^2 \leqslant C \ka \norm{\phi}_{H^s_{\ka ,y} \cH_{el}}^2, \label{Hbar_dissipative_8}
\end{equation}
for some constant $C > 0$. Looking at the second term on the right-hand side of \eqref{Hbar_dissipative_6}, we observe that the left side of the inner product is $O_{\cL_{s+1, 0}}(1)$ (since $P D_y^\alpha \oP$ is $O_{\cL_{s-1,0}}(\ka)$, and $\oP T \oP$ is $O_{\cL_{2,0}}(\ka)$), while the right side is only $O_{\cL_{s-1,0}}(\ka)$. We make both sides $O_{\cL_{s,0}}$ by commuting one derivative operator $D_{y_l}$, for some index $l$, from $P D_y^\alpha \oP$ and transposing it to the right side. In this manner we are able to estimate
\begin{equation}
    \rRe \langle \frac{i}{\ka} P D_y^{\alpha} \oP T \oP \phi, P D_y^{\alpha} \oP \phi \rangle \leqslant C \ka \norm{\phi}_{H^s_{\ka ,y} \cH_{el}}^2, 
\end{equation}
which combined with \eqref{Hbar_dissipative_8} yields the estimate
\begin{equation}
    \rRe I_2^\alpha = \rRe \langle P D_y^{\alpha} \oP (\frac{i}{\ka}\oH) \phi, P D_y^{\alpha} \oP \phi \rangle \leqslant C \ka \norm{\phi}_{H^s_{\ka ,y} \cH_{el}}^2,
\end{equation}
for some constant $C > 0$. Combining in turn this estimate with \eqref{Hbar_dissipative_6}, \eqref{Hbar_dissipative_2} and \eqref{Hbar_dissipative_decomposition}, and summing over all $\nrm{\alpha} = s$ we find 
\begin{equation}
    \rRe \langle \frac{i}{\ka} \oH \phi, \phi \rangle_{H^s_{\ka ,y} \cH_{el}} \leqslant \overline{C} \norm{\phi}_{H^s_{\ka ,y} \cH_{el}}^2,
\end{equation} for some constant $\overline{C} > 0$. Then as desired, $\frac{i}{\ka} \oH - \overline{C}$ is dissipative, and the existence of $\oU_t \big|_{\oP H^s_{\ka ,y} \cH_{el}}$ and estimate \eqref{Ubart_estimate} follows. 
\end{proof}

The second result in this section concerns the estimate for $U^P_t$, which acts on the subspace $\Ran P$ that has additional structure which we can exploit. This avoids working with the generator $P H_\ka P$, running into the same problems as in the case of $\oP H_\ka \oP$, and instead we look at the generator $\heff = T + E + \ka^2 v$ which can be dealt with in the same manner as $H_\ka = T + \Hbo$. 

\begin{theorem} \label{theorem_UPt_estimate}
Let Assumptions \ref{assumption1} - \ref{assumption4} hold. For all $\Psi \in \Ran P$, 
\begin{equation} \label{UPt_e-ihefft/ka_relation_appendixversion}
    U^P_t \Psi(x,y) = \psi_\circ(x,y) e^{- i \heff t/\ka} f(y), 
\end{equation}
where $f(y) = \lan \psi_\circ(x,y), \Psi(x,y) \ran_{\cH_{el}}$ and $\heff = T + E + \ka^2 v$. Moreover, if $\Psi \in H^s_{\ka, y} \cH_{el}$, then $f \in H^s_{\ka,y}$ and 
\begin{equation}
    \norm{U^P_t \phi}_{H^s_{\ka ,y} \cH_{el}} \leqslant C \lan t \ran^s \norm{\phi}_{H^s_{\ka ,y} \cH_{el}}. \label{UPt_estimate}
\end{equation}
\end{theorem}

\begin{proof}[Proof of Theorem \ref{theorem_UPt_estimate}]
The relation \eqref{UPt_e-ihefft/ka_relation_appendixversion} was shown in \ref{lemma_PKP_and_UPt_estimate} part b). By the smoothness assumptions, $E + \ka^2 v$ is differentiable with bounded derivatives (see Proposition \ref{smoothness_proposition} and Lemma \ref{smoothness_psicirc_lemma}). Then it is easy to see that $\heff = T + E + \ka^2 v$ is dissipative on the Sobolev spaces $H^{s,\ka}_y$, analogous to the dissipativity of $H = T + \Hbo$ (where $\Hbo$ is differentiable with bounded derivatives) shown in Step 1 in the proof of Theorem \ref{Ut_Ubart_estimate_theorem}. Similarly, the polynomial estimate 
\begin{equation} \label{heff_semigroup_estimate}
    \norm{e^{- i \heff t/\ka} \phi}_{H^s_{\ka,y}} \lesssim \lan t \ran^s \norm{\phi}_{H^s_{\ka,y}}
\end{equation}
follows by repeating the Duhamel-type argument in Step 2. \eqref{heff_semigroup_estimate} and \eqref{UPt_e-ihefft/ka_relation_appendixversion} together imply \eqref{UPt_estimate}. 
\end{proof}

\printbibliography

\end{document}